\documentclass[a4paper]{article}
\pdfoutput=1
\usepackage{jheppub}

\usepackage{amsmath}

\usepackage{amssymb}
\usepackage{graphicx,color}

\usepackage{amsmath,amsthm}
\usepackage{txfonts}

\usepackage{mathrsfs}

\usepackage{bm}

\newcommand{\Ref}[1]{(\ref{#1})}

\newtheorem{Theorem}{Theorem}[section]
\newtheorem{Definition}{Definition}[section]
\newtheorem{Lemma}[Theorem]{Lemma}

\newcommand{\Z}{\mathbb{Z}}
\newcommand{\R}{\mathbb{R}}
\newcommand{\C}{\mathbb{C}}

\newcommand{\half}{\frac{1}{2}}

\newcommand{\ccirc}{\kern0.2ex\vcenter{\hbox{$\scriptstyle\circ$}}\kern0.2ex}

\newcommand{\bA}{{\bar{A}}}

\newcommand{\Su}{\mathrm{SU}(2)}

\def\be{\begin{eqnarray}}
\def\ee{\end{eqnarray}}

\newcommand{\cc}{\mathcal C}

\newcommand{\cf}{\mathcal F}
\newcommand{\cg}{\mathcal G}
\newcommand{\ch}{\mathcal H}

\newcommand{\ck}{\mathcal K}

\newcommand{\cn}{\mathcal N}

\newcommand{\cs}{\mathcal S}

\newcommand{\sm}{\mathscr{M}}

  \newcommand{\Fn}{\mathfrak{N}}

\renewcommand{\a}{\alpha}
\renewcommand{\b}{\beta}
\newcommand{\g}{\gamma}
\newcommand{\G}{\Gamma}

\newcommand{\eps}{\varepsilon}

\newcommand{\sig}{\sigma}
\newcommand{\Sig}{\Sigma}
\renewcommand{\l}{\lambda}

\newcommand{\rmd}{\mathrm d}

\newcommand{\lt}{\left}
\newcommand{\rt}{\right}

\newcommand{\lag}{\left\langle}
\newcommand{\rag}{\right\rangle}

\newcommand{\tr}{\mathrm{tr}}

\title{Semiclassical Behavior of Spinfoam Amplitude with Small Spins and Entanglement Entropy}

\author[1,2]{Muxin Han}  

\affiliation[1]{Department of Physics, Florida Atlantic University, 777 Glades Road, Boca Raton, FL 33431-0991, USA}

\affiliation[2]{Institut f\"ur Quantengravitation, Universit\"at Erlangen-N\"urnberg, Staudtstr. 7/B2, 91058 Erlangen, Germany}

\emailAdd{hanm(At)fau.edu}

\abstract{Spinfoam amplitudes with \emph{small} spins can have interesting semiclassical behavior and relate to semiclassical gravity and geometry in 4 dimensions. We study the generalized spinfoam model (Spinfoams for all loop quantum gravity (LQG) \cite{KKL,generalize}) with small spins $j$ but a large number of spin degrees of freedom (DOFs), and find that it relates to the simplicial Engle-Pereira-Rovelli-Livine-Freidel-Krasnov (EPRL-FK) model with large spins and Regge calculus by coarse-graining spin DOFs. Small-$j$ generalized spinfoam amplitudes can be employed to define semiclassical states in the LQG kinematical Hilbert space. Each of these semiclassical states is determined by a 4-dimensional Regge geometry. We compute the entanglement R\'enyi entropies of these semiclassical states. The entanglement entropy interestingly coarse-grains spin DOFs in the generalized spinfoam model, and satisfies an analog of the thermodynamical first law. This result possibly relates to the quantum black hole thermodynamics in \cite{GP2011}. 
}


\begin{document}

\maketitle

\section{Introduction}

Loop Quantum Gravity (LQG) is a candidate of non-perturbative and background-independent theory of quantum gravity. A covariant approach of LQG is developed by the spinfoam formulation, in which the quantity playing the central role is the \emph{spinfoam amplitude} \cite{rovelli2014covariant,Perez2012}. 4-dimensional spinfoam ampliutdes give transition amplitudes of boundary 3d quantum geometry states in LQG, and formulate the LQG version of quantum gravity path-integral. The spinfoam formulation is a successful program for demonstrating the semiclassical consistency of LQG. The recent progresses on the semiclassical analysis reveal that spinfoam amplitudes relate to the semiclassical Einstein gravity in the large spin regime, e.g. \cite{CFsemiclassical,semiclassical,HZ,lowE,propagator3,frankflat,Han:2018fmu,Liu:2018gfc}.

Although the analysis of large spin spinfoam ampitude has been fruitful for demonstrating the semiclassical behavior, there are good reasons to expect that some even more interesting semiclassical behavior of spinfoams, or in general LQG, should appear in the regime where spins are all small. There are 2 motivations for the semiclassical analysis in small spin regime: 

Firstly, recall that the large spin semiclassicality is motivated by requiring the geometrical surface area $\mathbf{a}_S$ to be semiclassical, i.e. $\mathbf{a}_S\gg \ell_P^2$ where $\ell_P$ is the Planck length. The requirement leads to the spin $j\gg 1$ provided the area spectrum $\mathbf{a}_S=8\pi\g\ell_P^2 \sqrt{j(j+1)}$, if we assume that there is only a single spin-network link colored by $j$ intersecting the surface $S$. Large-j is a sufficient condition for the semiclassical area but clearly not necessary. Indeed if we relax this assumption and allow more than one intersecting links $l$, the area spectrum may become $\mathbf{a}_S=8\pi\g\ell_P^2 \sum_{l=1}^N\sqrt{j_l(j_l+1)}$ which sums ``area elements'' $8\pi\g\ell_P^2 \sqrt{j_l(j_l+1)}$ at $l$. $N$ is the total number of intersecting links. Then the semiclassical surface area $\mathbf{a}_S\gg \ell_P^2$ can be achieved not only by large $j$ and small $N$, but also by small $j$ and large $N$. For instance, all $j=1/2$ and $N\gg1$ lead to $\mathbf{a}_S\gg \ell_P^2$. Therefore we anticipate that small spins (with large number of intersecting links) should also lead to semiclassical behaviors of LQG.

The second motivation comes from the statistical interpretation of black hole entropy in LQG: The black hole horizon with a fixed total area punctured by a large number of spin-network links $l$. The punctures are colored by spins $j_l$, each of which contribute area element $8\pi\g\ell_P^2 \sqrt{j_l(j_l+1)}$ to the horizon. The black hole entropy counts the total number of microstates which give the same total horizon area \cite{Perez:2017cmj,Agullo:2008yv,GP2011}. It turns out that the total number of states is dominant by states at punctures with small $j_l$, while the number of states decays exponentially as $j$ becomes large. The fact that small $j$s dominate the semiclassical horizon area and entropy suggests that small spins should play an important role in the semiclassical analysis of LQG.  

This work takes the first step to study systematically the semiclassical behavior of LQG in the small spin regime, in particular in the spinfoam formulation. From the above motivation, given a surface $S$ punctured by $N$ spin-network links, the semiclassical area of $S$ can be given not only by small $N$ and large $j$, but also by large $N$ and small $j$. Section \ref{STFQPAPR} generalizes this observation to quantum polyhedra represented by intertwiners (SU(2) invariant tensors) at spin-network nodes. We find among intertwiners with a fixed large rank $N\gg1$ (quantum polyhedra with $N$ facets $f$), there are a subclass of small-$j$ and large-$N$ coherent intertwiners $||\{j_f\},\{\xi_f\}\rangle_N$ $(f=1,\cdots,N)$ relating to the large-$J$ and rank-4 coherent intertwiner $||\{J_\Delta\},\{\xi_\Delta\}\rangle_4$ $(\Delta=1,\cdots,4)$ and having the semiclassical behavior as geometrical flat tetrahedra. $\Delta$ are 4 groups of intertwiner legs $f$, and every $\Delta$ contains a large number $N_\Delta\gg1$ of $f$'s. The subclass of coherent intertwiners exhibiting semiclassical behaviors are defined by the \emph{parallel restriction} on $\xi_f$'s
\be
\xi_f=\xi_{f'}\equiv\xi_\Delta\quad\text{up to a phase}\quad\forall\ f,f'\in\Delta,\label{PR0}
\ee 
i.e. $\xi_f,\xi_{f'}$ give the same unit 3-vector $\vec{n}_\Delta=\langle\xi_\Delta|\vec{\sig}|\xi_\Delta\rangle$ where $\vec{\sig}$ are Pauli matrices. Geometrical tetrahedra resulting from these intertwiners has face area proportional to $J_\Delta=\sum_{f\in\Delta} j_f$ and face normals $\vec{n}_\Delta$. $J_\Delta$ is large since $N_\Delta\gg1$ and $j_f\neq0$. This result has a simple geometrical picture: Given a classical flat tetrahedron, we may partition every face $\Delta$ into $N_\Delta$ facets $f$, while the face area sums the facet areas and the facet normals are parallel among facets in a $\Delta$. By partitioning tetrahedron faces into facets, the tetrahedron becomes a polyhedron with a total number of $N=\sum_{\Delta=1}^4N_\Delta$ facets, each of which has a small area (see FIG.\ref{tetra}(a)). The correspondence between polyhedra and intertwiners in LQG \cite{shape} relates $f$ to intertwiner legs (and tetrahedron faces $\Delta$ to 4 groups of intertwiner legs) and facet areas and normals to coherent intertwiner labels (see FIG.\ref{tetra}(b)). These parallel normals motivates the above parallel restriction. Beyond the semiclassical behaviors of these intertwiners, quantum corrections to semiclassical tetrahedron geometries are of $O(1/J_\Delta)=O(1/N_\Delta)$ thus is suppressed by large rank $N$ (or $N_\Delta$). The above result demonstrates that at the level of quantum polyhedra, we can trade small $j_f$ and large rank $N\gg1$ for large $J_\Delta$ and small rank $N=4$ to obtain the semiclassicality. 

Note that the above semiclassical result still holds if we replace the tetrahedron by polyhedra in case their numbers of faces $\Delta$ are still small. A similar idea as the above is applied in \cite{HanHung} to relate LQG states to holographic tensor networks, and relates to \cite{Bodendorfer:2018csn}.

Section \ref{vertex} generalizes the small-$j$ semiclassical analysis to the spinfoam vertex amplitude in 4 dimensions. The vertex amplitude $A_v$ is associated to a 4-dimensional cell $B_4$ whose boundary are closed and made by gluing 5 polyhedra $\a=1,\cdots 5$, each of which has a large number $N_\a$ of facets (see FIG.\ref{tetra}(c)). Every pair of polyhedra share a large number $N_\Delta$ of facets, where $\Delta=\a\cap\b$ is the face made by facets shared by 2 polyhedra $\a,\b$. Ignoring the fine partition of $\Delta$, $B_4$ relates to a 4-simplex where $\Delta$ relates to triangles of the 4-simplex. $A_v$ depends on the boundary data which contains small spins $j_f$ and 5 intertwiners $||\{j_f\},\{\xi_{\a f}\}\rangle_{N_\a}$ of quantum polyhedra. To be concrete, we consider $A_v$ to be the generalized spinfoam vertex \cite{KKL,generalize} (in Euclidean signature with $0<\g<1$) which admits non-simplicial cells. We writing $A_v$ in terms of coherent intertwiners and impose the parallel restriction Eq.\Ref{PR0} to boundary data $\xi_{\a f}$ with $f\in\Delta$. We find that up to an overall phase, $A_v$ with small $j_f$ and large $N_\Delta$ is identical to the Engle-Pereira-Rovelli-Livine-Freidel-Krasnov (EPRL-FK) vertex amplitude of 4-simplex with large spins $J_\Delta\gg1$, where 10 $\Delta$ become triangles of the 4-simplex and $J_\Delta=\sum_{f\in\Delta} j_f$ similar to the case of polyhedra. Due to large $J_\Delta$, the same asymptotic analysis as in \cite{semiclassicalEu} can be applied to $A_v$ and gives the following asymptotic formula relating to the 4-simplex Regge action $\sum_{\Delta}\g J_{\Delta}\Theta_{\Delta}$ (The triangle area $\mathbf{a}_\Delta=8\pi\g\ell_P^2 J_\Delta$)
\be
A_v=\lt(\text{overall phase}\rt)\lt(\frac{2\pi}{N}\rt)^{12}\Bigg[2\cn_{+-}^\g\cos\lt(\sum_{\Delta}\g J_{\Delta}\Theta_{\Delta}\rt)+\cn^\g_{++}e^{\sum_{\Delta} J_{\Delta}\Theta_{\Delta}}+\cn^\g_{--}e^{-\sum_{\Delta} J_{\Delta}\Theta_{\Delta}}\Bigg]\lt(1+O\lt(\frac{1}{N}\rt)\rt).\label{Avexp0000}
\ee
We refer the reader to \cite{semiclassicalEu} for expressions of $\cn_{+-}^\g,\cn_{++}^\g,\cn_{--}^\g$. The expansion parameter $N$ is the order of magnitude of $N_\Delta\sim J_\Delta$. 

Section \ref{complex} generalizes the discussion to spinfoam amplitude $A(\ck)$ on cellular complexes $\ck$ in 4 dimensions. The 4d cell of $\ck$ is $B_4$ to define vertex amplitudes $A_v$ as above. We again apply the generalized spinfoam formulation to define the amplitude on $\ck$. By the above relation between $B_4$ and 4-simplex, $\ck$ relates to a unique simplicial complex $\ck_s$, where decomposing triangles $\Delta\in \ck_s$ into facets $f$ gives $\ck$. In the above analysis of a single $A_v$, the parallel restriction can be applied since $\xi_{\a f}$ are boundary data. However for the spinfoam amplitude $A(\ck)$ we do need to consider internal $\xi_{\a f}$ beyond the parallel restriction since individual $\xi_{\a f}$'s are integrated independently in $A(\ck)$. We write the spinfoam amplitude as a sum over spins $A(\ck)=\sum_{\{j_f\}}A_{\{j_f\}}(\ck)$ and focus on $A_{\{j_f\}}(\ck)$ in Section \ref{complex}. $A_{\{j_f\}}(\ck)$ has the standard integral expression:
\be 
A_{\{j_f\}}(\ck)=\prod_f A_\Delta({j_f})\int[\rmd \xi_{\a f}\rmd g^\pm_{v\a}]\,e^{S},\quad
S=\sum_{\pm}\sum_{v,f}2j_{f}^\pm\ln\lag \xi_{\a f}\lt|g_{v\a}^\pm{}^{-1}g_{v\b}^\pm \rt|\xi_{\b f}\rag,\label{Ajf000}
\ee
where the face amplitude $A_\Delta({j_f})$ is $2j_f+1$ to certain power only depending on $\Delta$. It turns out that the stationary phase analysis can be still applied to $A_{\{j_f\}}(\ck)$ with small nonzero $j_f$ but large $N_\Delta$. It is clear from the discussion in the last paragraph that $A_{\{j_f\}}(\ck)$ reduces to the simplicial EPRL-FK spinfoam amplitude with large spins $J_\Delta=\sum_{f\in\Delta} j_f$ if we impose by hand the parallel restriction to internal $\xi_{\a f}$'s. We prove that all critical points of large $J_\Delta$ simplicial EPRL-FK amplitude give critical points of $A_{\{j_f\}}(\ck)$ if we relate the critical data by $J_\Delta=\sum_{f\in\Delta} j_f$, internal $\xi_{\a\Delta}=\xi_{\a f}$ (up to a phase), and identifying $g^\pm_{v\a}$ between simplicial EPRL-FK and $A_{\{j_f\}}(\ck)$. We denote these critical points by $(g_{v\a}^\pm,\xi_{\a\Delta})_c[J_\Delta]$. Some of these critical points relate to Regge geometries in 4 dimensions similar to the simplicial EPRL-FK amplitude \cite{Han:2018fmu,HZ1}. At these critical points, $J_\Delta$ is identified to be the area of the triangle $\Delta$. The application of critical points to the stationary phase analysis is discussed in Section \ref{beyond}.

The relation between the simplicial EPRL-FK amplitude and $A(\ck)$ suggests a new viewpoint that the EPRL-FK model with spins $J_\Delta$ can be an effective theory emergent from a more fundamental theory formulated by $A(\ck)$ with $j_f$. The EPRL-FK model is obtained from $A(\ck)$ by coarse-graining from $j_f$ to $J_\Delta$ and imposing the parallel restriction (more rigorously, the EPRL-FK model appears as a partial amplitude in $A(\ck)$ after integrating out the non-parallel $\xi_{\a f}$ as shown in Section \ref{beyond}). The EPRL-FK amplitude with given $J_\Delta$ is a collection of a large number of micro-degrees of freedom $\{j_f\}$ satisfying $J_\Delta=\sum_{f\in\Delta} j_f$ at all $\Delta$. Critical points from EPRL-FK model and Regge geometries are ``macrostates'' which contain $\{j_f\}$ as ``microstates''. This picture is interesting and turns out to be important in the computation of entanglement entropy.

Before the analysis of the full amplitude $A(\ck)$ in Section \ref{beyond}, Sections \ref{LQGstate}-\ref{EE} make a modification of the amplitude by imposing weakly the parallel restriction to internal $\xi_{\a f}$'s, and applies the modified amplitude to the study of entanglement entropy in LQG (see e.g. \cite{HanHung,Bianchi:2018fmq,Bodendorfer:2014fua,Chirco:2019dlx,Hamma:2015xla,Feller:2017jqx,Gruber:2018lef} for some existing studies of entanglement entropy in LQG). The modified amplitude is used to define a class of states in the LQG Hilbert space: Given a 4-manifold $\sm_4$ with boundary $\Sig$ and consider $\ck$ (whose 4-cells are $B_4$) as a cellular decomposition of $\sm_4$ (e.g. FIG.\ref{manifold}). The boundary complex $\partial\ck\subset\Sig$ gives the dual graph $\partial\ck^*\subset\Sig$. $\ch_\Sig$ is defined as the LQG kinematical Hilbert space on $\partial\ck^*$ and is spanned by the spin-network states $|T_{\vec{j},\vec{i}}\rangle$ with spins $\vec{j}$ and intertwiners $\vec{i}$ on links and nodes of $\partial\ck^*$. In Section \ref{LQGstate}, we construct a class of states $|\psi\rangle\in\ch_{\Sig}$ as finite linear combinations of spin-networks $|T_{\vec{j},\vec{i}}\rangle$ weighted by spinfoam amplitudes whose boundary data are ${\vec{j},\vec{i}}$. In terms of coherent intertwiners,
\be
|\psi\rangle=\sideset{}{'}\sum_{\{j_f\}}\prod_{f}A_\Delta(j_f)\int_{\Fn_{g,\xi}}[\rmd g^\pm_{v\a}\rmd \xi_{\a f}]\,e^{S+N V }\, |T_{\vec{j},\vec{\xi}}\rangle,\label{psiintro}
\ee
where $|T_{\vec{j},\vec{\xi}}\rangle$ are spin-networks with coherent intertwiners. $V$ is a potential which imposes the parallel restriction when $N\to\infty$. $|\psi\rangle$ depends on a choice of the isolated critical point $(g^\pm_{v\a},\xi_{\a\Delta})_c[J_\Delta]$ where $\xi_{\a f}=\xi_{\a\Delta}$ (up to phases) satisfy the parallel restriction. $\sum_{\{j_f\}}'$ is constrained by $\sum_{f\in\Delta} j_f=J_\Delta$ thus is a finite sum. $\int[\rmd g^\pm_{v\a}\rmd \xi_{\a f}]$ is over a neighborhood $\Fn_{g,\xi}$ which contains a unique isolated critical point $(g^\pm_{v\a},\xi_{\a\Delta})_c[J_\Delta]$. $|\psi\rangle$ has nice semiclassical property: the weight of $|T_{\vec{j},\vec{\xi}}\rangle$ is peaked (in the space of boundary $\vec{\xi}$) at the boundary value $\vec{\xi}$ from the critical data $(g^\pm_{v\a},\xi_{\a\Delta})_c[J_\Delta]$. The implementation of the parallel restriction by $V$ makes the entanglement entropy of $|\psi\rangle$ computable with tools from the stationary phase approximation.

We subdivide $\Sig$ into 2 subregions $A$ and $\bA$, such that the boundary $\cs$ between $A$ and $\bA$ is triangulated by $\Delta\subset \ck_s$. Accordingly the Hilbert space is split by $\ch_{\Sig}\simeq \ch_A\otimes\ch_{\bA}$ (here $\ch_{\Sig}$ has to be suitably enlarged to include some non-gauge-invariant states in order to define the split and entanglement entropy, see Section \ref{EE} for details). The reduced density matrix $\rho_A$ and the $n$-th R\'enyi entanglement entropy $S_n(A)$ are defined by 
\be
\rho_A=\tr_\bA|\psi\rangle\langle\psi|,\quad S_n(A)=\frac{1}{1-n}\ln\frac{\tr(\rho_A^n)}{\tr(\rho_A)^n}
\ee
while the Von Neumann entropy is given by $S(A)=\lim_{n\to 1}S_n(A)$. Entanglement entropies characterize the amount of entanglement from $|\psi\rangle$ between degrees of freedom (DOFs) in $A$ and $\bA$. Section \ref{EE} computes the R\'enyi entropy $S_n(A)$ and shows that $S_n(A)$ is a function of ``macrostates'' $J_\Delta,N_\Delta$: 
\be
S_{n}(A)\simeq \sum_{\Delta\subset\cs}\Big[\l_\Delta (n) J_\Delta+\sig_\Delta(n) N_\Delta\Big],\label{SJN}
\ee
where $\l_\Delta (n),\sig_\Delta (n)$ depend on the ratio $J_\Delta/N_\Delta$. When $\ck$ and $\cs$ are chosen such that all $\Delta\in\cs$ are shared by the same number of $B_4$'s, $\l_\Delta (n)=\l (n),\ \sig_\Delta(n)=\sig(n)$ become independent of $\Delta$. In this case,
\be
S_{n}(A)\simeq \l(n)\, J_\cs+\sig(n)\, N_\cs,
\ee
where $J_\cs=\sum_{\Delta\subset\cs}J_\Delta$ and $N_\cs=\sum_{\Delta\subset\cs}N_\Delta$ are total area and total number of facets of $\cs$. 

Section \ref{Darwin-Fowler} demonstrates an important intermediate step toward $S_{n}(A)$: Computing $\tr(\rho_A^n)$ reduces to a quantity which can be interpreted as counting microstates $\{j_f\}$ in a statistical ensemble with fixed ``macrostate'' $J_\Delta,N_\Delta$ at a given $\Delta$. The computation has an interesting analog to the statistical ensemble of identical systems, in which $J_\Delta,N_\Delta$ are the total energy and total number of identical systems. This counting of microstates is similar to the black hole entropy counting in LQG \cite{GP2011}. 

Section \ref{1stlaw} points out that the resulting R\'enyi entanglement entropy $S_{n}(A)$ and its differential give an analog of the thermodynamical first law:
\be
&&\delta S_{n}(A)\simeq \sum_{\Delta\subset\cs}\Big[\l_\Delta (n)\, \delta J_\Delta+\sig_\Delta(n)\, \delta N_\Delta\Big],\label{themo0}\\
\text{or}&&\delta S_{n}(A)\simeq \l(n)\, \delta J_\cs+\sig(n)\, \delta N_\cs,\label{themo10}
\ee
where in Eq.\Ref{themo10}, $\ck$ and $\cs$ are chosen such that all $\Delta\in\cs$ are shared by the same number of $B_4$'s. Since $J_\cs$ is an analogs of the total energy, Eq.\Ref{themo10} suggests the analog between $\l(n)^{-1}$ and the temperature, as well as between $-\sig(n)/\l(n)$ and the chemical potential. In the most general situation Eq.\Ref{themo0}, the temperature and chemical potential are not constants over $\cs$. $\cs$ is in a non-equilibrium state, although every plaquette $\Delta$ are in equilibrium. Interestingly, Eq.\Ref{themo10} is very similar to the thermodynamical first law derived from the quantum isolated horizon in \cite{GP2011}, if we relate $S_{n}(A)$ to the black hole entropy, $J_\cs$ to the horizon area (proportional to the quasilocal energy observed by the near-horizon Unruh observer), and $N_\cs$ to the total number of spin-network punctures on the horizon.

The above analogy with thermodynamics is clearly a consequence from coarse-graining in the spinfoam model $A(\ck)$. The entanglement entropy effectively coarse-grains the micro-DOFs $\{j_f\}$ collected by the macrostate $J_\Delta,N_\Delta$.

The above discussion mostly focuses on the spinfoam small-$j$ amplitudes with the implementation of parallel restriction. Section \ref{beyond} studying the full amplitude $A(\ck)$ in Eq.\Ref{Ajf000} by removing parallel restrictions to all internal $\xi_{\a f}$'s, while integrating out explicitly all non-parallel DOFs of $\xi_{\a f}$ at every $\Delta$. As a result, the amplitude becomes a sum over Ising configurations at all $\Delta$, where at each $\Delta$ some $\xi_{\a f}$ are parallel $\xi_{\a f}=\xi_{\a \Delta}$ while others are anti-parallel $\xi_{\a f}=J \xi_{\a \Delta}$ ($J(\xi^1,\xi^2)^T=(-\bar{\xi}^2,\bar{\xi}^1)^T$, that $\xi$'s are anti-parallel means that $\vec{n}=\langle\xi |\vec{\sig}|\xi\rangle$ are anti-parallel). The amplitude constrained by the parallel restriction is identified as a partial amplitude in the sum and relates to the simplicial EPRL-FK amplitude, while all other partial amplitudes are made by flip a certain number of $\xi_{\a f}$ from $\xi_{\a \Delta}$ to $J\xi_{\a \Delta}$. Importantly, all partial amplitudes in the sum can be studied by stationary phases approximation. All partial amplitudes, whose numbers of anti-parallel $\xi_{\a f}$ are much less than the numbers of parallel $\xi_{\a f}$ at all $\Delta$'s, are dominated by contributions from critical points $(g^\pm_{v\a},\xi_{\a\Delta})_c[J_\Delta]$ satisfying the parallel restriction. In particular, 4d Regge geometries can still be realized as a subset of critical points in the full amplitude $A(\ck)$. However, for partial amplitudes whose numbers of anti-parallel $\xi_{\a f}$ are comparable to the numbers of parallel $\xi_{\a f}$ at certain $\Delta$'s, they give critical points corresponding to semiclassically degenerate tetrahedron geometries. 4d geometrical interpretations of these critical points are not clear at the moment.

\section{Quantum Polyhedron and Parallel Restriction}\label{STFQPAPR}

In LQG, polyhedron geometries are quantized by intertwiners $||i\rangle_N\in \mathrm{Inv}_{\Su}(j_1,\cdots,j_N)$ which are invariant in the tensor product of $N$ $\Su$ unitary irreps $\ch_{j_1}\otimes\cdots\otimes\ch_{j_N}$ (spins $j\geq 1/2$ label the irreps) \cite{LS,CF,Freidel:2009nu}. In this paper we always assume $j$s to be small but the rank $N$ to be large: $N\gg1$. Denoting by $\vec{L}_f$ SU(2) generators acting on the $f$th irrep $\ch_{j_f}$ ($f=1,\cdots,N$), every invariant tensor $||i\rangle$ satisfies $\sum_{f=1}^N \vec{L}_f\,|| i\rangle_N=0$, which is a quantum analog of the classical closure condition $\sum_{f=1}^N \mathbf{a}_f\vec{n}_f=0$ ($\mathbf{a}_f\in\R,\ \vec{n}_f$ unit 3-vectors). $\{\mathbf{a}_f, \vec{n}_f\}_{f=1}^N$ satisfying this condition uniquely determines a geometrical polyhedron with $N$ facets, such that $\mathbf{a}_f$ is the area of the facet $f$ while $\vec{n}_f$ is the unit normal vector of $f$ \cite{Minkowski}. 


An overcomplete basis of $\mathrm{Inv}_{\Su}(j_1,\cdots,j_N)$ can be chosen to be coherent intertwiners \cite{LS}
\be
||\{j_f\},\{\xi_f\}\rangle_N =\int_{\Su}\rmd h \bigotimes_{f=1}^N  h|j_f,\xi_f\rangle
\ee
where $\rmd h$ is the Haar measure, and $|j,\xi\rangle$ is the SU(2) coherent state in spin-$j$ irrep labelled by $\xi=(\xi^1,\xi^2)^T$ normalized by the Hermitian inner product
\be
|j,\xi\rangle= g(\xi) |j,j\rangle,\quad g(\xi)=\begin{pmatrix}
  \xi^1 & -\bar{\xi}^2 \\
  \xi^2 & \bar{\xi}^1
  \end{pmatrix}.
\ee
Suppose $j$ are all large, $||\{j_f\},\{\xi_f\}\rangle_N$ gives a semiclassical flat polyhedron geometry with $N$ facets, which have areas $\mathbf{a}_f\propto j_f$ and normals $\vec{n}_f=\lag\xi_f|\vec{\sig}|\xi_f\rag$ ($\vec{\sig}$ are Pauli matrices) \cite{shape,LS}. However when $j$ are small, this semiclassical geometry is lost, since the quantum fluctuation is of order $1/j$. However as we see below, some different semiclassical polyhedron geometries can still be found from some $||\{j_f\},\{\xi_f\}\rangle_N$ with small $j$.

\begin{figure} 
  \begin{center}
  \includegraphics[width = 0.5\textwidth]{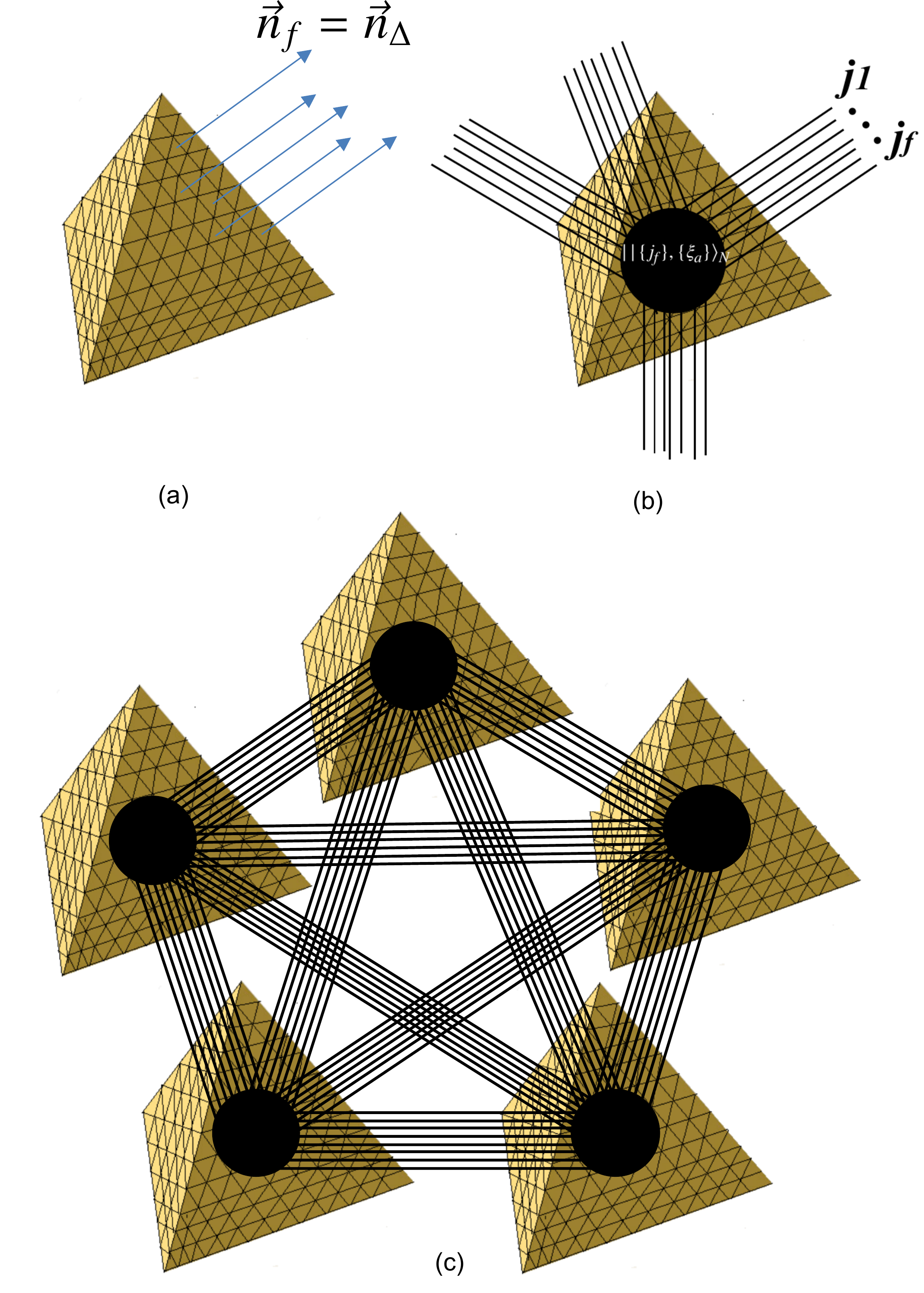}
  \end{center}
  \caption{(a) The classical tetrahedron geometry emergent from a rank-$N$ coherent intertwiner $||\{j_f\},\{\xi_\Delta e^{i\varphi_f}\}\rangle_N$ with small spins but large rank. The tetrahedron with 4 large face is also a polyhedron with $N$ small facets, while normals $\vec{n}_f$ of small polyhedron facets $f$s are parallel if $f$s are in the same large tetrahedron face. The flat large tetrahedron faces are composed by many small facets. Each tetrahedron face area $J_{a=1,\cdots,4}$ is a sum of small areas $j_f$. (b) The rank-$N$ coherent intertwiner $||\{j_f\},\{\xi_\Delta e^{i\varphi_f}\}\rangle_N$ with small spins $j_f$ can be illustrated as a spin-network node connecting to $N$ links, where each link is dual to a polyhedron facet $f$ and colored by $j_f$. (c) A spinfoam vertex amplitude defined by a spin-network with 5 nodes  $(\a=1,\cdots,5)$, connected as shown in the figure. Nodes are colored by intertwiners $||\{j_{f},\{\xi_{\a f}\}\rangle_{N_\a}$ of large rank but small spins. Geometrically, each node corresponds to a polyhedron of many facets as in (a), and the vertex amplitude glues 5 polyhedra to form a close boundary of a 4d region. $\{j_f\},\{\xi_{\a f}\}$ are boundary data of the vertex amplitude.  }
  \label{tetra}
  \end{figure}

An observation is that a subclass of small-spin and large-rank coherent intertwiners relate to large-spin coherent intertwiners with small-rank. Let's consider the small rank to be 4 as an example (generalizations to other small ranks is trivial): we make a partition of $\{1,\cdots,N\}$ into 4 sets, say $\{1,\cdots,N_1\},\{N_1+1,\cdots,N_1+N_2\},\{N_1+N_2+1,\cdots N_1+N_2+N_3\},\{N_1+N_2+N_3+1,\cdots N\} $, where each set has a large number $N_{\Delta}\gg1$ elements, and we use $\Delta=1,\cdots,4$ to label these 4 sets. We restrict to a subclass of coherent states denoted by $||\{j_f\},\{\xi_\Delta\}\rangle_N$, asking $\xi_f$s are parallel up to a phase when $f\in \Delta$:
\be
\text{Parallel restriction:}\quad \xi_f=\xi_\Delta e^{i\varphi_f},\quad \text{or}\quad \vec{n}_{f}=\vec{n}_\Delta,\quad \forall\ f\in \Delta.\label{parallel}
\ee
Parallel $\xi_f$s up to phases make parallel normals $\vec{n}_f$s. Intuitively, this restriction makes a tetrahedron with 4 large flat faces from a polyhedron with many small facets (see FIG.\ref{tetra}(a)).


The squared norm of $||\{j_f\},\{\xi_\Delta e^{i\varphi_f}\}\rangle_N$ is computed by factorizations of coherent states $|j,\xi\rangle=|\xi\rangle^{\otimes 2j}$ and Eq.\Ref{parallel}:
\be
\lt|\lt|\,||\{j_f\},\{\xi_\Delta e^{i\varphi_f}\}\rangle_N\rt|\rt|^2=\int \rmd h \prod_{\Delta=1}^4\lag\xi_\Delta|\, h\, |\xi_\Delta\rag^{2J_\Delta},\quad J_\Delta\equiv\sum_{f\in \Delta} j_f.\label{norm1}
\ee
Although $j_f$ are small, but $J_\Delta\gg 1$ because $N_\Delta\gg 1$ and $j_f\geq\half$. When above $J_\Delta$s satisfy triangle inequalities, Eq.\Ref{norm1} is of the same expression as the square norm of the rank-4 coherent intertwiner $||\{J_\Delta\},\{\xi_\Delta\}\rangle_4$ if we relate the above $J_\Delta$ to the large spins of the rank-4 intertwiner. Thus the same stationary phase analysis in \cite{LS} can be applied to Eq.\Ref{norm1} and shows that Eq.\Ref{norm1} is exponentially suppressed unless the following closure condition holds for the coherent state labels
\be
\sum_{\Delta=1}^4 J_\Delta\vec{n}_\Delta=\sum_{f=1}^N j_f\vec{n}_f=0,\label{closure}
\ee
where $\vec{n}_\Delta=\lag \xi_\Delta|\vec{\sig}|\xi_\Delta\rag$ thus $\vec{n}_f=\vec{n}_\Delta$ for all $f\in \Delta$. Comparing to the classical closure condition of polyhedron, Eq.\Ref{closure} uniquely determines a classical flat geometrical tetrahedron, whose face areas are proportional to $J_{\Delta} \gg 1$ and face normals are $\vec{n}_\Delta$. However here $J_\Delta$ emerges from summing many small $j_f$s. Eq.\Ref{closure} may still be interpreted as a classical closure condition of a polyhedron with $N$ facets with small areas $j_f$s, while facets composes large flat faces of the tetrahedron. The quantum correction of the classical geometry is of $O(1/J_\Delta)$ thus suppressed by the large-rank.

The above demonstrates that the classical tetrahedron geometry can emerge from intertwiners with small $j$s but large rank $N_\Delta\to \infty$. The geometrical picture of the tetrahedron/polyhedron is illustrated in FIG.\ref{tetra}(a).

Importantly, rank-$N$ intertwiners have much more degrees of freedom (DOFs) than tetrahedron. There are coherent intertwiners with $\xi_f$s beyond the parallel restriction, while $||\{j_f\},\{\xi_\Delta\}\rangle_N$ only span a subspace. In addition, the same tetrahedron geometry with areas $J_\Delta$ may come from different spin configurations $\{j_f\}$ satisfying $J_\Delta\equiv\sum_{f\in \Delta} j_f$. 

\begin{Lemma}\label{triangle inequality}

Given 4 $J_\Delta$ satisfying the triangle inequality such that $\otimes_{\Delta=1}^4\ch_{J_\Delta}$ has a nontrivial invariant subspace, any spin configuration $\{j_f\}_{f\in\Delta}$ satisfying $J_\Delta\equiv\sum_{f\in \Delta} j_f$ leads to a nontrivial invariant subspace in $\otimes_{\Delta=1}^4\otimes_{f\in\Delta}\ch_{j_f}$.

\end{Lemma}

\textbf{Proof:} It is convenient to consider coherent intertwiners satisfying the parallel restriction Eq.\Ref{parallel} and use the factorization property $|j,\xi\rangle=|\xi\rangle^{\otimes 2j}$
\be
||\{j_f\},\{\xi_\Delta e^{i\varphi_f}\}\rangle_N &=&e^{i\sum_f2j_f\varphi_f}\int_{\Su}\rmd h \bigotimes_{\Delta=1}^4\bigotimes_{f\in\Delta}  h|j_f,\xi_\Delta\rangle=e^{i\sum_f2j_f\varphi_f}\int_{\Su}\rmd h  \bigotimes_{\Delta=1}^4\bigotimes_{f\in\Delta} \lt( h|\xi_\Delta\rangle\rt)^{\otimes 2j_f}\nonumber\\
&=&e^{i\sum_f2j_f\varphi_f}\int_{\Su}\rmd h  \bigotimes_{\Delta=1}^4 \lt( h|\xi_\Delta\rangle\rt)^{\otimes 2J_\Delta}
=e^{i\sum_f2j_f\varphi_f}\int_{\Su}\rmd h  \bigotimes_{\Delta=1}^4 h|J_\Delta, \xi_\Delta\rangle
\ee
The right-hand side gives up to a phase the rank-4 coherent intertwiner, which is nonzero by the assumption that $J_\Delta$ satisfying the triangle inequality. Therefore $||\{j_f\},\{\xi_\Delta e^{i\varphi_f}\}\rangle_N $ is nonzero thus the invariant subspace in $\otimes_{\Delta=1}^4\otimes_{f\in\Delta}\ch_{j_f}$ is nontrivial. \\
$\Box$



\section{Spinfoam Vertex Amplitude}\label{vertex}

We exend our discussion of small-$j$ semiclassicality to LQG dynamics in the spinfoam formulation. We firstly focus on a class of spinfoam vertex amplitudes asssociated to a 4d spacetime region $B_4$ whose closed boundary is made by gluing 5 polyhedra (labelled by $\a,\b=1,\cdots,5$) through facets. Each polyhedron has $N_\a\gg1$ facets, and every pair of polyhedra $\a,\b$ share a large number $N_{\Delta}\gg1$ facets. $\Delta$ denotes the interface between $\a,\b$ made by $N_\Delta$ facets $f$. 

We apply the generalized spinfoam formulation to construct amplitude on non-simplicial $B_4$ \cite{generalize,KKL}. The vertex amplitude of $B_4$ evaluates a spin-network with 5 nodes (dual to polyhedra), and each pair of nodes $\a,\b$ are connected by $N_{\Delta}$ links. See FIG.\ref{tetra}(c) for an illustration. Links connecting nodes are dual to $f$s shared by polyhedra and colored by spins $j_f$. We color every node $\a$ by rank-$N_\a$ coherent intertwiners $||\{j_{f}\},\{\xi_{\a f}\}\rangle_{N_\a}$ studied above ($j_f\neq0$ but small), while making the parallel restriction as in Eq.\Ref{parallel}:
\be
\xi_{\a f}=\xi_{\a \Delta}e^{i\varphi_{\a f}}\quad \forall\ f\in\Delta.\label{parallel1}
\ee 
The vertex amplitude $A_v(j_f,\xi_{\a f})$ (in Euclidean singature) describes a local transition in $B_4$ of boundary geometrical states $\otimes_{\a=1}^5||\{j_{f}\},\{\xi_{\a f}\}\rangle_{N_\a}$:
\be
A_v=\int[\rmd g^\pm_\a]\prod_{\pm}\prod_{\a<\b}\prod_{f\in(\a,\b)}\lag j_f^\pm,\xi_{\a f}\lt|g_\a^\pm{}^{-1}g_\b^\pm \rt|j_f^\pm,\xi_{\b f}\rag=\int[\rmd g^\pm_\a]\, e^{\sum_{\pm}\sum_{f}2j_f^\pm\ln\lag \xi_{\a f}\lt|g_\a^\pm{}^{-1}g_\b^\pm \rt|\xi_{\b f}\rag}.\label{Av0}
\ee
where $(g^+_\a,g^-_\a)\in\mathrm{Spin}(4)$ associates to each node, and $j^\pm_f=(1\pm\g)j_f/2$ with $\g<1$. We have applied the factorization property of coherent state in the above. 
By the parallel restriction,
\be
A_v=\prod_{\Delta,f} e^{2ij_f\lt(\varphi_{\b f}-\varphi_{\a f}\rt)}\int[\rmd g^\pm_\a] \,e^{\sum_{\pm}\sum_{\Delta}2J_{\Delta}^\pm\ln\lag \xi_{\a\Delta}\lt|g_\a^\pm{}^{-1}g_\b^\pm \rt|\xi_{\b\Delta}\rag},\ \ J_{\Delta}^\pm=\sum_{f\in\Delta}j^\pm_f.\label{Av1}
\ee
where ten $J_{\Delta}=\sum_{f\in(\a,\b)}j_f\gg 1$ emerges as summing $j_f$ over facets $f\in\Delta$. $J_{\Delta}$ are all large since $N_{\Delta}\gg1$ and $j_f\geq\half$. $\prod_{\Delta,f} e^{2ij_f\lt(\varphi_{\b f}-\varphi_{\a f}\rt)}$ is an overall phase since Eq.\Ref{parallel1} restrict $\xi_{\a f}$ parallel up to a phase. 

Although $A_v$ is a generalized spinfoam vertex with boundary polyhedra and small spins, the integral Eq.\Ref{Av1} has the same expression as the EPRL-FK 4-simplex amplitude (boundary states are rank-4 intertwiners) \cite{EPRL,FK,semiclassicalEu} if we relates $J_{\Delta}$ to actual spins in the EPRL-FK amplitude.

\begin{Definition}

Given an integral $\int_D\rmd^n x\, e^{S(x)}$, its stationary points $x_0$ are solution of $\vec{\nabla} S(x_0)=0$, and its critical points are stationary points with $\mathrm{Re}(S(x_0))=0$.

\end{Definition}

Since $\mathrm{Re}\ln\lag\xi'|\xi \rag=\ln|\lag\xi'|\xi \rag|\leq \ln(||\xi'||\cdot||\xi ||)=0$ by Schwarz inequality, the exponents in Eqs.\Ref{Av0} and \Ref{Av1} are non-positive. The critical points of $A_v$ in Eq.\Ref{Av0} are solutions of
\be
\hat{g}_{v\b}^\pm \vec{n}_{\b f}= \hat{g}^\pm_{v\a}\vec{n}_{\a f},\quad \sum_{f\subset\a} j_{f}\kappa_{\a \Delta}\ \vec{n}_{\a f}=0\label{criticalsingle}
\ee
where the 1st equation comes from $\mathrm{Re}(S)=0$. $\kappa_{\a\Delta}=\pm1$ appears when $\partial_{g_{v\a}^\pm}$ acts on $g_{v\a}^\pm$ or $g_{v\a}^\pm{}^{-1}$. $\hat{g}^\pm_{v\a}\in\mathrm{SO(3)}$ is the 3-dimensional irrep of $g_{v\a}^\pm$. When the parallel restriction is imposed to boundary data. The critical equations Eq.\Ref{criticalsingle} reduce to
\be
\hat{g}_{v\b}^\pm \vec{n}_{\b \Delta}= \hat{g}^\pm_{v\a}\vec{n}_{\a \Delta},\quad \sum_{\Delta\subset\a} J_{\Delta}\kappa_{\a \Delta}\ \vec{n}_{\a \Delta}=0,\label{criticalsingle1}
\ee
which are also critical equations from Eq.\Ref{Av1}.

The same asymptotic analysis as in \cite{semiclassicalEu} is valid for Eq.\Ref{Av1} as $J_{\Delta}\gg 1$. Here we adapt results in \cite{semiclassicalEu} to our $A_v$: When the boundary data $j_f,\xi_{\Delta}$ satisfy the closure condition as in Eq.\Ref{closure}, and give flat geometrical tetrahedra that are glued (with $\Delta$ matching in shapes and orientation-matching) to form a closed boundary of a flat nondegenerate 4-simplex, the asymptotics of $A_v$ relates to the Regge action of the 4-simplex: If we define $N$ to be the order of magnitude of $N_{\Delta}$ ($N\sim N_{\Delta}\sim J_{\Delta}$ since all $j_f\sim O(1)$), then $A_v$ has the following asymptotic formula:
\be
A_v=\lt(\text{overall phase}\rt)\lt(\frac{2\pi}{N}\rt)^{12}\Bigg[2\cn_{+-}^\g\cos\lt(\sum_{\Delta}\g J_{\Delta}\Theta_{\Delta}\rt)+\cn^\g_{++}e^{\sum_{\Delta} J_{\Delta}\Theta_{\Delta}}+\cn^\g_{--}e^{-\sum_{\Delta} J_{\Delta}\Theta_{\Delta}}\Bigg]\lt(1+O\lt(\frac{1}{N}\rt)\rt).\label{asymp}
\ee
We refer the reader to \cite{semiclassicalEu} for expressions of $\cn_{+-}^\g,\cn_{++}^\g,\cn_{--}^\g$. The asymptotics is dominant by contributions from 4 critical points $(g_{v\a}^+,g_{v\a}^-)$, $(g_{v\a}^-,g_{v\a}^+)$, $(g_{v\a}^+,g_{v\a}^+)$, $(g_{v\a}^-,g_{v\a}^-)$ solving Eq.\Ref{criticalsingle1} with the boundary condition. $\Theta_{\Delta}$ is the 4d dihedral angle between a pair of tetrahedra in the geometrical 4-simplex. The quantity inside the cosine is the Regge action of classical gravity when we identify the tetrahedron face area $\mathbf{a}_{\Delta}$ as 
\be
\mathbf{a}_{\Delta}=8\pi \g J_{\Delta}\ell_P^2=\sum_{f\in\Delta}\mathbf{a}_{f},\quad \mathbf{a}_{f}=8\pi\g j_f\ell_P^2.\label{area}
\ee
The large tetrahedron face area is given by summing small areas of polyhedron facets. $\ell_P$ is the Planck length.



\section{Spinfoam Amplitudes on Complexes}\label{complex}

Our semiclassical analysis with small spins can be generalized to spinfoam amplitudes on cellular complexes with arbitrarily many cells. We construct a generalized spinfoam amplitude on a complex $\ck$ whose cells $\cc$ are similar to $B_4$ (every $\partial\cc$ are made by 5 polyhedra $\a$ of large numbers of facets $f$, though different $\cc$ may have different number of facets). $N\sim N_\Delta\gg 1$ are assumed. $\cc$s are glued in $\ck$ by sharing boundary polyhedra. $\ck$ determines a simplicial complex $\ck_s$ by substituting all polyhedra and $\cc$ with tetrahedra and 4-simplices. We associates $A_v$ to every $\cc$, and write the spinfoam amplitude on $\ck$ by \cite{HZ1,CF}
\be
\quad A(\ck)
&=&\sum_{\{j_f\}}\prod_f A_f({j_f})\int[\rmd \xi_{\a f}\rmd g^\pm_{v\a}]\,e^{S},\label{Z}\\
S&=&\sum_{\pm}\sum_{v,f}2j_{f}^\pm\ln\lag \xi_{\a f}\lt|g_{v\a}^\pm{}^{-1}g_{v\b}^\pm \rt|\xi_{\b f}\rag,
\ee  
where $A_f$ is the face amplitude given by \cite{face} (see Appendix \ref{face} for explanations) 
\be
&A_f({j_f})=A_\Delta(j_f)=(2j_f+1)^{n_v(\Delta)+1}& \text{for internal $f$},\nonumber\\
&A_f({j_f})=A_\Delta(j_f)=(2j_f+1)^{n_v(\Delta)+2}& \text{for boundary $f$},
\ee
$n_v(\Delta)$ is the number of $B_4$ sharing $f\in\Delta$ in $\ck$ and equals the number of 4-simplices sharing $\Delta$ in $\ck_s$. $A_f$ depends on $n_v(\Delta)$ in the coherent state formulation since $(2j+1)\int\rmd\xi |j, \xi\rangle\langle j, \xi|=1$ where $\rmd\xi$ is the standard normalised measure on the unit sphere. $\sum_{\{j_f\}}$ and $ \int[\rmd \xi_{\a f}]$ sums coherent state labels of all internal facets $f$. Each $\int\rmd \xi_{\a f}$ is over $S^2$. Different from $A_v$ where we can apply the parallel restriction to boundary data, $A(\ck)$ sums independently $\xi_{\a f}$s at different internal $f$s, so we need to take into account fluctuations beyond the parallel restriction. When $\ck$ has boundary, we still make the parallel restriction to boundary $\xi_{\a f}$s.

$S$ has the following gauge symmetry:
\begin{itemize}

\item \emph{Continuous}: (1) A diagonal Spin(4) action at $\sig$, $g_{v\a}^\pm\to h^\pm_v g_{v\a}^\pm $ for all $\a$ at $v$ by $(h^+_v,h^-_v) \in\mathrm{Spin(4)}$; (2) At any internal $\a$, $|\xi_{\a f}\rangle\to h_{\a}|\xi_{\a f}\rangle$ and $g^\pm_{v \a}\to g^\pm_{v \a}h_{\a}^{-1}$ for all $v$ having $\a$ at boundaries; and (3) $|\xi_{\a f}\rangle\to e^{i\theta_{\a f}}|\xi_{\a f}\rangle$ at any internal $|\xi_{\a f}\rangle$.

\item \emph{Discrete}: $g^+_{v\a}\to\pm g^+_{ v\a}$ and independently $g^-_{v\a}\to\pm g^-_{v\a}$. 

\end{itemize}


If we expand $S$ at ${\xi}_{\a f}$ satisfying the parallel restriction, i.e. $e^{-i\varphi_{\a f}}{\xi}_{\a f}=\xi_{\a\Delta}+\delta\xi_{\a f},\ \forall f\in\Delta$. $\delta\xi_{\a f}$ are fluctuations of ${\xi}_{\a f}$ away from the parallel restriction. Notice that $\xi_{\a f}\to e^{i\varphi_{\a f}}{\xi}_{\a f}$ at internal $f$s are gauge symmetries of $S$,
\be
S&=&S_0+\sum_f 2j_f \Phi_\Delta[\xi_{\a f}],\quad 
S_0=\sum_{\pm}\sum_{v,{\Delta}}2J_{\Delta}^\pm\ln\lag \xi_{\a \Delta}\lt|g_{v\a}^\pm{}^{-1}g_{v\b}^\pm \rt|\xi_{\b \Delta}\rag\nonumber\\
\Phi_\Delta&=&\sum_{\pm}\frac{1\pm\g}{2}\sum_v\lt[\ln\lag \xi_{\a f}\lt|g_{v\a}^\pm{}^{-1}g_{v\b}^\pm \rt|\xi_{\b f}\rag-\ln\lag \xi_{\a \Delta}\lt|g_{v\a}^\pm{}^{-1}g_{v\b}^\pm \rt|\xi_{\b \Delta}\rag\rt]=O(\delta\xi)
\ee
where $J_{\Delta}^\pm$ is the same as in Eq.\Ref{Av1} and is large by $N_{\Delta}\gg1$. $J_\Delta$ are assumed to satisfy the triangle inequality. $S_0$ reduces to Eq.\Ref{Av1} at each $v$ and is the same as the EPRL-FK spinfoam action used for large spin asymptotics on the simplicial complex $\ck_s$. 

\subsection{Critical points satisfying parallel restriction}

Critical points of $S_0$, denoted by $(g^\pm_{v\a},\xi_{\a\Delta})_c[J_\Delta]$, are gauge equivalence classes of solutions of criticial equations $\mathrm{Re} (S_0)=\partial_{g_{v\a}^\pm} S_0=\partial_{\xi_{\a \b}} S_0=0$. These critical equations have been well studied in \cite{HZ1,CF,semiclassicalEu} and reduce to (it is straightforward to check that $\partial_{\xi_{\a \b}} S_0=0$ follows from $\mathrm{Re} (S_0)=0$) 
\be
g_{v\b}^\pm|\xi_{\b\Delta}\rangle=e^{i\phi^\pm_{\a v\b}} g^\pm_{v\a}|\xi_{\a\Delta}\rangle,\quad \sum_{\Delta\subset\a} J_{\Delta}\kappa_{\a\Delta}(v)\ \vec{n}_{\a\Delta}=0\label{critical}
\ee
where $\kappa_{\a\Delta}(v)=\pm1$ when $\partial_{g_{v\a}^\pm}$ acts on $g_{v\a}^\pm$ or $g_{v\a}^\pm{}^{-1}$.

\begin{Theorem}\label{SS0}

Critical points of $S_0$ are also critical points of $S$.

\end{Theorem}

\textbf{Proof:} We check that $\mathrm{Re}(S)=\partial S/\partial g_{v\a}^\pm=\partial S/\partial \xi_{\a f}=0$ at all critical points of $S_0$. First of all, at any critical point of $S_0$,
\be
\mathrm{Re}(S)\big|_c=\mathrm{Re}(S_0)\big|_c=0
\ee
where $|_c$ means evaluating at any critical point $(g^\pm_{v\a},\xi_{\a\Delta})_c[J_\Delta]$ of $S_0$ where $\xi_{\a f}=\xi_{\a\Delta}$, $\forall f\in\Delta$. 

If we write $\xi=(\xi^1,\xi^2)$ and define $J\xi=(-\bar{\xi}^{2},\bar{\xi}^{1})$, $\xi,J\xi$ form an orthonormal basis in $\C^2$ with the Hermitian inner product. When we perturbing $S$, we write $\delta\xi_{\a f}=\eps_{\a f}J\xi_{\a f}+i\eta_{\a f} \xi_{\a f}$ where $\eps_{\a f}\in\C$ and $\eta_{\a f}\in\R$. The coefficient in front of $\xi_{\a f}$ is purely imaginary because $\xi_{\a f}$ is normalized. Since every $\xi_{\a f}$ is shared by 2 terms with neighboring $v$s
\be
\delta_{\xi_{\a f}}S \big|_c
&=&\sum_{\pm}(1\pm\g)j_f\lt[\eps_{\a f} \frac{\lag{{\xi_{\b' f}}\big|(g^\pm_{v' \b'})^{-1}g^\pm_{v'\a}\big|J{\xi_{\a f}}}\rag}{\lag{{\xi_{\b' f}}\big|(g^\pm_{v'\b'})^{-1}g^\pm_{v'\a }\big|{\xi_{\a f}}}\rag}+{\eps}^*_{\a f}\frac{\lag J{\xi_{\a f}}\big|(g^\pm_{v\a})^{-1}g^\pm_{v\b}\big|{\xi_{\b f}}\rag}{\lag{\xi_{\a f}}\big|(g^\pm_{v\a})^{-1}g^\pm_{v\b}\big|{\xi_{\b f}}\rag}\rt]_c=0.
\ee
At the critical point, $\xi_{\a f}=\xi_{\a\Delta}$, $\xi_{\b f}=\xi_{\b\Delta}$ at $v$ and satisfying Eq.\Ref{critical}, similarly $\xi_{\b' f}=\xi_{\b'\Delta}$ and satisfy Eq.\Ref{critical} at $v'$. Then $\delta_{\xi_{\a f}}S=0$ by the orthogonality between $\xi,J\xi$.

For derivative in $g_{v\a}^\pm$, we use $\delta {g_{v\a}^\pm}= \frac{i}{2} \theta_{v\a}^\pm\vec{\sig} {g_{v\a}^\pm}$ ($\theta_{v\a}\in\R$). At the critical point and by Eq.\Ref{critical},
\be
\delta_{g_{v\a}^\pm} S \big|_c=\frac{i}{2} \theta_{v\a}^\pm\sum_{\Delta}\kappa_{\a \Delta}(1\pm\g) \sum_{f\in\Delta} j_f \frac{\lag{{\xi_{\a f}}\big|(g^\pm_{v\a})^{-1}\vec{\sig} {g_{v\b}^\pm}\big|{\xi_{\b f}}}\rag}{\lag{{\xi_{\a f}}\big|(g^\pm_{v\a})^{-1}g^\pm_{v\b}\big|{\xi_{\b f}}}\rag}\Bigg|_c=\frac{i}{2} \theta_{v\a}^\pm\,(1\pm\g)\, g^\pm_{v\a}\cdot\sum_{\Delta}\kappa_{\a\Delta}J_{\Delta}\hat{n}_{\a\Delta}\big|_c=0
\ee
where $\hat{n}_{\a\Delta}=\lag{{\xi_{\a\Delta}}\big|\vec{\sig}\big|{\xi_{\a\Delta}}}\rag$ is a unit 3-vector. $\kappa_{\a\Delta}=\pm 1$ relates to orientations of links in FIG.\ref{tetra}(c). We have chosen orientations such that all links connecting $\a,\b$ are oriented parallel. \\
$\Box$\\

Critical points of $S_0$ has been completely classified in case that all tetrahedra reconstructed from the closure condition are nondegenerate. We refer the reader to \cite{Han:2018fmu,hanPI,CF,HZ1} for details of the classification. When $J_{\Delta}$ are areas relating to edge-lengths on $\ck_s$ by ($\ell_{ij},\ell_{jk},\ell_{ik}$ are 3 edge-lengths of a triangle $\Delta$)
\be
\g {J}_\Delta(\ell)=\frac{1}{4}\sqrt{2(\ell^2_{ij}\ell^2_{jk}+\ell^2_{ik}\ell^2_{jk}+\ell^2_{ij}\ell^2_{ik})-\ell_{ij}^4-\ell_{ik}^4-\ell_{jk}^4},\label{area-length}
\ee  
there are a subset $\mathscr{G}$ of critical points $(g^\pm_{v\a},\xi_{\a\Delta})_c[J_\Delta]$ of $S_0$ that can be interpreted as nondegenerate 4d Regge geometries, if the boundary condition of $\xi_{\a\Delta}$ gives the boundary 3d Regge geometry. Defining $N_{\a}(v)$ by $N^0_\a(v)\mathbf{1}+iN^i_\a(v)\sig_i=g_{v\a}^{-}(g_{v\a}^{+})^{-1}$ ($\sig_i$ are Pauli matrices), $\mathscr{G}$ is defined by critical points $(g^\pm_{v\a},\xi_{\a\Delta})_c[J_\Delta]$ with 
\be
\det\lt(N_{\a_1}(v),N_{\a_2}(v),N_{\a_3}(v),N_{\a_4}(v)\rt)\neq0, \label{NNNN}
\ee
for all $v\subset\ck_s$ and all 4 $\a_1,\a_2,\a_3,\a_4$ out of 5 $\a$'s at $v$. We have the following 1-to-1 correspondence \cite{Han:2018fmu,hanPI}:
\be
&\text{Critical points }(g^\pm_{v\a},\xi_{\a\Delta})_c[J_\Delta]\in \mathscr{G}&\nonumber\\
&\updownarrow&\nonumber\\
&\text{4d nondegenerate Regge geometry on $\ck_s$}&\nonumber\\
&  \text{and 4-simplex orientations}.& \label{equiv0}
\ee 
Triangles $\Delta$ in Regge geometries are made by polyhedron facets as FIG.\Ref{tetra}(c), and $\g J_\Delta$ is the area of $\Delta$. Different critical points may give the same Regge geometry but different 4d orientations $\mu(v)=\pm 1$ at individual $v$. We focus on critical points $(g^\pm_{v\a},\xi_{\a\Delta})_c[J_\Delta]\in \mathscr{G}$ that are isolated. 

Consider infinitesimal deformations $(g^\pm_{v\a},\xi_{\a \Delta})\mapsto (g^\pm_{v\a}+\delta g^\pm_{v\a},\xi_{\a \Delta}+\delta \xi_{\a  \Delta})$ (including boundary data $\xi_{\a  \Delta}$) from $(g^\pm_{v\a},\xi_{\a\Delta})_c[J_\Delta]\in \mathscr{G}$ with fixed $J_\Delta$, and ask whether the deformation can reach another critical point (solution of Eq.\Ref{critical}). Any infinitesimal deformation cannot break the condition Eq.\Ref{NNNN}, so cannot reach critical points outside $\mathscr{G}$. Moreover the deformation cannot flip the orientation \cite{Han:2018fmu}. Therefore if the deformation reaches another critical point $({g'}^\pm_{v\a},\xi'_{\a\Delta})_c[J_\Delta]$, $({g'}^\pm_{v\a},\xi'_{\a\Delta})_c[J_\Delta]$ must still belong to $\mathscr{G}$, and correspond to a different non-degenerate Regge geometry with the same set of areas $\g J_\Delta$. In other words, $({g}^\pm_{v\a},\xi_{\a\Delta})_c[J_\Delta]$ and $({g'}^\pm_{v\a},\xi'_{\a\Delta})_c[J_\Delta]$ correspond to 2 different non-degenerate Regge geometries with the same set of areas. At any 4-simplex, Eq.\Ref{area-length} with 10 fixed areas gives 10 quadratic equations for 10 squared edge-lengths. These 2 different Regge geometries correspond to 2 different solutions of these 10 quadratic equations with fixed $J_\Delta$ at at least one 4-simplex. And these 2 different solutions are infinitesimally close to each other, since one comes from the infinitesimal deformation from the other. Then it implies the $10\times10$ matrix $\partial J_\Delta^2/\partial \ell^2_{ij}$ is degenerate at $({g}^\pm_{v\a},\xi_{\a\Delta})_c[J_\Delta]$. As a result, If $({g}^\pm_{v\a},\xi_{\a\Delta})_c[J_\Delta]$ gives Regge geometry with non-degenerate $\partial J_\Delta^2/\partial \ell^2_{ij}$ at all 4-simplices, $({g}^\pm_{v\a},\xi_{\a\Delta})_c[J_\Delta]$ is an isolated critical point. Note that the deformations considered above includes deformations of boundary data $\xi_{\a  \Delta}$, so $({g}^\pm_{v\a},\xi_{\a\Delta})_c[J_\Delta]$ is isolated in a larger space of $g^\pm_{v\a},\xi_{\a  \Delta}$ including boundary $\xi_{\a  \Delta}$. It is easy to find isolated critical points by numerically check the determinant of $\partial J_\Delta^2/\partial \ell^2_{ij}$. Some experience from numerics suggests that degenerate $\partial J_\Delta^2/\partial \ell^2_{ij}$ might only happen at degenerate 4-simplices.

A critical point $(g^\pm_{v\a},\xi_{\a\Delta})_c[J_\Delta]\in\mathscr{G}$ with a uniform orientation $\mu(v)=\mu$ at all $v$'s evaluates 
\be
S_0\Big|_c=\mu\lt(\sum_{\Delta\in\ck_s}\g J_{\Delta}\eps_{\Delta}+\sum_{\Delta\in\partial\ck_s} \g J_{\Delta}\Theta_{\Delta}\rt)=\frac{i\mu}{{8\pi\ell_P^2}}\lt(\sum_{\Delta\in\ck_s} {\mathbf{a}_{\Delta}\eps_{\Delta}}+\sum_{\Delta\in\partial\ck_s} \mathbf{a}_{\Delta}\Theta_{\Delta}\rt),\quad \mu=\pm1\label{regge}
\ee 
is the Regge action on $\ck_s$ \cite{CF,HZ1,Han:2018fmu,Han:2017xwo,regge}. $|_c$ means evaluating at any critical point $(g^\pm_{v\a},\xi_{\a\Delta})_c[J_\Delta]$ of $S_0$. $\eps_{\Delta},\ \Theta_{\Delta}$ are the deficit angles and dihedral angles hinged at internal and boundary $\Delta$s. $\g J_{\Delta}$ are interpreted as triangle areas $\mathbf{a}_{\Delta}=\sum_{f\in\Delta}\mathbf{a}_{f}$ made by facet areas $\mathbf{a}_{f}$ as in Eq.\Ref{area}. The validity of Eq.\Ref{regge} has some topological requirements on $\ck_s$: (1) all internal $\Delta$ are shared by an even number of 4-simplices, and (2) $\ck_s$ is a triangulation of manifold $\sm$ with trivial 2nd cohomology $H^2(\sm,\Z_2)=0$ \cite{Han:2018fmu}. The 1st requirement is generically satisfied by triangulations used in Regge calculus, see. e.g. \cite{Han:2018fmu, 0264-9381-5-12-007} for examples. The above result applies to e.g. $\sm$ is $S^4$, $S^3\times I$ (where $I$ is an interval in $\R$), or a topologically trivial region in $\R^4$.

Beyond the subset $\mathscr{G}$, there are other critical points with the BF-type and/or vector geometry critical data \cite{Han:2018fmu,HZ1,semiclassicalEu}. Each of these critical points has critical data of $g_{v\a}^\pm$ satisfy $g_{v\a}^+=g_{v\a}^-$ or equivalently $\det\lt(N_{\a_1}(v),N_{\a_2}(v),N_{\a_3}(v),N_{\a_4}(v)\rt)=0$ at certain $v$'s. The difference between the BF-type and vector geometry critical data is that the BF-type data still associate to nondegnerate 4-simplices, while vector geometries are degenerate 4-simplices.

\subsection{Critical points violating parallel restriction}

The converse of Theorem \ref{SS0} is not true. There exist critical points of $S$ which are not critical points of $S_0$. Critical points of $S$ satisfy 
\be
\hat{g}_{v\b}^\pm \vec{n}_{\b f}= \hat{g}^\pm_{v\a}\vec{n}_{\a f},\quad \sum_{f\subset\a} j_{f}\kappa_{\a \Delta}\ \vec{n}_{\a f}=0\label{critical22}
\ee

\begin{Theorem}\label{SSSSS}

Every critical point of $S$ that are not critical point of $S_0$ either (1) relates to a critical point of $S_0$, $(g_{v\a}^\pm,\xi_{\a\Delta})_c[J_\Delta]$, by $g_{v\a}^+\neq g_{v\a}^-$ and $\xi_{\a f}=J\xi_{\a\Delta}$ up to a phase at some internal $f\in\Delta$, or (2) satsifies $g_{v\a}^+=g_{v\a}^-$ for all $v,\a$ modulo discrete gauge.

\end{Theorem}

\textbf{Proof:} We write $\hat{g}_{\a\b}^\pm\equiv (\hat{g}_{v\a}^\pm)^{-1} (\hat{g}_{v\b}^\pm)$, the 1st equation in \Ref{critical22} gives $\hat{g}_{\a\b}^+ \vec{n}_{\b f}=\vec{n}_{\a f}$ and $\hat{g}_{\a\b}^- \vec{n}_{\b f}=\vec{n}_{\a f}$, and implies $(\hat{g}_{\a\b}^+)^{-1}\hat{g}_{\a\b}^- \vec{n}_{\b f}=\vec{n}_{\b f}$ for all $f\in\Delta$, i.e. $\vec{n}_{\b f}$ at all $f\in\Delta$ are eigenvectors of $(\hat{g}_{\a\b}^+)^{-1}\hat{g}_{\a\b}^- $ with unit eigenvalue. It doesn't constrain $\vec{n}_{\b f}$ if $(\hat{g}_{\a\b}^+)^{-1}\hat{g}_{\a\b}^- =1$. But when the SO(3) matrix $(\hat{g}_{\a\b}^+)^{-1}\hat{g}_{\a\b}^-\neq1 $, its eigenspace with the unit eigenvalue is at most 1-dimensional. Therefore in this case, all $\vec{n}_{\b f}$ are co-linear thus $\vec{n}_{\b f}=\pm \vec{n}_{\b f'}$ for any pair of $f,f'\in\Delta$, and Eq.\Ref{critical22} reduces to Eq.\Ref{critical} whose solution gives $(g_{v\a}^\pm,\xi_{\a\Delta})_c[J_\Delta]$. Hence $\vec{n}_{\b f}=\pm\vec{n}_{\a\Delta}$ i.e. $\xi_{\b f}=\xi_{\a\Delta}$ or $J\xi_{\a\Delta}$ up to a phase. At each $v$, we have to gauge fix $g_{v\a}^\pm=1$ at a certain $\a$, then requiring $(\hat{g}_{\a\b}^+)^{-1}\hat{g}_{\a\b}^-\neq1 $ is equivalent to $g_{v\b}^+\neq g_{v\b}^-$ for all $\b\neq \a$ ($g_{v\b}^+= - g_{v\b}^-$ still implies $(\hat{g}_{\a\b}^+)^{-1}\hat{g}_{\a\b}^-=1 $, but it is gauge equivalent to $g_{v\b}^+= g_{v\b}^-$ by a discrete gauge transformation).\\
$\Box$

We may generalize the definition Eq.\Ref{NNNN} of the subclass $\mathscr{G}$ to include all critical points of $S$. It contains critical points of $S_0$, $(g_{v\a}^\pm,\xi_{\a\Delta})_c[J_\Delta]\in\mathscr{G}$, and critical points of $S$ which flip some internal or boundary $\xi_{\a f}\to J\xi_{\a\Delta}$. Critical points in either class (1) or (2) in Theorem \ref{SSSSS} are isolated from $(g_{v\a}^\pm,\xi_{\a\Delta})_c[J_\Delta]\in\mathscr{G}$ because an infinitesimal deformation from $(g_{v\a}^\pm,\xi_{\a\Delta})_c[J_\Delta]$ at fixed $J_\Delta$ cannot flip $\xi_{\a f}\to J\xi_{\a\Delta}$, and cannot break the condition Eq.\Ref{NNNN}.

Although we find critical points of $S_0$ and $S$, we cannot apply the stationary phase approximation of the integral at the present stage since all $j_f$'s are small. The critical points in Theorem \ref{SSSSS} seem useless. But we come back to the computation of the integral in Section \ref{beyond} and see why these critical points are useful to the stationary phase approximation of integrals.

\section{Semiclassical States from Spinfoam Amplitude}\label{LQGstate}

\begin{figure}[t]
  \begin{center}
  \includegraphics[width = 0.3\textwidth]{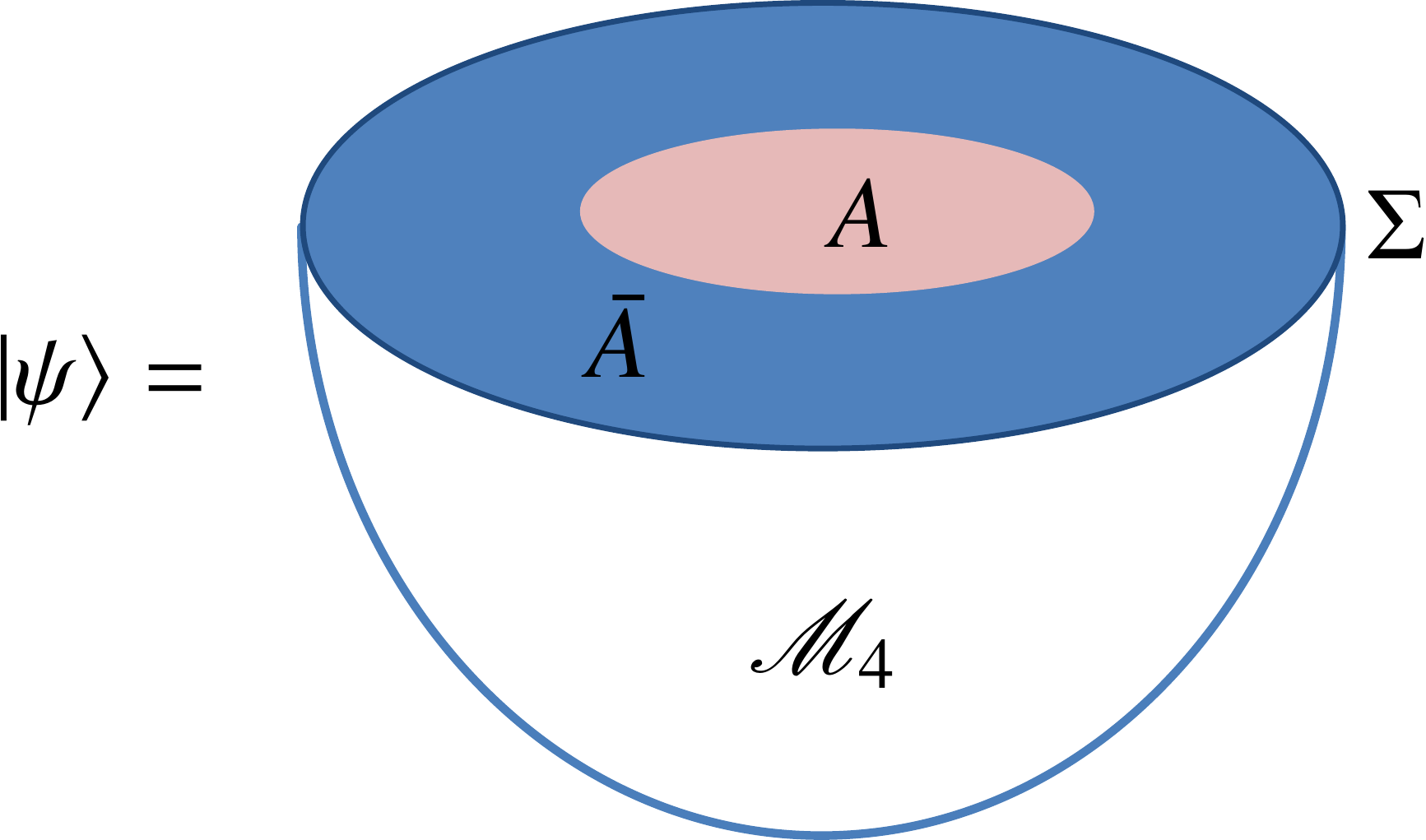}
  \end{center}
  \caption{A 4-manifold (viewed from 5 dimensions) with a boundary 3-manifold $\Sig$. The state $|\psi\rangle$ given by Eq.\Ref{psi0} is constructed by the spinfoam amplitude on a cellular partition $\ck$ of $\sm_4$. The boundary $\Sig$ is subdivided into region $A$ and its complement $\bA$. The subdivision $A$ and $\bA$ is adapted to $\ck$, in the sense that the boundary $\cs$ between $A$ and $\bA$ is triangulated by $\Delta$'s, each of which is made by a large number of facets $f$ in $\ck$.}
  \label{manifold}
  \end{figure}

Spinfoam amplitudes can be used to construct quantum states in LQG Hilbert space. Given a 4-manifold $\sm_4$ with a spatial boundary $\Sig$ as in FIG.\ref{manifold}, we make an arbitrary cellular decomposition of $\sm_4$. The cellular complex is denoted by $\ck$. Spinfoam amplitudes can be defined on $\ck$ and denoted by $A(\ck)_{\vec{j},\vec{i}}$ where $\vec{j},\vec{i}$ are spins and intertwiners coloring the boundary dual complex $\partial\ck^*$. On the other hand, $\Sig$ associates a LQG kinematical Hilbert space $\ch_\Sig$ in which spin-network states $T_{\cg,\vec{j},\vec{i}}(\vec{U})$ for all graphs $\cg$ colored by $\vec{j},\vec{i}$. $\vec{U}$ are SU(2) holonomies along links of $\cg$. We define a linear combination of $T_{\cg,\vec{j},\vec{i}}$ by identifying $\cg=\partial\ck^*$ and letting the coefficients are $A(\ck)_{\vec{j},\vec{i}}$:
\be
\Psi_\ck(\vec{U}) =\sum_{\vec{j},\vec{i}}A(\ck)_{\vec{j},\vec{i}} T_{\partial\ck^*,\vec{j},\vec{i}}(\vec{U}).\label{Psi1}
\ee
One may even consider to sum over the cellular decomposition and define $\Psi(\vec{U})=\sum_{\ck} \Psi_\ck(\vec{U})$. If we truncate the sum in $\Psi_\ck$ (or $\Psi$) to be finite,  $\Psi_\ck$ (or $\Psi$) is a state in the kinematical Hilbert space $\ch_\Sig$. If the sum in $\Psi_\ck$ are kept infinite, $\Psi_\ck$ may be not normalizable in $\ch_\Sig$, but one may anticipate that $\Psi_\ck$ is a physical state living in the dual space of a dense subspace in $\ch_\Sig$. $\Psi_{\ck}$ may be viewed as a spinfoam analog of the Hartle-Hawking wave function. 

When $\sm_4$ has several disconnected boundaries $\Sig_1,\Sig_2,\cdots,\Sig_n$ in additional to $\Sig$, a cellular decomposition $\ck$ of $\sm_{4}$ induces boundary dual complexes $\partial\ck^*_{1},\cdots, \partial\ck^*_{n}$. A state $\Psi_\ck(\vec{U})$ on $\Sig$ can be defined by choices of (initial) states $\phi_a\in \ch_{\Sig_a}$ $(a=1,\cdots,n)$, whose spin-network decompositions are $\phi_a=\sum_{\vec{j}_a,\vec{i}_a}\prod_l (2j_l+1)(\phi_a)_{\vec{j}_a,\vec{i}_a}T_{\partial\ck_a^*,\vec{j}_a,\vec{i}_a}$. $\phi_a$ is based on a single graph $\partial\ck_a^*$. $\Psi_\ck(\vec{U})$ can be constructed as
\be
\Psi_\ck(\vec{U}) =\sum_{\vec{j},\vec{i}}\sum_{\{\vec{j}_a,\vec{i}_a\}_{a=1}^n}\prod_{a=1}^n (\phi_a)_{\vec{j}_a,\vec{i}_a}\,A(\ck)_{\vec{j},\vec{i},\{\vec{j}_a,\vec{i}_a\}_{a=1}^n} T_{\partial\ck^*,\vec{j},\vec{i}}(\vec{U}). \label{Psi2}
\ee

It is useful to write Eqs.\Ref{Psi1} and \Ref{Psi2} in terms of coherent intertwiners. For instance, if we consider $\ck$ whose cells are $B_4$ as in Section \ref{complex}, and apply the spinfoam amplitude $A(\ck)$ as in Eq.\Ref{Z}, $\Psi_{\ck}(\vec{U})$ in Eq.\Ref{Psi1} can be written as
\be
\Psi_{\ck}(\vec{U})=\sum_{\{j_f\}}\prod_{f}A_\Delta(j_f)\int[\rmd g^\pm_{v\a}\rmd \xi_{\a f}]\,e^{S}\, T_{\partial\ck^*,\vec{j},\vec{\xi}}(\vec{U}),\label{Psi3}
\ee
while Eq.\Ref{Psi2} can be written analogously. In Eq.\Ref{Psi3}, $\sum_{\{j_f\}}$ and $\int\rmd \xi_{\a f}$ integrate all internal and boundary $j_f$'s and $\xi_{\a f}$'s. gauge symmetries of the integrand 
\be
g_{v\a}^\pm\sim g_{v\a}^\pm h_\a,\quad \xi_{\a f}\sim h_\a^{-1}\xi_{\a f},\quad \forall\ h_\a\in\Su,\quad \xi_{\a f}\sim e^{i\varphi }\xi_{\a f}\label{gaugeb}
\ee 
apply to both internal and boundary $\a$. $T_{\vec{j},\vec{\xi}}(\vec{U})$ are spin-network states with coherent intertwiners (see Appendix \ref{face} for convention):
\be
T_{\partial\ck^*,\vec{j},\vec{\xi}}(\vec{U})=\tr\lt[\bigotimes_{f\in\partial\ck} R^{j_f}(U_f) \bigotimes_{\a\in\partial\ck}\lt|\big|\{j_f\},\{\xi_{\a f}\}\rt\rangle\bigotimes_{\b\in\partial\ck }\lt\langle\{j_f\},\{\xi_{\b f}\}\big|\rt|\rt]
\ee
where $||\{j_f\},\{\xi_{\a f}\}\rangle $ are coherent intertwiners at polyhedra $\a\in\partial\ck$, and are bras or kets depending on the orientation of spin-network graph. $R^{j_f}(U_f)$ satisfies the following normalization:
\be
\int_{\Su}\rmd U\,\overline{ R_{mn}^j(U)}\,R_{m'n'}^{j'}(U)=\frac{1}{\dim(j)}\delta_{j,j'}\delta_{mm'}\delta_{nn'}.\label{normalizationR}
\ee

\subsection{Truncated states $|\psi\rangle$ with parallel restriction}

In the following we always consider states constructed by spinfoam amplitudes on a fixed cellular complex $\ck$, plus certain truncations. The resulting states are inside $\ch_\Sig$. We again focus on $\ck$ whose 4-cells are $B_4$. The boundary $\partial\ck$ is a polyhedral decomposition of $\Sig$. 

We apply the following truncations to $\Psi_\ck$: (1) The sum $\sum_{\{j_f\}}$ is constrained by $\sum_{f\in\Delta}j_f=J_\Delta$ with fixed $J_\Delta$ at every $\Delta$; (2) The integral of $\int[\rmd g^\pm_{v\a}\rmd \xi_{\a f}]$ is over a neighborhood $\Fn_{g,\xi}$ (of both internal and boundary variables) at an isolated critical point $(g^\pm_{v\a},\xi_{\a\Delta})_c[J_\Delta]\in\mathscr{G}$ of $S_0$ (the critical point is isolated in the space of $g^\pm_{v\a},\xi_{\a f}$ including boundary $\xi_{\a f}$). $\Fn_{g,\xi}$ only contains a single critical point\footnote{$\psi$ contains integral over boundary $\vec{\xi}$, different boundary data $\vec{\xi}$ might lead to different critical points for the integral over $g_{v\a}^\pm$ and internal $\xi_{\a f}$. Here the assumption that $\Fn_{g,\xi}$ only contains a single critical point means that arbitrary changes of boundary data $\vec{\xi}$ within $\Fn_{g,\xi}$ do not lead to any other critical point in $\Fn_{g,\xi}$ different from $(g^\pm_{v\a},\xi_{\a\Delta})_c[J_\Delta]$. $\Fn_{g,\xi}$ satisfying the requirement is nontrivial. Indeed if an infinitesimal change of boundary data $\vec{\xi}$ leads to another critical point in $\Fn_{g,\xi}$ different from $(g^\pm_{v\a},\xi_{\a\Delta})_c[J_\Delta]$. The new critical point has to be infinitesimally close to $(g^\pm_{v\a},\xi_{\a\Delta})_c[J_\Delta]$ otherwise this new critical point can be excluded by redefining $\Fn_{g,\xi}$. But it violates the assumption that $(g^\pm_{v\a},\xi_{\a\Delta})_c[J_\Delta]$ is isolated. 
}. The critical data $(g^\pm_{v\a},\xi_{\a\Delta})_c[J_\Delta]$ is a gauge equivalence class by Eq.\Ref{gaugeb} and other gauge transformations mentioned in Section \ref{complex}. $(g^\pm_{v\a},\xi_{\a\Delta})_c[J_\Delta]$ include the data of boundary $J_\Delta,\xi_{\a\Delta}$. (3) We impose the parallel restriction to $\xi_{\a f}$ by a real gauge invariant potential $V_{\a,\Delta}(\xi_{\a f})$ at every pair of internal and boundary $\a,\Delta$, such that the minimum $V_{\a,\Delta}(\xi_{\a f})=0$ gives the parallel restriction. The truncated state is denoted by $\psi$:
\be
\psi(\vec{U})=\sideset{}{'}\sum_{\{j_f\}}\prod_{f}A_\Delta(j_f)\int_{\Fn_{g,\xi}}[\rmd g^\pm_{v\a}\rmd \xi_{\a f}]\,e^{S}\prod_{\a,\Delta} e^{-N V_{\a,\Delta}}\, T_{\partial\ck^*,\vec{j},\vec{\xi}}(\vec{U}),\label{psi0}
\ee
An example of $V_{\a,\Delta}(\xi_{\a f})$ may be an analog of the 2d spin-chain Hamiltonian: $V_{\a,\Delta}(\xi_{\a f})=\sum_{\langle f,f'\rangle}(1-\vec{n}_{\a f}\cdot\vec{n}_{\a f'})$ where $\langle f,f'\rangle$ are close-neighbor pairs. Our following discussion doesn't reply on details of $V_{\a,\Delta}$. $N$ is of the same order of magnitude as $J_\Delta$. $\sum_{\{j_f\}}'$ only sums nonzero $j_f$ in Eq.\Ref{psi0}. $\sum_{\{j_f\}}'$ constrained by $\sum_{f\in\Delta}j_f=J_\Delta$ is a finite sum, so $\psi\in\ch_{\Sig}$.

Sending the coupling constant of $V_{\a,\Delta}$ to infinity $N\to\infty$ independent of $J_\Delta$ imposes strongly the parallel restriction which reduces the vertex amplitude used in Eq.\Ref{psi0} to the EPRL-FK 4-simplex amplitude. Eq.\Ref{psi0} is a generalization from the following analog using large-$J$ EPRL-FK amplitudes on the simplicial complex $\ck_s$:
\be
\psi_{\text{EPRL-FK}}(\vec{U})=\prod_{\Delta}A_\Delta(J_\Delta)\int_{\Fn_{g,\xi}}[\rmd g^\pm_{v\a}\rmd \xi_{\a \Delta}]\,e^{S_0}\, T_{\partial\ck^*_s,\vec{J},\vec{\xi}}(\vec{U}).\label{psiEPRL}
\ee
The generalization from $\psi_{EPRL-FK}$ to $\psi$ releases mildly the DOFs of non-parallel $\xi_{\a f}$'s in $\Delta$, but releases a large number of micro-DOFs of small $j_f$'s at every $\Delta$. Spinfoam amplitude with the parallel restriction imposed by $V_{\a\Delta}$ is constructed for the purpose of defining $\psi$ which has the semiclassical property discussed below and gives interesting entanglement entropy (see Section \ref{EE}). The computation of the amplitude without the parallel restriction is discussed in Section \ref{beyond}.

Given that $\psi$ associates to a unique critical point $(g^\pm_{v\a},\xi_{\a\Delta})_c[J_\Delta]$, when $(g^\pm_{v\a},\xi_{\a\Delta})_c[J_\Delta]\in\mathscr{G}$ corresponds to a Regge spacetime geometry, $\psi$ may be viewed as a semiclassical state associated to the Regge spacetime geometry. Indeed if the boundary $\vec{j},\vec{\xi}$ in $T_{\partial\ck^*,\vec{j},\vec{\xi}}$ are consistent with the boundary data of $(g^\pm_{v\a},\xi_{\a\Delta})_c[J_\Delta]$, its coefficient gives 
\be
&&\sideset{}{'}\sum_{\{j_f\}}\prod_{f}A_\Delta(j_f)\int_{\Fn_{g,\xi}}[\rmd g^\pm_{v\a}\rmd \xi_{\a f}]\,e^{S}\prod_{\a,\Delta} e^{-N V_{\a,\Delta}}\nonumber\\
&=& \int_{\Fn_{g,\xi}}[\rmd g^\pm_{v\a}\rmd \xi_{\a \Delta}]\lt(\frac{2\pi}{N}\rt)^{\sum_{\Delta\in i(\ck)}\lt(N_\Delta-1\rt)}\frac{e^{S_0} }{\sqrt{\det H_V(\xi_{\a\Delta})}} \sideset{}{'}\sum_{\{j_f\}}\prod_{f}A_\Delta(j_f) \lt[1+O\lt(\frac{1}{N}\rt)\rt]\nonumber\\
&=& \lt(\frac{2\pi}{N}\rt)^{24 N_v+2\sum_{\Delta\in\ck}N_\Delta}\frac{e^{S_0|_c}}{\sqrt{\det\lt( H_V|_c\rt)\det \lt(-H_{0}|_c\rt)}} \sideset{}{'}\sum_{\{j_f\}}\prod_{f}A_\Delta(j_f) \lt[1+O\lt(\frac{1}{N}\rt)\rt]
\ee
In the 1st step we choose a $f_0$ in every $\Delta$ and define $\xi_{\a f_0}\equiv \xi_{\a\Delta}$, then integrate out $\xi_{\a f}$'s ($f\neq f_0$) by $N\gg1$, and reduce $S$ to $S_0$ which depends on $j_f$ only through $J_\Delta$. In the 2nd step we apply the stationary phase approximation of the integral with $S_0$ in $\Fn_{g,\xi}$ which contains a single critical point. $H_V,H_{0}$ are Hessian matrices of $\sum_{\a,\Delta}V_{\a,\Delta}$ and $S_0$, and are assumed to be nondegenerate. If the boundary $\vec{\xi}$ in $T_{\partial\ck^*,\vec{j},\vec{\xi}}$ are away from the boundary data of $(g^\pm_{v\a},\xi_{\a\Delta})_c[J_\Delta]$, critical equations from $S_0$ have no solution in $\Fn_{g,\xi}$, so the integral is suppressed exponentially by large $J_\Delta$. It shows that coefficients in Eq.\Ref{psi0} as a function of boundary $\vec{\xi}$ is peaked at the boundary data of $(g^\pm_{v\a},\xi_{\a\Delta})_c[J_\Delta]$. $\psi$ is a spinfoam analog of Hartle-Hawking state. 

In addition, $\psi$ also explicitly depends on the size of the neighborhood $\Fn_{g,\xi}$. But as we are going to see in a moment, the squared norm of $\psi$ and entanglement entropy only mildly depend on the size $\Fn_{g,\xi}$ through the subleading order. 




\subsection{Squared norm of $|\psi\rangle$}

The squared norm of $|\psi\rangle$ is computed as follows:
\be
\lag\psi|\psi\rag&=&\sideset{}{'}\sum_{\{j_f\},\{j'_f\}}\prod_{f}A_\Delta(j_f)A_\Delta(j'_f)\int_{{\Fn_{g,\xi}\times\Fn_{g,\xi}}}[\rmd g^\pm_{v\a}\rmd g{'}^\pm_{v\a}\rmd \xi_{\a f}\rmd \xi'_{\a f}]\,e^{S+\overline{S}} \prod_{\a,\Delta} e^{-N\lt[ V_{\a,\Delta}(\xi_{\a f})+ V_{\a,\Delta}(\xi_{\a f}')\rt]}\nonumber\\
&&\times\ \prod_{f\subset\Sig}\frac{\delta_{j_f,j_f'}}{2j_f+1}\prod_{\a\subset\Sig}\int_{\Su} \rmd g_{\a}\, e^{\sum_{f\in \partial\a}2j_f\ln\lag\xi_{\a f}'\lt| g_{\a}\rt|\xi_{\a f} \rag}\prod_{\b\subset\Sig}\int_{\Su} \rmd g_{\a}\, e^{\sum_{f\in \partial\a}2j_f\ln\lag\xi_{\b f}\lt| g_{\a}\rt|\xi_{\b f}' \rag}
\ee
where $j_f',g{'}_{v\a}^\pm,\xi'_{\a f}$ denotes variables from $\langle\psi|$. $S $ and $ \overline{S}$ are from $|\psi\rangle$ and $\langle\psi|$, thus depend on unprimed and primed variables respectively. $j'_f=j_f$ for $f\subset\Sig$. $2j_f+1$ in the denominator comes from the normalization Eq.\Ref{normalizationR}. We have applied the integral expressions of inner products between coherent intertwiners:
\be
\lt\langle\{j_f\},\{\xi'_{\a f}\}\rt|\!\lt|\{j_f\},\{\xi_{\a f}\}\rag=\int_{\Su} \rmd g_{\a}\, e^{\sum_{f\in \partial\a}2j_f\ln\lag\xi_{\a f}'\lt| g_{\a}\rt|\xi_{\a f} \rag}.
\ee
$h_\a$ in the integrand of $\lag\psi|\psi\rag$ can be removed by a gauge transformation Eq.\Ref{gaugeb}. 

We may define a total action by collecting all exponents in the integrand:
\be
S_{tot}&=&S\lt[j_f,g_{v\a}^\pm,\xi_{\a f}\rt]+\overline{S\lt[j'_f, {g'}_{v\a}^\pm,\xi'_{\a f}\rt]}-N\sum_{\a,\Delta}\lt[ V_{\a,\Delta}(\xi_{\a f}')+ V_{\a,\Delta}(\xi_{\a f})\rt]\nonumber\\
&&+\sum_{\a\subset\Sig}\sum_{f\in \partial\a}2j_f\ln\langle\xi_{\a f}'|g_\a| \xi_{\a f} \rangle+\sum_{\b\subset\Sig}\sum_{f\in \partial\b}2j_f\ln\langle\xi_{\b f}|g_\b|\xi_{\b f}' \rangle.
\ee
We may choose a $f_0$ in every $\Delta$ and define $\xi_{\a f_0}\equiv \xi_{\a\Delta}$. The large $N$ implements the parallel restriction, and reduces $S$ to $S_0$, $\sum_{f\in \Delta }2j_f\ln\langle\xi_{\a f}'|\xi_{\a f} \rangle$ to $2J_\Delta\ln\langle\xi_{\a \Delta}'|\xi_{\a \Delta} \rangle$ up to $O(1/N)$ after integrating out non-parallel $\xi_{\a f}$'s. The integral in $\lag\psi|\psi\rag$ reduces to
\be
&&\int_{\Fn_{g,\xi}\times\Fn_{g,\xi}}[\rmd g^\pm_{v\a}\rmd g{'}^\pm_{v\a}\rmd \xi_{\a\Delta }\rmd \xi'_{\a\Delta } \rmd g_{\a}\rmd g_{\b}]\frac{\lt({2\pi}/{N}\rt)^{2\sum_{\Delta\in\ck}\lt(N_\Delta-1\rt)}}{\sqrt{\det H_V(\xi_{\a\Delta})\,\det H_V(\xi'_{\a\Delta})}}e^{S_{tot}'}\lt[1+O\lt(\frac{1}{N}\rt)\rt]\label{intparall}\\
&&S_{tot}'=S_0+\overline{S_0}+\sum_{\a\subset\Sig}\sum_{\Delta\subset \a}2J_\Delta\ln\langle\xi_{\a \Delta}'|g_\a|\xi_{\a \Delta} \rangle+\sum_{\b\subset\Sig}\sum_{\Delta\subset \b}2J_\Delta\ln\langle\xi_{\b \Delta}|g_\b|\xi_{\b \Delta}' \rangle\nonumber
\ee
where $H_V(\xi_{\a\Delta})$ is the Hessian matrix of $\sum_{\a,\Delta} V_{\a,\Delta}(\xi_{\a f})$ evaluated at the minimum. Eq.\Ref{intparall} can be computed by stationary phase approximation. The critical equation of this integral is given by Eq.\Ref{critical} from $S_0$ and in addition 
\be
g_\b|\xi_{\b \Delta}'\rangle= e^{i\varphi_{\b\Delta}}|\xi_{\b \Delta} \rangle,\quad\forall\ \Delta\subset\b,\ \b\subset\Sig.\label{additioncritical}
\ee
from $\mathrm{Re}(2J_\Delta\ln\langle\xi_{\b \Delta}|g_\b|\xi_{\b \Delta}' \rangle) =0$. Eq.\Ref{additioncritical} implies that $|\xi_{\b \Delta}'\rangle$ and $|\xi_{\b \Delta} \rangle$ are related by a gauge transformation. A critical point $(g^\pm_{v\a},\xi_{\a\Delta})_c[J_\Delta]$ of $S_0$ gives rise to a critical point of $S_{tot}'$ by double copying, i.e. $({g'}^\pm_{v\a},{\xi'}_{\a\Delta})_c[J_\Delta]=(g^\pm_{v\a},\xi_{\a\Delta})_c[J_\Delta]$ modulo gauge equivalence. A gauge transformation $|\xi_{\b \Delta} \rangle\mapsto  e^{-i\varphi_{\b\Delta}} g_\b|\xi_{\b \Delta} \rangle$, $g_{v\b}^\pm\mapsto g_{v\b}^\pm g_{\b}^{-1}$ identifies $|\xi_{\b \Delta} \rangle=|\xi'_{\b \Delta} \rangle$ by Eq.\Ref{additioncritical}. $\Fn_{g,\xi}$ contains a single critical point $(g^\pm_{v\a},\xi_{\a\Delta})_c[J_\Delta]$ implies that $\Fn_{g,\xi}\times\Fn_{g,\xi}$ contains a single critical point made by double copying. $S_{tot}'$ vanishes at the critical point\footnote{At the critical point, we apply the gauge transformation $|\xi_{\b \Delta} \rangle\mapsto  g_\b|\xi_{\b \Delta} \rangle$ to boundary $\xi_{\b\Delta}$'s and set phase conventions such that $|\xi_{\b \Delta} \rangle=|\xi'_{\b \Delta} \rangle$ (set $\varphi_{\b\Delta}=0$ by gauge transformation). They make $\ln\langle\xi_{\b \Delta}|g_\b|\xi_{\b \Delta}' \rangle$ vanishes, and identify the complex conjugate of $S_0$ to be $\overline{S_0}$. $S_0+\overline{S_0}$ vanishes since $S_0$ is purely imaginary at the critical point.}, so Eq.\Ref{intparall} is estimated by
\be
\lt(\frac{2\pi}{N}\rt)^{24 N_v+2\sum_{\Delta\in\ck}N_\Delta}\frac{1}{\det\lt( H_V|_c\rt)}\frac{1}{\sqrt{\det \lt(-H'_{tot}|_c\rt)}}\lt[1+O\lt(\frac{1}{N}\rt)\rt]\label{asympt}
\ee 
where $N_v$ is the total number of $B_4$ in $\ck$, and $24 N_v+2\sum_{\Delta\in\ck}N_\Delta$ is the total number of integration variables in $\psi(\vec{U})$. $H'_{tot}|_c$ is the Hessian matrix of $S_{tot}'$ evaluated at the critical point, and is assumed to be nondegenerate.  

We observe that the leading order in Eq.\Ref{asympt} depends on $\{j_f\}$ only through their sum $J_\Delta$, so is a constant in the sum over $\{j_f\}$ in $\lag\psi|\psi\rag$. Therefore inserting the above estimate of the integral,
\be
\lag\psi|\psi\rag&=&\lt(\frac{2\pi}{N}\rt)^{24 N_v+2\sum_{\Delta\in\ck}N_\Delta}\frac{1}{\det\lt( H_V|_c\rt)}\frac{1}{\sqrt{\det \lt(-H'_{tot}|_c\rt)}}\prod_{\Delta\in i(\ck_s)} \G_\Delta[J_\Delta]^2\prod_{\Delta\subset\Sig}\G'_\Delta[J_\Delta]\lt[1+O\lt(\frac{1}{N}\rt)\rt],\label{norm1}
\ee
where $i(\ck_s)$ is the interior of the simplicial complex $\ck_s$ determined by $\ck$, and $\G_\Delta,\G_\Delta'$ are given by
\be
\G_\Delta[J_\Delta]&=&\sideset{}{'}\sum_{\{j_{f\in\Delta}\}}\prod_{f}A_\Delta(j_f)=\sideset{}{'}\sum_{\{j_{f\in\Delta}\}}\prod_{f}(2j_f+1)^{n_v(\Delta)+1},\nonumber\\
\G'_\Delta[J_\Delta]&=&\sideset{}{'}\sum_{\{j_{f\in\Delta}\}}\prod_{f}\frac{A_\Delta(j_f)^2}{2j+1}=\sideset{}{'}\sum_{\{j_{f\in\Delta}\}}\prod_{f}(2j_f+1)^{2n_v(\Delta)+3}.
\ee

\section{Analog with Microstate Counting}\label{Darwin-Fowler}

Interestingly, $\G_\Delta[J_\Delta]$ and $\G'_\Delta[J_\Delta]$ are two analogs of counting microstates corresponding to the macrostate ($J_\Delta$,$N_\Delta$), where the microstates are $\{j_f\}$ with degeneracy $(2j_f+1)^{n_v(\Delta)+1}$ and $(2j_f+1)^{2n_v(\Delta)+3}$ at the level $j_f$. Here we list quantities in $\G_\Delta[J_\Delta]$ or $\G'_\Delta[J_\Delta]$ as analogs with quantities in a statistical ensemble of identical systems:
\be
N_\Delta \quad &\leftrightarrow &\quad \text{total number of identical systems in the ensemble}\nonumber\\
J_\Delta \quad &\leftrightarrow &\quad \text{total energy of the ensemble}\nonumber\\
j\quad &\leftrightarrow &\quad \text{energy levels of the system}\nonumber\\
(2j+1)^{2n_v(\Delta)+1}\ \text{or}\ (2j+1)^{2n_v(\Delta)+3} \quad &\leftrightarrow &\quad \text{degeneracy at each energy level}\nonumber\\
\G_\Delta[J_\Delta]\ \text{or}\ \G'_\Delta[J_\Delta]\quad &\leftrightarrow &\quad \text{total number of microstates in the ensemble.}
\ee
$\G_\Delta[J_\Delta]$ and $\G'_\Delta[J_\Delta]$ is similar to the black hole entropy counting in LQG \cite{GP2011}

Here we focus on computing the boundary contribution $\G'_\Delta[J_\Delta]$. We define $n_j$ to be the number of facets $f$ carrying the nonzero spin $j$. 
\be
\G'_\Delta\lt[J_\Delta\rt]=\sideset{}{'}\sum_{\{j_{f\in\Delta}\}}\prod_{f\in\Delta}g_\Delta(j_f)=\sideset{}{'}\sum_{\{n_j\}}N_\Delta ! \prod_{j\neq 0}\frac{g_\Delta(j)^{n_j}}{n_j!},\quad 
g_\Delta(j)=\frac{A_\Delta(j)^2}{2j+1}=(2j+1)^{2n_v(\Delta)+3}\label{Gammaprime0}
\ee
where $\sum_{j=1/2}^\infty j n_j=J_\Delta$ and $\sum_{j=1/2}^\infty n_j=N_\Delta$ is imposed to $\sum_{\{n_j\}}'$. $\G_\Delta[J_\Delta]$ is computed by simply replacing $g_\Delta(j)$ by $(2j+1)^{2n_v(\Delta)+1}$. Following the Darwin-Fowler method in statistical mechanics (see e.g. \cite{kerson}), we define the generating functional 
\be
\sum_{J_\Delta=1/2}^\infty\G'_\Delta\lt[J_\Delta\rt] z^{2J_\Delta}=\sum_{\{n_j\}}N_\Delta ! \prod_{j=1/2}^\infty\frac{g_\Delta(j)^{n_j}z^{2n_jj}}{n_j!}=\Bigg[\sum_{j=1/2}^\infty z^{2j} g_\Delta(j)\Bigg]^{N_\Delta}
\ee
where $\sum_{J_\Delta=1/2}^\infty$ relaxes the constraint $\sum_{j=1/2}^\infty j n_j=J_\Delta$ on $\sum_{\{n_j\}}$. $\sum_{\{n_j\}}$ only satisfies one constraint $\sum_{j=1/2}^\infty n_j=N_\Delta$. $\sum_{j=1/2}^\infty z^{2j} g_\Delta(j)$ has a nonzero radius of convergence, so is an analytic function of $z$ at a neighborhood at $z=0$. $\G'_\Delta\lt[J_\Delta\rt]$ is given by a contour integral
\be
\G'_\Delta\lt[J_\Delta\rt]&=&\frac{1}{2\pi i} \oint_{z=0}\rmd z \frac{1}{z^{2J_\Delta+1}}\Bigg[\sum_{j=1/2}^\infty z^{2j} g_\Delta(j)\Bigg]^{N_\Delta}\nonumber\\
&=&\frac{1}{2\pi i} \oint_{z=0} \rmd z\, \exp\lt(N_\Delta\ln\Bigg[\sum_{j=1/2}^\infty z^{2j} g_\Delta(j)\Bigg]-(2J_\Delta+1)\ln(z) \rt).
\ee
The integration contour is a circle inside the domain where the generating function is analytic. The exponent in the integrand is bounded along the contour. Given that both $N_\Delta,J_\Delta\gg 1$, the above integral can be computed by the method of steepest descent: If we denote the exponent by 
\be
N_\Delta f(z)\simeq N_\Delta\ln\Bigg[\sum_{j=1/2}^\infty z^{2j} g_\Delta(j)\Bigg]-2J_\Delta\ln(z)
\ee
The variational principle $\partial_z f(z_0)=0$ gives
\be
\frac{\sum_{j=1/2}^\infty j\, z_0^{2j} g_\Delta(j)}{\sum_{j=1/2}^\infty z_0^{2j} g_\Delta(j)}=\frac{J_\Delta}{N_\Delta},\label{vari}
\ee
There is always a solution on the positive real axis, $z_0>0$, which maximizes the integrand on the circle \cite{kerson}. We denote by
\be
z_0= e^{-\b_\Delta/2},\quad\text{and}\quad e^{\mu_\Delta}=\sum_{j=1/2}^\infty z_0^{2j} g_\Delta(j).
\ee
The integral can be approximated by 
\be
\Gamma'_\Delta\lt[J_\Delta\rt]= e^{N_\Delta f(z_0)}\lt(\frac{1}{{2\pi N_\Delta f''(z_0)}}\rt)^{\half}\lt[1+O\lt(\frac{1}{N_\Delta}\rt)\rt],\quad N_\Delta f(z_0)\equiv\mu_\Delta N_\Delta+\b_\Delta J_\Delta \label{sumz0}
\ee
where 
\be
f''(z_0)\simeq \frac{\sum_{j=1/2}^\infty 2j(2j-1)\, z_0^{2j-2 }g_\Delta(j)}{\sum_{j=1/2}^\infty z_0^{2j }g_\Delta(j)}-4\frac{{J_\Delta}^2/{N_\Delta}^2-{J_\Delta}/{N_\Delta}}{z^2_0}
\ee
In all following numerical computations of $z_0$, we always check that $f''(z_0)\neq 0$. The following gives examples of solutions $z_0$ at different $n_v(\Delta)$ and $J_\Delta/N_\Delta$:

\begin{table}[h]
\caption{Solutions $z_0$ maximizing $f(z_0)$ at different $n_v(\Delta)$ and $J_\Delta/N_\Delta$ ($f''(z_0)$ are all nonzero).}
\begin{center}
\begin{tabular}{|c|c|c|c|c|c|}
\hline
                  & $J_\Delta/N_\Delta$ = 0.6 & $J_\Delta/N_\Delta$ = 0.7 & $J_\Delta/N_\Delta$ = 0.8 &  $J_\Delta/N_\Delta$ = 0.9 & $J_\Delta/N_\Delta$ = 1\\
\hline
$n_v(\Delta)=1$   & $z_0=0.0257781 $   &  $z_0=0.0505039$   &   $z_0=0.0742575$  & $z_0=0.0971007$    &  $z_0=0.119083 $           \\
\hline
$n_v(\Delta)=2$   & $z_0=0.0119832$   &  $z_0=0.0244767$  & $z_0= 0.0374077$  &  $z_0=0.0506988$ & $z_0= 0.0642717$ \\
\hline
$n_v(\Delta)=3$   & $z_0=0.00552678$   &  $z_0=0.0117148$  & $z_0= 0.0185671$  &  $z_0=0.0260657$ & $z_0= 0.0341736$ \\
\hline
\end{tabular}
\end{center}
\label{}
\end{table}%

\section{Entanglement R\'enyi Entropy}\label{EE}

\subsection{Second R\'enyi entropy}

\begin{figure}[t]
  \begin{center}
  \includegraphics[width = 0.7\textwidth]{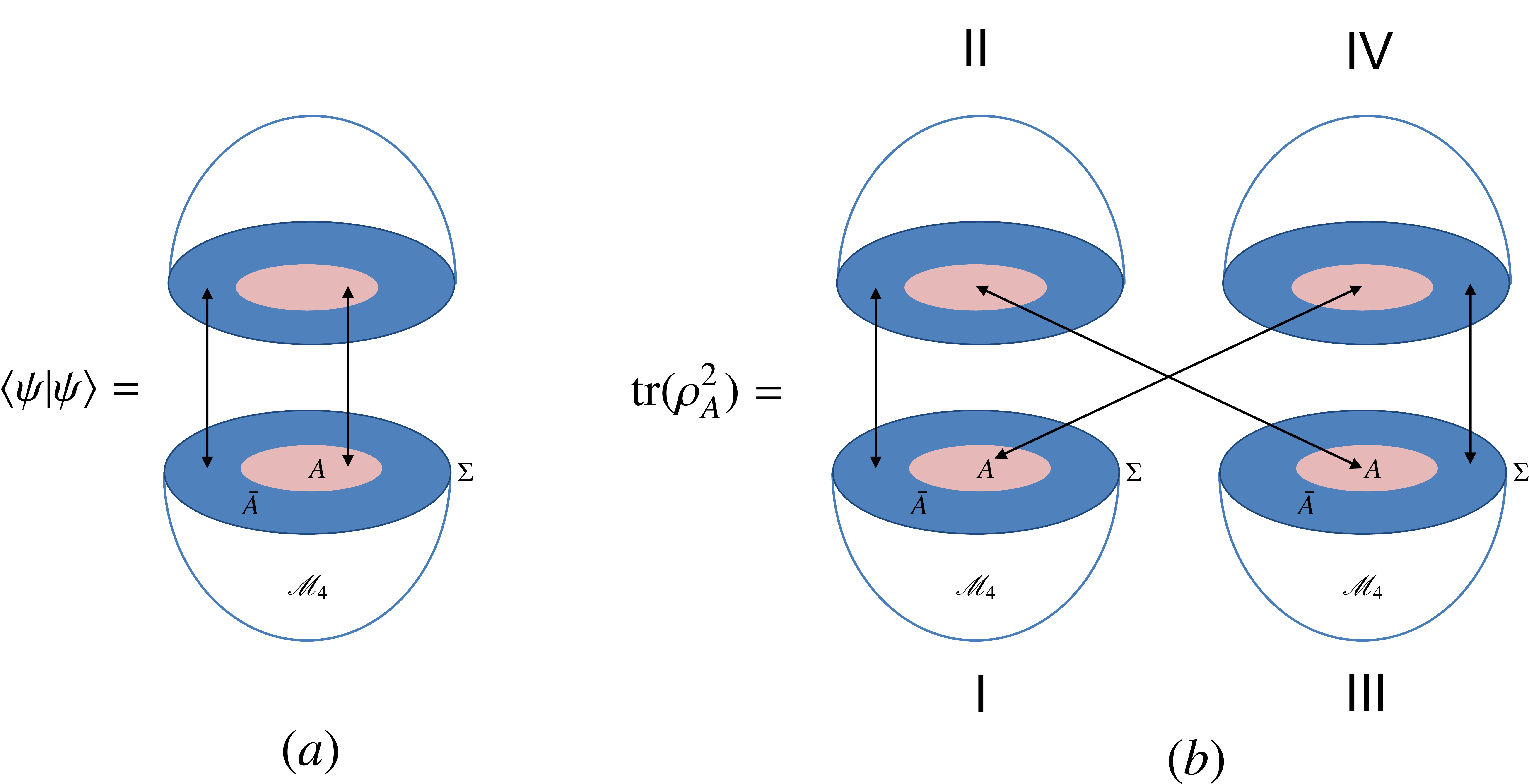}
  \end{center}
  \caption{(a) The inner product $\langle\psi|\psi\rangle$ is taken in both $\ch_A$ and $\ch_{\bA}$ between 2 copies of $|\psi\rangle$; (b) In $\tr(\rho_A^2)$, the inner products in $\ch_{\bA}$ are taken between copies I and II and between III and IV of $|\psi\rangle$, while the inner products in $\ch_A$ are taken between copies I and IV and between II and III. If the inner products are understood as gluing manifolds and their path integrals, the manifold for $\tr(\rho_A^2)$ has a branch cut whose branch points make the boundary $\cs$ between $A$ and $\bA$. }
  \label{replica}
  \end{figure}

  \begin{figure}[t]
  \begin{center}
  \includegraphics[width = 0.8\textwidth]{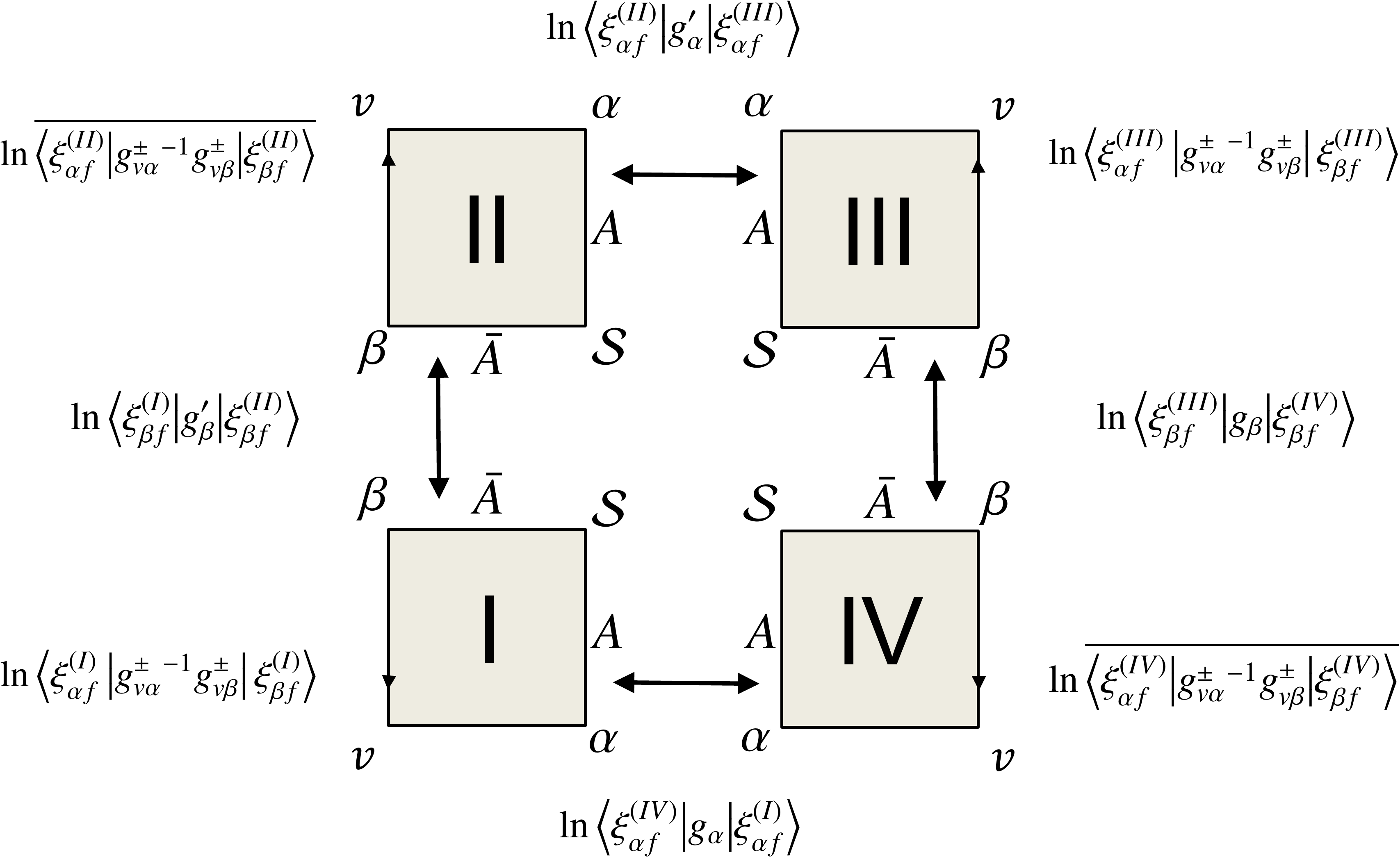}
  \end{center}
  \caption{The situation that $f\subset\cs$ is contained in a single $B_4$ in $\ck$, the figure draws 4 copies of faces in $\ck^*$ dual to a $f\subset\cs$ from 4 copies of $\psi$ in computing the second R\'enyi entropy. $U_A, U_\bA, U_A', U_\bA'$ in Eq.\Ref{RRRR} are holonomies along links labeled by $A$, $\bA$. Integrating these holonomies glues 4 copies of dual faces.}
  \label{square}
  \end{figure}

We subdivide the boundary slice $\Sig$ into 2 subregions $A$ and $\bA$ (FIG.\Ref{manifold}). The subdivision is assumed to compatible to the complexes $\ck$ and $\ck_s$, in the sense that the boundary $\cs$ between $A$ and $\bA$ are triangulated by triangles $\Delta\in\ck_s$, each of which is made by a large number of facets $f\in\ck$. Thus the spin-network functions $T_{\vec{j},\vec{\xi}}(\vec{U})$ in the definition of $\psi(\vec{U})$ are defined on graph $\cg_0=\partial\ck^*$ which have (many) links intersecting $\cs$, while $\cs$ doesn't intersect the spin-network nodes. 

We improve the spin-network graph $\cg_0$ by including all intersecting points $n_\cs=l\cap\cs$ between $\cs$ and links. $n_\cs$ breaks $l$ into 2 links $l_1,l_2$. The improved graph is denoted by $\cg$. By the cylindrical consistency, all $T_{\vec{j},\vec{\xi}}(\vec{U})$ are also spin-networks on the improved graph $\cg$, since all $U_l$ along links intersecting $\cs$ can be decomposed into $U_l=U_{l_1}U_{l_2}$.

The boundary Hilbert space $\ch_\Sig$ is defined as follows: We denote by $L(\cg)$, $L(\cg_ A)$ and $L(\cg_\bA)$ the set of links in $\cg$, $\cg_A=\cg\cap A$, and $\cg_\bA=\cg\cap \bA$, 
\be
&&\ch_\Sig=\ch_A\otimes\ch_{\bA},\quad\text{where}\quad \ch_\Sig=L^2(\Su)^{\otimes |L(\cg)|}/\text{gauge}(\cg_0),\nonumber\\
&&\ch_{A}=L^2(\Su)^{\otimes |L(\cg_A)|}/\text{gauge}(\cg_A),\quad  \ch_\bA=L^2(\Su)^{\otimes |L(\cg_\bA)|}/\text{gauge}(\cg_\bA).
\ee
Here $\text{gauge}(\cg_0)$ only includes gauge transformations acting on nodes in $\cg_0$ (without bivalent nodes $n_\cs$'s). $\text{gauge}(\cg_A)$, $\text{gauge}(\cg_\bA)$ only include gauge transformations acting on nodes in the interior of $A$ and $\bA$. $T_{\vec{j},\vec{\xi}}(\vec{U})$ and $\psi(\vec{U})$ are also gauge invariant at all $n_\cs$'s thus belong to a proper Hilbert subspace in $\ch_\Sig$. However this subspace does not admit a factorization into Hilbert spaces associated to $A$ and $\bA$. Therefore in our discussion of quantum entanglement in $|\psi\rangle$, we view $|\psi\rangle$ as a state in the larger Hilbert space $\ch_\Sig$, although some states in $\ch_\Sig$ are not gauge invariant at bivalent nodes $n_\cs$'s.

We define a reduced density matrix $\rho_A$ from $|\psi\rangle\in \ch_{\Sig}$ by tracing out the DOFs in $\ch_{\bA}$:
\be
\rho_A=\tr_\bA(\rho),\quad \rho=|\psi\rangle\langle\psi|.
\ee
The quantum entanglement in $|\psi\rangle$ can be quantified by the $n$-th R\'enyi entanglement entropy associated to $A$:
\be
S_n(A)=\frac{1}{1-n}\ln\frac{\tr(\rho_A^n)}{\tr(\rho_A)^n}
\ee
The Von Neumann entanglement entropy is given by $S(A)=\lim_{n\to 1}S_n(A)$.

$\tr(\rho_A)=\langle\psi|\psi\rangle$ has been computed above. The following task is to compute $\tr(\rho_A^n)$. Let us firstly focus on the second R\'enyi entropy at $n=2$. The computation is illustrated graphically in FIG.\ref{replica}. $\tr(\rho_A^2)$ is made by inner products among 4 copies of $\psi$. The inner products in $\ch_{\bA}$ take place between copies I and II and between III and IV, while the inner products in $\ch_A$ take place between copies I and IV and between II and III. The inner products of $\tr(\rho_A^2)$ are computed in the same way as the above derivation for $\langle\psi|\psi\rangle$:
\be
\tr(\rho_A^2)&=&\sideset{}{'}\sum_{\{j_f^{(I)}\},\{j_f^{(II)}\},\{j_f^{(III)}\},\{j_f^{(IV)}\}}
\prod_{f}\prod_{a=I}^{IV}A_\Delta(j_f^{(a)}) \int_{\Fn^{\times 4}_{g,\xi}}\lt[\prod_{a=I}^{IV}\rmd g^{(a)\pm}_{v\a}\rmd \xi^{(a)}_{\a f}\rt]\, e^{S^{(I)}+\overline{S^{(II)}}+S^{(III)}+\overline{S^{(IV)}}}\nonumber\\
&&\times\  
\prod_{\a,\Delta} e^{-N\lt[ V_{\a,\Delta}(\xi^{(I)}_{\a f})+ V_{\a,\Delta}(\xi_{\a f}^{(II)})+V_{\a,\Delta}(\xi^{(III)}_{\a f})+ V_{\a,\Delta}(\xi_{\a f}^{(IV)})\rt]} \prod_{f\subset\cs}\frac{1}{(2j_f+1)^3}\prod_{f\subset\Sig\setminus\cs}\frac{1}{2j_f+1}\nonumber\\
&&\times\prod_{\a\subset A}
\lag\{j^{(IV)}_f\},\{\xi^{(IV)}_{\a f}\}\rt|\!\lt|\{j^{(I)}_f\},\{\xi^{(I)}_{\a f}\}\rt\rangle\lag\{j_f^{(II)}\},\{\xi^{(II)}_{\a f}\}\rt|\!\lt|\{j_f^{(III)},\{\xi^{(III)}_{\a f}\}\rt\rangle\nonumber\\
&&\times\prod_{\b\subset\bA}\lt\langle\{j^{(III)}_f\},\{\xi^{(III)}_{\b f}\}\rt|\!\lt|\{j^{(IV)}_f\},\{\xi^{(IV)}_{\b f}\}\rag\lt\langle\{j_f^{(I)}\},\{\xi^{(I)}_{\b f}\}\rt|\!\lt|\{j_f^{(II)}\},\{\xi^{(II)}_{\b f}\}\rag\label{rhoA2111}
\ee
where $j_f^{(a)}$, $g^{(a)\pm}_{v\a}$, and $\xi^{(a)}_{\a f}$ are variables in the $a$-th copy of $\psi$ ($a=I,\cdots,IV$), and $S^{(a)}$ depends on the variables labelled by $a$. We apply the convention in the above formula that $\langle\{j\},\{\xi\}||\{j'\},\{\xi'\}\rangle=\delta^{jj'}\langle\{j\},\{\xi\}| |\{j\},\{\xi'\}\rangle$. A factor ${1}/{(2j_f+1)^3}$ appearing for each $f\subset\cs$ comes from the following inner products at $f$:
\be
&& \int\rmd U_A\rmd U_\bA\rmd U_A'\rmd U_\bA'\sum_{k^{(I)},k^{(II)},k^{(III)},k^{(IV)}} R^{j_f^{(I)}}_{m^{(I)}k^{(I)}}(U_A)R^{j_f^{(I)}}_{k^{(I)}n^{(I)}}(U_\bA)\  \overline{R^{j_f^{(II)}}_{m^{(II)}k^{(II)}}(U_A')}\overline{R^{j_f^{(II)}}_{k^{(II)}n^{(II)}}(U_\bA)}\nonumber\\
&&\quad\quad\quad\quad\quad\quad\quad\quad R^{j_f^{(III)}}_{m^{(III)}k^{(III)}}(U'_A)R^{j_f^{(III)}}_{k^{(III)}n^{(III)}}(U'_\bA)\  \overline{R^{j_f^{(IV)}}_{m^{(IV)}k^{(IV)}}(U_A)}\overline{R^{j_f^{(IV)}}_{k^{(IV)}n^{(IV)}}(U'_\bA)}\nonumber\\
&=&\lt(\frac{1}{2j_f+1}\rt)^4\delta^{j_f^{(I)}j_f^{(II)}}\delta^{j_f^{(II)}j_f^{(III)}}\delta^{j_f^{(III)}j_f^{(IV)}}\delta^{j_f^{(IV)}j_f^{(I)}}\sum_{k^{(I)},k^{(II)},k^{(III)},k^{(IV)}} \delta_{k^{(I)}k^{(II)}}\delta_{k^{(II)}k^{(III)}}\delta_{k^{(III)}k^{(IV)}}\delta_{k^{(IV)}k^{(I)}}\nonumber\\
&& \delta_{n^{(I)}n^{(II)}}\delta_{m^{(II)}m^{(III)}} \delta_{n^{(III)}n^{(IV)}}\delta_{m^{(I)}m^{(IV)}}\nonumber\\
&=&\lt(\frac{1}{2j_f+1}\rt)^3 \delta^{j_f^{(I)}j_f^{(II)}}\delta^{j_f^{(II)}j_f^{(III)}}\delta^{j_f^{(III)}j_f^{(IV)}}\delta^{j_f^{(IV)}j_f^{(I)}} \delta_{n^{(I)}n^{(II)}}\delta_{m^{(II)}m^{(III)}} \delta_{n^{(III)}n^{(IV)}}\delta_{m^{(I)}m^{(IV)}}\label{RRRR}
\ee
where $U_AU_\bA$ is the holonomy along the link intersecting $\cs$ and dual to $f$ in $\Sig$ (see FIG.\ref{square}). The above inner products identify 4 spins of $f$ from 4 different copies of $\psi$: $j_f^{(I)}=j_f^{(II)}=j_f^{(III)}=j_f^{(IV)}=j_f$. The total action in Eq.\Ref{rhoA2111} is given by 
\be
S^{(2)}_{tot}&=&S^{(I)}+\overline{S^{(II)}}+S^{(III)}+\overline{S^{(IV)}}-N\sum_{\a,\Delta}\lt[ V_{\a,\Delta}(\xi^{(I)}_{\a f})+ V_{\a,\Delta}(\xi_{\a f}^{(II)})+V_{\a,\Delta}(\xi^{(III)}_{\a f})+ V_{\a,\Delta}(\xi_{\a f}^{(IV)})\rt]\nonumber\\
&&+\sum_{\a\subset A}\sum_{f\subset\a}2j^{(IV)}_f\ln
\lag \xi^{(IV)}_{\a f}\rt|g_\a\lt|\xi^{(I)}_{\a f}\rt\rangle+\sum_{\a\subset A}\sum_{f\subset\a}2j^{(II)}_f\ln\lag \xi^{(II)}_{\a f}\rt|g_\a'\lt| \xi^{(III)}_{\a f} \rt\rangle\nonumber\\
&&+\sum_{\b\subset\bA}\sum_{f\subset\a}2j^{(III)}_f\ln\lt\langle \xi^{(III)}_{\b f} \rt|g_\b \lt| \xi^{(IV)}_{\b f} \rag++\sum_{\b\subset\bA}\sum_{f\subset\a}2j^{(I)}_f\ln\lt\langle \xi^{(I)}_{\b f} \rt|g_\b'\lt| \xi^{(II)}_{\b f} \rag.\label{stot4}
\ee
The situation at $f\subset\cs$ is illustrated in FIG.\ref{square}. The large $N$ again imposes the parallel restriction to $\xi_{\a f}$ and reduces $S_{tot}^{(2)}$ to
\be
{S'}_{tot}^{(2)}&=&S_0^{(I)}+\overline{S_0^{(II)}}+S_0^{(III)}+\overline{S_0^{(IV)}}\nonumber\\
&&+\sum_{\a\subset A}\sum_{\Delta\subset\a}2J^{(IV)}_\Delta\ln
\lag \xi^{(IV)}_{\a \Delta}\rt|g_\a\lt|\xi^{(I)}_{\a\Delta}\rt\rangle+\sum_{\a\subset A}\sum_{\Delta\subset\a}2J^{(II)}_\Delta\ln\lag \xi^{(II)}_{\a \Delta}\rt|g_\a'\lt| \xi^{(III)}_{\a\Delta} \rt\rangle\nonumber\\
&&+\sum_{\b\subset\bA}\sum_{\Delta\subset\a}2J^{(III)}_\Delta\ln\lt\langle \xi^{(III)}_{\b \Delta} \rt|g_\b \lt| \xi^{(IV)}_{\b \Delta} \rag++\sum_{\b\subset\bA}\sum_{\Delta\subset\a}2J^{(I)}_\Delta\ln\lt\langle \xi^{(I)}_{\b \Delta} \rt|g_\b'\lt| \xi^{(II)}_{\b \Delta} \rag.
\ee
A large-$J_\Delta$ stationary phase analysis similar to $\langle\psi|\psi\rangle$ shows that the integration domain of Eq.\Ref{rhoA2111} again only contain a single critical point, which is 4 copies of $(g_{v\a}^\pm,\xi_{\a \Delta})_c[J_\Delta]$ with their boundary data identified according to FIG.\ref{replica}. $S^{(2)}_{tot}$ vanishes at the critical point. 

The asymptotic behavior of the integral depends on $j_f$ only through their sum $J_\Delta$, so similar to the computation of $\langle\psi|\psi\rangle$, 
\be
\tr(\rho_A^2)&\simeq& \lt(\frac{2\pi}{N}\rt)^{48 N_v+4\sum_{\Delta\in\ck}N_\Delta}\frac{1}{\det\lt( H_V|_c\rt)^2}\frac{1}{\sqrt{\det \lt(-{H'}^{(2)}_{tot}|_c\rt)}} \nonumber\\
&&\times\ \prod_{\Delta\in i(\ck_s)} {\G_\Delta}[J_\Delta]^4\prod_{\Delta\subset i(A)}{\G'_\Delta}[J_\Delta]^2\prod_{\Delta\subset i(\bA)}\G'_\Delta[J_\Delta]^2\prod_{\Delta\subset\cs}{\G}^{(2)}_\Delta[J_\Delta]\,\lt[1+O\lt(\frac{1}{N}\rt)\rt]\label{trrho2}
\ee 
where ${H'}^{(2)}_{tot}|_c$ is the Hessian matrix of ${S'}_{tot}^{(2)}$ evaluated at the critical point and is assumed to be nondegenerate. $\Delta\subset\cs$ are special because they are shared by all 4 copies of $\psi$ in $\tr(\rho_A^2)$. ${\G}^{(2)}_\Delta$ for $\Delta\subset\cs$ is given by
\be
{\G}^{(2)}_\Delta\lt[J_\Delta\rt]=\sideset{}{'}\sum_{\{j_{f\in\Delta}\}}\prod_{f\in\Delta}g^{(2)}_\Delta(j_f)
,\quad 
 g^{(2)}_\Delta(j)=\frac{A_\Delta(j)^4}{(2j+1)^3}=(2j+1)^{4n_v(\Delta)+5},\label{Gammaprime2}
\ee
Similar to $\G'_\Delta$, ${\G}^{(2)}_\Delta$ can also be viewed as an analog of microstate counting, where $g^{(2)}_\Delta(j)$ corresponds to the degeneracy of microstates at the level $j$. The label $(2)$ indicates that it is for computing the second R\'enyi entropy. 
\be
{\G}^{(2)}_\Delta\lt[J_\Delta\rt]\simeq e^{N_\Delta f^{(2)}\lt(z_0^{(2)}\rt)}\lt(\frac{1}{{2\pi N_\Delta {f^{(2)}}''\lt(z_0^{(2)}\rt)}}\rt)^{\half}\lt[1+O\lt(\frac{1}{N_\Delta}\rt)\rt],\quad N_\Delta f^{(2)}\lt(z_0^{(2)}\rt)\equiv\mu^{(2)}_\Delta N_\Delta+\b_\Delta^{(2)} J_\Delta \label{sumz02}
\ee
where $f^{(2)}(z)$ and $z^{(2)}_0$ are given by 
\be
N_\Delta f^{(2)}(z)\simeq N_\Delta\ln\Bigg[\sum_{j=1/2}^\infty z^{2j} g^{(2)}_\Delta(j)\Bigg]-2J_\Delta\ln(z),\ \quad \frac{\sum_{j=1/2}^\infty j\, \lt[z_0^{(2)}\rt]^{2j} g^{(2)}_\Delta(j)}{\sum_{j=1/2}^\infty \lt[z_0^{(2)}\rt]^{2j} g^{(2)}_\Delta(j)}=\frac{J_\Delta}{N_\Delta}.\label{vari2}
\ee
The second equation in Eq.\Ref{vari2} comes from the variation principle of $ f^{(2)}(z)$. We denote
\be
z_0^{(2)}=e^{-\b_\Delta^{(2)}/2},\quad e^{\mu^{(2)}_\Delta}=\sum_{j=1/2}^\infty \lt[z_0^{(2)}\rt]^{2j} g^{(2)}_\Delta(j).
\ee
Table \ref{2} gives examples of solutions $z_0$ at different $n_v(\Delta)$ and $J_\Delta/N_\Delta$.

\begin{table}[h]
\caption{Solutions $z_0$ maximizing $f^{(2)}(z_0)$ at different $n_v(\Delta)$ and $J_\Delta/N_\Delta$ (${f^{(2)}}''(z_0)$ are all nonzero).}
\begin{center}
\begin{tabular}{|c|c|c|c|c|}
\hline
                  & $J_\Delta/N_\Delta$ = 0.6 & $J_\Delta/N_\Delta$ = 0.7 & $J_\Delta/N_\Delta$ = 0.8 &  $J_\Delta/N_\Delta$ = 0.9 
                  \\
\hline
$n_v(\Delta)=1$   & $z_0=0.00552678 $   &  $z_0=0.0117148$   &   $z_0=0.0185671$  & $z_0=0.0260657$    
       \\
\hline
$n_v(\Delta)=2$   & $z_0=0.0011542$   & $z_0= 0.00260368$  &  $z_0= 0.00441412$ & $z_0=0.00664713$ 
\\
\hline
$n_v(\Delta)=3$   & $z_0=0.000236694$   &  $z_0=0.000560573$  & $z_0=0.00100989$  &  $z_0=0.00163299$ 
\\
\hline
\end{tabular}
\end{center}
\label{2}
\end{table}%


Combining Eq.\Ref{trrho2} with Eq.\Ref{sumz02} for $\tr(\rho_A^2)$ and Eq.\Ref{norm1} for $\langle\psi|\psi\rangle=\tr(\rho_A)$ gives the following second R\'enyi entropy:
\be
S_2(A)&=&-\ln\frac{\tr(\rho_A^2)}{\tr(\rho_A)^2}=-\ln\frac{\prod_{\Delta\subset\cs}{\G}^{(2)}_\Delta}{\prod_{\Delta\subset\cs}\G'_\Delta{}^2}\frac{\det(-H_{tot}'|_c)}{\sqrt{\det(-H'{}^{(2)}_{tot}|_c)}}\lt[1+O\lt(\frac{1}{N}\rt)\rt]\nonumber\\
&\simeq&\sum_{\Delta\subset\cs}N_\Delta\lt[2 f(z_0)-f^{(2)}(z_0^{(2)})\rt]=\sum_{\Delta\subset\cs}\lt[\lt(2 \b_\Delta-\b^{(2)}_\Delta\rt) J_\Delta+\lt(2 \mu_\Delta-\mu^{(2)}_\Delta\rt) N_\Delta\rt].
\ee
where $\ln\frac{\det(-H_{tot}'|_c)}{\sqrt{\det(-H'{}^{(2)}_{tot}|_c)}}$ is subleading and negligible as $J_\Delta\sim N_\Delta\gg1$. 

$z_0,z_0^{(2)}$ or $\b_\Delta,\mu_\Delta, \b_\Delta^{(2)},\mu_\Delta^{(2)}$ clearly depends on $J_\Delta,N_\Delta$. If we fix $J_\Delta$ and let $N_\Delta$ vary, 
\be
\frac{\partial \lt[N_\Delta f(z_0)\rt]}{\partial N_\Delta}&=&\mu_\Delta+N_\Delta\lt(\frac{J_\Delta}{N_\Delta}+\frac{\partial\mu_\Delta}{\partial\b_\Delta}\rt)\frac{\partial\b_\Delta}{\partial N_\Delta}=\mu_\Delta,\quad \frac{\partial\mu_\Delta}{\partial\b_\Delta}=\frac{\sum_j e^{-\b_\Delta j}(-j)g_\Delta(j)}{\sum_j e^{-\b_\Delta j}g_\Delta(j)}=-\frac{J_\Delta}{N_\Delta}\nonumber\\
\frac{\partial \lt[N_\Delta f^{(2)}(z_0)\rt]}{\partial N_\Delta}&=&\mu^{(2)}_\Delta+N_\Delta\lt(\frac{J_\Delta}{N_\Delta}+\frac{\partial\mu^{(2)}_\Delta}{\partial\b^{(2)}_\Delta}\rt)\frac{\partial\b^{(2)}_\Delta}{\partial N_\Delta}=\mu^{(2)}_\Delta,\quad \frac{\partial\mu^{(2)}_\Delta}{\partial\b^{(2)}_\Delta}=\frac{\sum_j e^{-\b_\Delta^{(2)} j}(-j)g^{(2)}_\Delta(j)}{\sum_j e^{-\b_\Delta^{(2)} j}g^{(2)}_\Delta(j)}=-\frac{J_\Delta}{N_\Delta}.
\ee
Therefore,
\be
\frac{\partial \lt[2 N_\Delta f(z_0)-N_\Delta f^{(2)}(z_0^{(2)})\rt]}{\partial N_\Delta}&=&2\mu_\Delta-\mu^{(2)}_\Delta
\ee
$S_2(A)$ is extremized at the value of the ratio $J_\Delta/N_\Delta$ which gives $2\mu_\Delta=\mu^{(2)}_\Delta$ at every $\Delta$. The extremal value of $S_2(A)$ gives
\be
S_2(A)\simeq \sum_{\Delta\subset\cs}\lt(2 \b_\Delta-\b^{(2)}_\Delta\rt) J_\Delta.\label{wal}
\ee
If the complex $\ck$ and the entangling surface $\cs$ are chosen such that $n_v(\Delta)$ is a constant for all $\Delta\subset\cs$ (every $\Delta\subset \cs$ is shared by the same number of $B_4$'s), $\b_\Delta,\b^{(2)}_\Delta$ are constants independent of $\Delta$, in this case, $S_2(A)$ satisfies the area-law
\be
S_2(A)\simeq c \sum_{\Delta\subset\cs} J_\Delta =\frac{c}{8\pi\g\ell_P^2}\mathbf{a}_\cs, \quad c=2 \b_\Delta-\b^{(2)}_\Delta,
\ee
where $\mathbf{a}_\cs=8\pi\g\ell_P^2\sum_{\Delta\subset\cs} J_\Delta$ is the total area of $\cs$. The relation between $\mathbf{a}_\cs$ and $J_\Delta$ is given by the geometrical interpretation of the critical point $(g_{v\a}^\pm,\xi_{\a\Delta})_c[J_\Delta]\in\mathscr{G}$. But in general the extremal $S_2(A)$ may satisfy a weighted area-law Eq.\Ref{wal} with different weights $2 \b_\Delta-\b^{(2)}_\Delta$ at different $\Delta$.

\begin{figure}[h]
  \begin{center}
  \includegraphics[width = 0.8\textwidth]{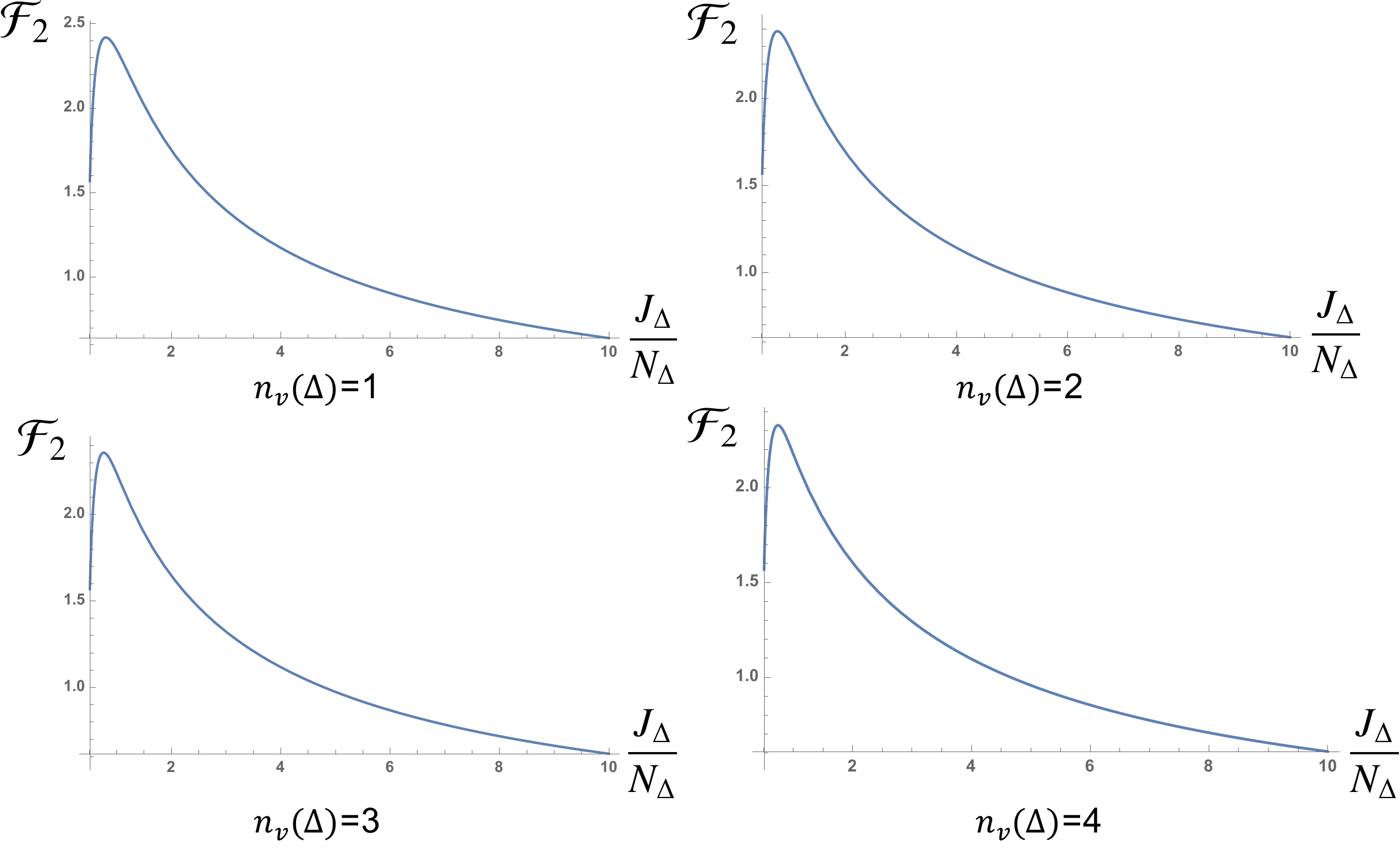}
  \end{center}
  \caption{Plots of $\cf_2[n_v(\Delta),{J_\Delta}/{N_\Delta}]$ in Eq.\Ref{cf} at $n_v(\Delta)=1,\cdots,4$ and ${J_\Delta}/{N_\Delta}\in[0.51,10]$.}
  \label{plot}
  \end{figure}

To see if $2\mu_\Delta=\mu^{(2)}_\Delta$ maximizes $S_2(A)$, we compute the second derivative:
\be
&&\frac{\partial^2 \lt[N_\Delta f(z_0)\rt]}{\partial N_\Delta^2}=\frac{\partial\mu_\Delta}{\partial\b_\Delta}\frac{\partial\b_\Delta}{\partial N_\Delta}=-\frac{1}{N_\Delta}\frac{1}{\frac{N^2_\Delta}{J_\Delta^2}\lag j^2\rag-1},\quad\quad\quad \lag j^2\rag\equiv\frac{\sum_{j=1/2}^\infty j^2\, e^{-\b j} g_\Delta(j)}{\sum_{j=1/2}^\infty  e^{-\b j} g_\Delta(j)}\nonumber\\
&&\frac{\partial^2 \lt[N_\Delta f^{(2)}(z_0^{(2)})\rt]}{\partial N_\Delta^2}=\frac{\partial\mu^{(2)}_\Delta}{\partial\b^{(2)}_\Delta}\frac{\partial\b^{(2)}_\Delta}{\partial N_\Delta}=-\frac{1}{N_\Delta}\frac{1}{\frac{N^2_\Delta}{J_\Delta^2}\lag j^2\rag^{(2)}-1},\quad\quad\quad \lag j^2\rag^{(2)}\equiv\frac{\sum_{j=1/2}^\infty j^2\, e^{-\b^{(2)}_\Delta j} g^{(2)}_\Delta(j)}{\sum_{j=1/2}^\infty  e^{-\b^{(2)}_\Delta j} g^{(2)}_\Delta(j)}\nonumber\\
&&\frac{\partial^2 \lt[2N_\Delta f(z_0)-N_\Delta f^{(2)}(z_0^{(2)})\rt]}{\partial N_\Delta^2}=\frac{1}{N_\Delta}\lt(\frac{1}{\frac{N^2_\Delta}{J_\Delta^2}\lag j^2\rag^{(2)}-1}-\frac{2}{\frac{N^2_\Delta}{J_\Delta^2}\lag j^2\rag-1}\rt)
\ee 
The following list some values of $J_\Delta/N_\Delta$ which give $2\mu_\Delta=\mu^{(2)}_\Delta$ at different $n_v(\Delta)$:
\be
n_v(\Delta)=1:&&  J_\Delta/N_\Delta=0.802182,\quad 2 \b_\Delta-\b^{(2)}_\Delta=2.41769,\quad N_\Delta\frac{\partial^2 \lt[2N_\Delta f(z_0)-N_\Delta f^{(2)}(z_0^{(2)})\rt]}{\partial N_\Delta^2}= -10.3142\nonumber\\
n_v(\Delta)=2:&&  J_\Delta/N_\Delta=0.782484,\quad 2 \b_\Delta-\b^{(2)}_\Delta=2.38741,\quad N_\Delta\frac{\partial^2 \lt[2N_\Delta f(z_0)-N_\Delta f^{(2)}(z_0^{(2)})\rt]}{\partial N_\Delta^2}=-11.0869\nonumber\\
n_v(\Delta)=3:&&  J_\Delta/N_\Delta=0.762613,\quad 2 \b_\Delta-\b^{(2)}_\Delta=2.35677, \quad N_\Delta\frac{\partial^2 \lt[2N_\Delta f(z_0)-N_\Delta f^{(2)}(z_0^{(2)})\rt]}{\partial N_\Delta^2}= -12.0193\nonumber
\ee
The negative second derivative implies that $2\mu_\Delta=\mu^{(2)}_\Delta$ gives the maximum of $S_2(A)$. FIG.\Ref{plot} plots 
\be
\cf_2\lt[n_v(\Delta),\frac{J_\Delta}{N_\Delta}\rt]:=\frac{N_\Delta}{J_\Delta}\lt[2 f(z_0)- f^{(2)}(z_0^{(2)})\rt],\quad S_2(A)=\sum_{\Delta\subset\cs}J_\Delta\cf_2\lt[n_v(\Delta),\frac{J_\Delta}{N_\Delta}\rt]\label{cf}
\ee 
at different $n_v(\Delta)$, and suggests that when $J_\Delta$ is fixed, $2\mu_\Delta=\mu^{(2)}_\Delta$ indeed give the global maximum of $S_2(A)$.

The above result shows that fixing $J_\Delta$, the second R\'enyi entropy $S_2(A)$, as a function of $N_\Delta$, is in general bounded by an (weighted) area-law:
\be
S_2(A)\leq \sum_{\Delta\subset\cs}\lt(2 \b_\Delta-\b^{(2)}_\Delta\rt) J_\Delta.
\ee
where the bound is saturated at ${J_\Delta}/{N_\Delta}$ which gives $2\mu_\Delta=\mu^{(2)}_\Delta$. The bound becomes an area-law if $n_v(\Delta)$ is a constant for all $\Delta\subset\cs$.

\subsection{Higher R\'enyi entropy}

The computation of higher R\'enyi entropy $S_n(A)$ with $n>2$ is a simple generalization of the second R\'enyi entropy computation. $\tr(\rho_A^n)$ includes $2n$ copies of $|\psi\rangle$ or $\langle\psi|$ in the computation illustrated by FIGs.\ref{replica} and \ref{square}. Eq.\Ref{trrho2} is modified to
\be
\tr(\rho_A^n)&\simeq& \lt(\frac{2\pi}{N}\rt)^{24n N_v+2n\sum_{\Delta\in\ck}N_\Delta}\frac{1}{\det\lt( H_V|_c\rt)^n}\frac{1}{\sqrt{\det \lt(-{H'}^{(n)}_{tot}|_c\rt)}} \nonumber\\
&&\times\ \prod_{\Delta\in i(\ck_s)} {\G_\Delta}[J_\Delta]^{2n}\prod_{\Delta\subset i(A)}{\G'_\Delta}[J_\Delta]^n\prod_{\Delta\subset i(\bA)}\G'_\Delta[J_\Delta]^n\prod_{\Delta\subset\cs}{\G}^{(n)}_\Delta[J_\Delta]\,\lt[1+O\lt(\frac{1}{N}\rt)\rt]\label{trrhon}
\ee 
Here ${\G}^{(n)}_\Delta$ for $\Delta\subset\cs$ is computed similar to ${\G}^{(2)}_\Delta$
\be
{\G}^{(n)}_\Delta\lt[J_\Delta\rt]=\sideset{}{'}\sum_{\{j_{f\in\Delta}\}}\prod_{f\in\Delta}g^{(n)}_\Delta(j_f),\quad
 g^{(n)}_\Delta(j)=\frac{A_\Delta(j)^{2n}}{(2j+1)^{2n-1}}=(2j+1)^{2n(n_v(\Delta)+1)+1}.\label{Gammaprimen}
\ee
As a result,
\be
S_n(A)&=&\frac{1}{1-n}\ln\frac{\tr(\rho_A^n)}{\tr(\rho_A)^n}=\frac{1}{1-n}\ln\frac{\prod_{\Delta\subset\cs}{\G}^{(n)}_\Delta}{\prod_{\Delta\subset\cs}\G'_\Delta{}^n}\frac{\det(-H_{tot}'|_c)^n}{\sqrt{\det(-H'{}^{(n)}_{tot}|_c)}}\lt[1+O\lt(\frac{1}{N}\rt)\rt]\nonumber\\
&\simeq&\sum_{\Delta\subset\cs}\lt[\frac{\b^{(n)}_\Delta-\b_\Delta n}{{1-n}} J_\Delta+\frac{\mu^{(n)}_\Delta-\mu_\Delta n}{{1-n}} N_\Delta\rt]
\ee
where $\b_\Delta^{(n)},\mu^{(n)}_\Delta$ satisfies
\be
e^{-\b_\Delta^{(n)}/2}=z_0^{(n)},\quad e^{\mu^{(n)}_\Delta}=\sum_{j=1/2}^\infty \lt[z_0^{(n)}\rt]^{2j} g^{(n)}_\Delta(j).
\ee
and $z_0^{(n)}\in(0,1)$ solves
\be
\frac{\sum_{j=1/2}^\infty j\, \lt[z_0^{(n)}\rt]^{2j} g^{(n)}_\Delta(j)}{\sum_{j=1/2}^\infty \lt[z_0^{(n)}\rt]^{2j} g^{(n)}_\Delta(j)}=\frac{J_\Delta}{N_\Delta}.\label{varin}
\ee
Similar to $S_2(A)$, if we fix $J_\Delta$ and let $N_\Delta$ vary, $S_n(A)$ maximizes at $\mu^{(n)}_\Delta=\mu_\Delta n$ thus is bounded by a weighted area law. 
\be
S_n(A)\leq \sum_{\Delta\subset\cs}\frac{\b^{(n)}_\Delta-\b_\Delta n}{{1-n}} J_\Delta.
\ee
where $J_\Delta$ relates to the area of $\Delta$ by the geometrical interpretation of the critical point $(g_{v\a}^\pm,\xi_{\a \Delta})_c[J_\Delta]$ in defining $|\psi\rangle$. FIG.\Ref{plotS3} plots 
\be
\cf_n\lt[n_v(\Delta),\frac{J_\Delta}{N_\Delta}\rt]:=\frac{N_\Delta}{J_\Delta}\lt[\frac{\b^{(n)}_\Delta-\b_\Delta n}{{1-n}} \frac{J_\Delta}{N_\Delta}+\frac{\mu^{(n)}_\Delta-\mu_\Delta n}{{1-n}} \rt],\quad S_n(A)=\sum_{\Delta\subset\cs}J_\Delta\cf_n\lt[n_v(\Delta),\frac{J_\Delta}{N_\Delta}\rt]\label{cf}
\ee 
at $n=3$ and $n_v(\Delta)=1,\cdots,4$. FIG.\Ref{plotS3} plots $\cf_n$ at ${J_\Delta}/{N_\Delta}=1$, $n_v(\Delta)=1,2$, and $n=2,\cdots,7$.

\begin{figure}[h]
  \begin{center}
  \includegraphics[width = 0.75\textwidth]{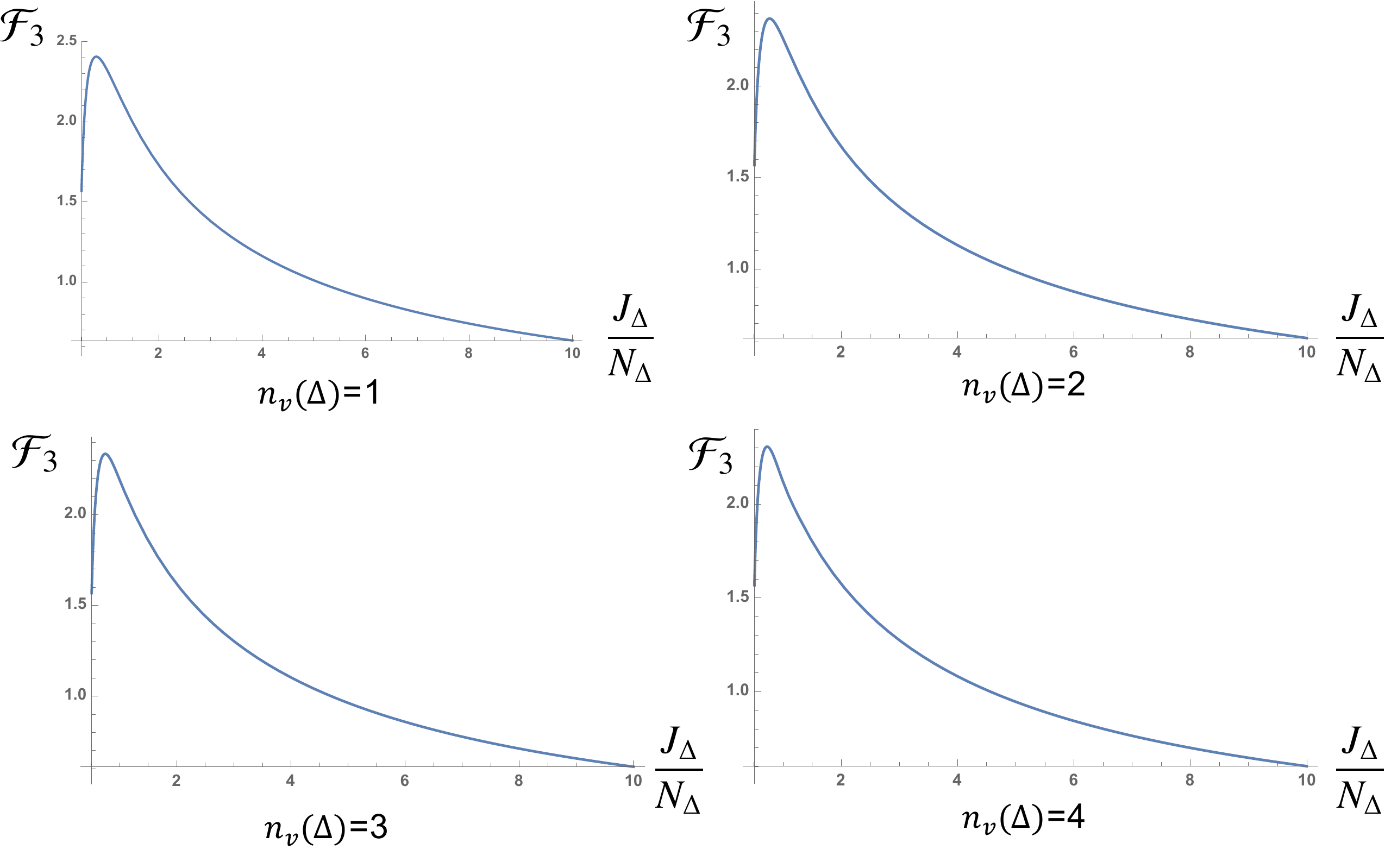}
  \end{center}
  \caption{Plots of $\cf_3[n_v(\Delta),{J_\Delta}/{N_\Delta}]$ at $n_v(\Delta)=1,\cdots,4$ and ${J_\Delta}/{N_\Delta}\in[0.51,10]$.}
  \label{plotS3}
  \end{figure}

\begin{figure}[h]
  \begin{center}
  \includegraphics[width = 0.75\textwidth]{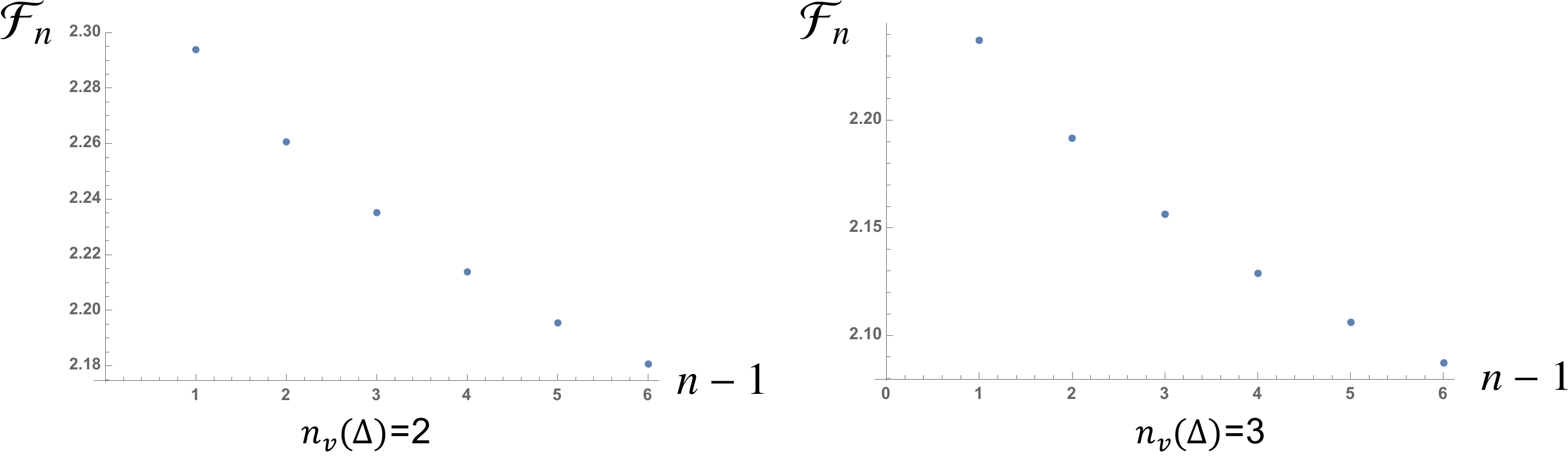}
  \end{center}
  \caption{Plots of $\cf_n[n_v(\Delta),{J_\Delta}/{N_\Delta}=1]$ at $n_v(\Delta)=1,2$, and $n=2,\cdots,7$.}
  \label{plotS3}
  \end{figure}

\section{Analogous Thermodynamical First Law}\label{1stlaw}

The R\'enyi entanglement entropy $S_n(A)$ derived in the last section is a function of the ``macrostate'' $J_\Delta,N_\Delta$ has interesting analog with entropy in thermodynamics. In Section \ref{Darwin-Fowler}, we give an analog between $J_\Delta,N_\Delta$ and the total energy and total number of identical systems of a statistical ensemble. 

\begin{Theorem}
The differential of $S_n(A)$ with respect to $J_\Delta,N_\Delta$ gives the following analog of the thermodynamical first law:
\be
\delta S_n(A) =\sum_{\Delta\subset\cs}\Big[\,\l_\Delta(n)\, \delta J_\Delta+\sig_\Delta(n)\, \delta N_\Delta\,\Big],\label{themo}
\ee
where $\l_\Delta(n)=\frac{\b^{(n)}_\Delta-\b_\Delta n}{{1-n}}$ and $\sig_\Delta(n)=\frac{\mu^{(n)}_\Delta-\mu_\Delta n}{{1-n}}$. When all $\Delta\in\cs$ have the same $n_v(\Delta)$, $\b^{(n)}_\Delta,\b_\Delta,\mu^{(n)}_\Delta,\mu_\Delta$ become independent of $\Delta$. In this case $\l_\Delta(n)\equiv \l(n)$ and $\sig_\Delta(n)\equiv \sig(n)$ becomes independent of $\Delta$, $\delta S_n(A)$ reduces to
\be
\delta S_n(A) =\l(n)\, \delta J_\cs+\sig(n)\, \delta N_\cs,\label{themo1}
\ee
where $J_\cs=\sum_{\Delta\subset\cs}J_\Delta$ and $N_\cs=\sum_{\Delta\subset\cs}N_\Delta$ are total area and total number of facets in $\cs$. 

\end{Theorem} 

\textbf{Proof:} Eq.\Ref{themo} can be checked by computing $\partial S_n(A)/\partial J_\Delta$ and $\partial S_n(A)/\partial N_\Delta$:
\be
\frac{\partial S_n(A)}{\partial J_\Delta}&=&\frac{1}{1-n}\lt(\frac{\partial\b^{(n)}_\Delta}{\partial J_\Delta}J_\Delta+\frac{\partial\mu^{(n)}_\Delta}{\partial\b^{(n)}_\Delta}\frac{\partial\b^{(n)}_\Delta}{\partial J_\Delta}N_\Delta+\b^{(n)}_\Delta\rt)-\frac{n}{1-n}\lt(\frac{\partial\b_\Delta}{\partial J_\Delta}J_\Delta+\frac{\partial\mu_\Delta}{\partial\b_\Delta}\frac{\partial\b_\Delta}{\partial J_\Delta}N_\Delta+\b_\Delta\rt)\nonumber\\
\frac{\partial S_n(A)}{\partial N_\Delta}&=&\frac{1}{1-n}\lt(\frac{\partial\b^{(n)}_\Delta}{\partial N_\Delta}J_\Delta+\frac{\partial\mu^{(n)}_\Delta}{\partial\b^{(n)}_\Delta}\frac{\partial\b^{(n)}_\Delta}{\partial N_\Delta}N_\Delta+\mu^{(n)}_\Delta\rt)-\frac{n}{1-n}\lt(\frac{\partial\b_\Delta}{\partial N_\Delta}J_\Delta+\frac{\partial\mu_\Delta}{\partial\b_\Delta}\frac{\partial\b_\Delta}{\partial N_\Delta}N_\Delta+\mu_\Delta\rt).\label{pd}
\ee
The definitions $\mu^{(n)}_\Delta=\ln [\sum_{j=1/2}^\infty e^{-\b^{(n)}_\Delta j} g^{(n)}_\Delta(j)]$ and $\mu_\Delta=\ln [\sum_{j=1/2}^\infty e^{-\b_\Delta j} g_\Delta(j)]$ imply
\be
\frac{\partial\mu^{(n)}_\Delta}{\partial\b^{(n)}_\Delta}=\frac{\sum_{j=1/2}^\infty(-j) e^{-\b^{(n)}_\Delta j} g^{(n)}_\Delta(j)}{\sum_{j=1/2}^\infty e^{-\b^{(n)}_\Delta j} g^{(n)}_\Delta(j)}=-\frac{J_\Delta}{N_\Delta},\quad \frac{\partial\mu_\Delta}{\partial\b_\Delta}=\frac{\sum_{j=1/2}^\infty(-j) e^{-\b_\Delta j} g_\Delta(j)}{\sum_{j=1/2}^\infty e^{-\b_\Delta j} g_\Delta(j)}=-\frac{J_\Delta}{N_\Delta}
\ee
Inserting in Eq.\Ref{pd}, we obtain
\be
\frac{\partial S_n(A)}{\partial J_\Delta}=\frac{\b^{(n)}_\Delta-\b_\Delta n}{{1-n}}=\l_\Delta(n),\quad \frac{\partial S_n(A)}{\partial N_\Delta}=\frac{\mu^{(n)}_\Delta-\mu_\Delta n}{{1-n}}=\sig_\Delta(n).
\ee
$\Box$

Eq.\Ref{themo1} suggests the analog between $\l(n)^{-1}$ and the temperature, as well as between $-\sig(n)/\l(n)$ and the chemical potential. In the most general situation Eq.\Ref{themo}, the temperature and chemical potential are not constants over the surface $\cs$. So $\cs$ are in a non-equilibrium state, although every $\Delta$ are in equilibrium.

Interestingly Eq.\Ref{themo1} shares similarities with the thermodynamical first law of the LQG black hole proposed in \cite{GP2011}. There the authors propose that the quantum isolated horizon is a statistical ensemble of identical spin-network punctures (quantum hairs) on the horizon, and the quasilocal energy of the horizon observed by the near-horizon Unruh observer is proportional to the total area $\mathbf{a}_{BH}$ of the horizon. Then a thermodynamical first law is derived by statistics on the quantum isolated horizon
\be
\delta S_{BH}= {\l}\, \delta J_{BH}+\sig\,\delta N_{BH},\quad J_{BH}\equiv\frac{\mathbf{a}_{BH}}{8\pi \g \ell_P^2}
\ee
where $S_{BH}$ is the black hole entropy, and $N_{BH}$ is the total number of punctures on the horizon, $\l$ relates to the Unruh temperature of the observer, and $\sig$ relates to the chemical potential. We immediately see the similarity between Eq.\Ref{themo1} and the above $\delta S_{BH}$ by relating the entangling surface $\cs$ to the black hole horizon, $S_{n}(A)$ to $S_{BH}$, $J_\cs$ to $J_{BH}$, and $N_\cs$ to $N_{BH}$.

\section{Removing the Parallel Restriction}\label{beyond}

Most of the above discussions relies on the parallel restriction on $\xi_{\a f}$ in spinfoam amplitude. In this section, we relax parallel restrictions to internal $\xi_{\a f}$'s, and compute the spinfoam amplitude 
\be
\quad A(\ck)=\sum_{\{j_f\}}\prod_f A_\Delta({j_f})\int[\rmd \xi_{\a f}\rmd g^\pm_{v\a}]\,\prod_{f,v,\pm}\lag \xi_{\a f}\lt|g_{v\a}^\pm{}^{-1}g_{v\b}^\pm \rt|\xi_{\b f}\rag^{2j_{f}^\pm}.\label{Z2}
\ee
Instead of imposing the potential $V_{\a \Delta}(\xi_{\a f})$ to suppress the non-parallel $\xi_{\a f}$'s, we are going to integrate out democratically all non-parallel $\xi_{\a f}$'s in the following analysis.

We again assume all $j_f\neq 0$, at a polyhedron $\a$ and among the facets $f\in \Delta$ ($\Delta$ is internal), we choose one $f_0$ and set
\be
|\xi_{\a f_0}\rangle\equiv| \xi_{\a\Delta}\rangle,
\ee
for all $\a$ containing $f_0$.

For any other $f\in\Delta$ and $f\neq f_0$, we write
\be
|\xi_{\a f}\rangle=a_{\a f}|\xi_{\a\Delta}\rangle+b_{\a f} |J\xi_{\a\Delta}\rangle,\quad a_{\a f}=\cos\lt(\frac{\theta_{\a f}}{2}\rt)\,e^{i\phi_{\a f}/2},\quad b_{\a f}=i\sin\lt(\frac{\theta_{\a f}}{2}\rt)\,e^{-i\phi_{\a f}/2} 
\ee
since $|\xi_{\a f}\rangle\in\C^2$ where $|\xi_{\a\Delta}\rangle,|J\xi_{\a\Delta}\rangle$ is a basis. $\phi_{\a f}\in[0,2\pi)$ and $\theta_{\a f}\in[0,\pi)$. we have the gauge equivalence $|\xi_{\a f}\rangle\sim e^{i\varphi}|\xi_{\a f}\rangle$. We insert the above relation into the following building block of the integrand in $A(\ck)$:
\be
&&\lag \xi_{\a f}\lt|g_{v\a}^\pm{}^{-1}g_{v\b}^\pm \rt|\xi_{\b f}\rag^{2j_f^\pm}\nonumber\\
&=&\Big(\bar{a}_{\a f}a_{\b f}\lag \xi_{\a \Delta}\lt|g_{v\a}^\pm{}^{-1}g_{v\b}^\pm \rt|\xi_{\b \Delta}\rag
+\bar{b}_{\a f}b_{\b f}\lag J\xi_{\a \Delta}\lt|g_{v\a}^\pm{}^{-1}g_{v\b}^\pm \rt|J\xi_{\b \Delta}\rag+\nonumber\\
&&\quad+ \bar{a}_{\a f}b_{\b f}\lag \xi_{\a \Delta}\lt|g_{v\a}^\pm{}^{-1}g_{v\b}^\pm \rt|J\xi_{\b \Delta}\rag
+\bar{b}_{\a f}a_{\b f}\lag J\xi_{\a \Delta}\lt|g_{v\a}^\pm{}^{-1}g_{v\b}^\pm \rt|\xi_{\b \Delta}\rag \Big)^{2j_f^\pm}
\ee
Applying the multinomial expansion to $\langle \xi_{\a f} |g_{v\a}^\pm{}^{-1}g_{v\b}^\pm |\xi_{\b f}\rangle^{2j_f^\pm}$ gives
\be
&=&\sum_{k_f^\pm(v)+l_f^\pm(v)+m_f^\pm(v)+n_f^\pm(v)=2j_f^\pm}\frac{2j^\pm_f !}{k_f^\pm(v)!l_f^\pm(v)!m_f^\pm(v)!n_f^\pm(v)!}\ \bar{a}_{\a f}^{k_f^\pm(v)+m_f^\pm(v)}
\bar{b}_{\a f}^{l_f^\pm(v)+n_f^\pm(v)}
a_{\b f}^{k_f^\pm(v)+n_f^\pm(v)}
b_{\b f}^{l_f^\pm(v)+m_f^\pm(v)}\nonumber\\
&&\lag \xi_{\a \Delta}\lt|g_{v\a}^\pm{}^{-1}g_{v\b}^\pm \rt|\xi_{\b \Delta}\rag^{k_f^\pm(v)}\lag J\xi_{\a \Delta}\lt|g_{v\a}^\pm{}^{-1}g_{v\b}^\pm \rt|J\xi_{\b \Delta}\rag^{l_f^\pm(v)}\lag \xi_{\a \Delta}\lt|g_{v\a}^\pm{}^{-1}g_{v\b}^\pm \rt|J\xi_{\b \Delta}\rag^{m_f^\pm(v)}\lag J\xi_{\a \Delta}\lt|g_{v\a}^\pm{}^{-1}g_{v\b}^\pm \rt|\xi_{\b \Delta}\rag^{n_f^\pm(v)}
\ee
where $k_f^\pm(v),l_f^\pm(v),m_f^\pm(v),n_f^\pm(v)\in\Z_+\cup\{0\}$. Applying the product over $\pm$ and all $f\neq f_0\in\Delta$, 
\be\label{klmn}
&&\prod_{f\neq f_0,\pm}\lag \xi_{\a f}\lt|g_{v\a}^\pm{}^{-1}g_{v\b}^\pm \rt|\xi_{\b f}\rag^{2j_f^\pm}
=\mathop{\sum_{\{k_f^\pm(v)\}_f,\{l_f^\pm(v)\}_f,\{m_f^\pm(v)\}_f,\{n_f^\pm(v)\}_f}}_{k_f^\pm(v)+l_f^\pm(v)+m_f^\pm(v)+n_f^\pm(v)=2j_f^\pm}\prod_{f\neq f_0;\pm}\frac{2j^\pm_f !}{k_f^\pm(v)!l_f^\pm(v)!m_f^\pm(v)!n_f^\pm(v)!}\times\\
&&\quad\times\  \prod_{f\neq f_0}
\bar{a}_{\a f}^{\sum_{\pm }k_f^\pm(v)+\sum_{\pm }m_f^\pm(v)}
\bar{b}_{\a f}^{\sum_{\pm }l_f^\pm(v)+\sum_{\pm }n_f^\pm(v)}
a_{\b f}^{\sum_{\pm }k_f^\pm(v)+\sum_{\pm }n_f^\pm(v)}
b_{\b f}^{\sum_{\pm }l_f^\pm(v)+\sum_{\pm }m_f^\pm(v)}\times \nonumber\\
&&\quad\times\ \lag \xi_{\a \Delta}\lt|g_{v\a}^\pm{}^{-1}g_{v\b}^\pm \rt|\xi_{\b \Delta}\rag^{K_\Delta^\pm(v)}\lag J\xi_{\a \Delta}\lt|g_{v\a}^\pm{}^{-1}g_{v\b}^\pm \rt|J\xi_{\b \Delta}\rag^{L_\Delta^\pm(v)}\lag \xi_{\a \Delta}\lt|g_{v\a}^\pm{}^{-1}g_{v\b}^\pm \rt|J\xi_{\b \Delta}\rag^{M_\Delta^\pm(v)}\lag J\xi_{\a \Delta}\lt|g_{v\a}^\pm{}^{-1}g_{v\b}^\pm \rt|\xi_{\b \Delta}\rag^{N_\Delta^\pm(v)}\nonumber
\ee
where 
\be
K_\Delta^\pm(v)=\sum_{f\neq f_0}k_f^\pm(v),\quad L_\Delta^\pm(v)=\sum_{f\neq f_0}l_f^\pm(v),\quad M_\Delta^\pm(v)=\sum_{f\neq f_0}m_f^\pm(v),\quad N_\Delta^\pm(v)=\sum_{f\neq f_0}n_f^\pm(v)
\ee
satisfying
\be
K_\Delta^\pm(v)+L_\Delta^\pm(v)+M_\Delta^\pm(v)+N_\Delta^\pm(v)=2\lt(J^\pm_\Delta- j^\pm_{f_0}\rt)\gg 1.
\ee
Therefore at least one of $K_\Delta^\pm(v),L_\Delta^\pm(v),M_\Delta^\pm(v),N_\Delta^\pm(v)$ has to be large.

We integrate non-parallel $\xi_{\a f}$ ($f\neq f_0$) by integrating $\theta_{\a f}$ and $\phi_{\a f}$ with the standard unit-sphere measure. Explicitly, 
\be
&&\frac{1}{4\pi}\int_{0}^{2\pi}\rmd\phi_{\a f}\int_{0}^\pi\rmd\theta_{\a f}\,\sin(\theta_{\a f})\,
\bar{a}_{\a f}^{\sum_{\pm }k_f^\pm(v)+\sum_{\pm }m_f^\pm(v)}
\bar{b}_{\a f}^{\sum_{\pm }l_f^\pm(v)+\sum_{\pm }n_f^\pm(v)}
a_{\a f}^{\sum_{\pm }k_f^\pm(v')+\sum_{\pm }n_f^\pm(v')}
b_{\a f}^{\sum_{\pm }l_f^\pm(v')+\sum_{\pm }m_f^\pm(v')}\nonumber\\
&=&e^{i\frac{\pi}{2}\lt[\sum_{\pm }l_f^\pm(v')+\sum_{\pm }m_f^\pm(v')-\lt(\sum_{\pm }l_f^\pm(v)+\sum_{\pm }n_f^\pm(v)\rt)\rt]}\nonumber\\
&&\quad \times\ \frac{1}{4\pi}\int_{0}^{2\pi}\rmd\phi_{\a f} \, e^{i\frac{\phi_{\a f}}{2}\lt[
\sum_{\pm }k_f^\pm(v')+\sum_{\pm }n_f^\pm(v')
+\sum_{\pm }l_f^\pm(v)+\sum_{\pm }n_f^\pm(v)-\lt(\sum_{\pm }k_f^\pm(v)+\sum_{\pm }m_f^\pm(v)+\sum_{\pm }l_f^\pm(v')+\sum_{\pm }m_f^\pm(v')\rt)\rt]}
\nonumber\\
&&\quad \times\ \int_{0}^\pi\rmd\theta_{\a f}\,\sin(\theta_{\a f})\,
\lt[\cos\lt(\frac{\theta_{\a f}}{2}\rt)\rt]^{\sum_{\pm }k_f^\pm(v)+\sum_{\pm }m_f^\pm(v)+\sum_{\pm }k_f^\pm(v')+\sum_{\pm }n_f^\pm(v')}
\lt[\sin\lt(\frac{\theta_{\a f}}{2}\rt)\rt]^{\sum_{\pm }l_f^\pm(v)+\sum_{\pm }n_f^\pm(v)+\sum_{\pm }l_f^\pm(v')+\sum_{\pm }m_f^\pm(v')}.\label{angleint}
\ee
Recall that $\sum_{\pm }k_f^\pm(v')+\sum_{\pm }l_f^\pm(v)+\lt(\sum_{\pm }k_f^\pm(v)+\sum_{\pm }l_f^\pm(v')\rt)=4\sum_\pm j^\pm_f=4j_f$ is even, thus $\sum_{\pm }k_f^\pm(v')+\sum_{\pm }l_f^\pm(v)-\lt(\sum_{\pm }k_f^\pm(v)+\sum_{\pm }l_f^\pm(v')\rt)$ is also even. Therefore the $\phi_{\a f}$-integral constraints
\be
\sum_{\pm }k_f^\pm(v')+\sum_{\pm }n_f^\pm(v')-\lt[\sum_{\pm }l_f^\pm(v')+\sum_{\pm }m_f^\pm(v')\rt]
=\sum_{\pm }k_f^\pm(v)+\sum_{\pm }m_f^\pm(v)-\lt[\sum_{\pm }l_f^\pm(v)+\sum_{\pm }n_f^\pm(v)\rt]
\ee
Recall that $\sum_{\pm }k_f^\pm(v)+\sum_{\pm }n_f^\pm(v)+\sum_{\pm }l_f^\pm(v)+\sum_{\pm }m_f^\pm(v)=2j_f$ independent of $v$, we obtain
\be
\sum_{\pm }k_f^\pm(v')+\sum_{\pm }n_f^\pm(v')=\sum_{\pm }k_f^\pm(v)+\sum_{\pm }m_f^\pm(v)\equiv k_f\quad 
\sum_{\pm }l_f^\pm(v')+\sum_{\pm }m_f^\pm(v')=\sum_{\pm }l_f^\pm(v)+\sum_{\pm }n_f^\pm(v)\equiv l_f.\nonumber
\ee
with $k_f+l_f=2j_f$. The integral \Ref{angleint} reduces to
\be
\half\int_{0}^\pi\rmd\theta_{\a f}\,\sin(\theta_{\a f})\,
\lt[\cos\lt(\frac{\theta_{\a f}}{2}\rt)\rt]^{2k_f}
\lt[\sin\lt(\frac{\theta_{\a f}}{2}\rt)\rt]^{2l_f}=\frac{k_f!l_f!}{(k_f+l_f+1)!}.
\ee

Inserting the results into Eqs.\Ref{klmn} and \Ref{Z2}, we write the integral as a sum of partial amplitudes
\be
&&\int[\rmd g_{v\a}^\pm\rmd\xi_{\a f}]\prod_{f,v,\pm}\lag \xi_{\a f}\lt|g_{v\a}^\pm{}^{-1}g_{v\b}^\pm \rt|\xi_{\b f}\rag^{2j_f^\pm}\label{klmn}\\
&=&\mathop{\sum_{\{k_f^\pm(v)\},\{l_f^\pm(v)\}}}_{\{m_f^\pm(v)\},\{n_f^\pm(v)\}}\int[\rmd g_{v\a}^\pm\rmd\xi_{\a \Delta}]\prod_{\Delta, v,\pm}\mathop{\prod_{f\in\Delta}}_{f\neq f_0}\frac{2j^\pm_f !}{k_f^\pm(v)!l_f^\pm(v)!m_f^\pm(v)!n_f^\pm(v)!}
\frac{k_f!l_f!}{(2j_f+1)!}
 \lag \xi_{\a \Delta}\lt|g_{v\a}^\pm{}^{-1}g_{v\b}^\pm \rt|\xi_{\b \Delta}\rag^{\tilde{K}_\Delta^\pm(v)}\nonumber\\
&&\quad\times\ \lag J\xi_{\a \Delta}\lt|g_{v\a}^\pm{}^{-1}g_{v\b}^\pm \rt|J\xi_{\b \Delta}\rag^{L_\Delta^\pm(v)}\lag \xi_{\a \Delta}\lt|g_{v\a}^\pm{}^{-1}g_{v\b}^\pm \rt|J\xi_{\b \Delta}\rag^{M_\Delta^\pm(v)}\lag J\xi_{\a \Delta}\lt|g_{v\a}^\pm{}^{-1}g_{v\b}^\pm \rt|\xi_{\b \Delta}\rag^{N_\Delta^\pm(v)}.\nonumber
\ee
where 
\be
\tilde{K}_\Delta^\pm(v)={K}_\Delta^\pm(v)+2j_{f_0}^\pm.
\ee

We introduce short-hand notations to write 
\be
\int[\rmd g_{v\a}^\pm\rmd\xi_{\a f}]\prod_{f,v,\pm}\lag \xi_{\a f}\lt|g_{v\a}^\pm{}^{-1}g_{v\b}^\pm \rt|\xi_{\b f}\rag^{2j_f^\pm}&\equiv&\mathop{\sideset{}{'}\sum_{\{\tilde{K}_\Delta^\pm(v)\},\{L_\Delta^\pm(v)\}}}_{\{M_\Delta^\pm(v)\},\{N_f^\pm(v)\}}\prod_\Delta w_\Delta\int[\rmd g_{v\a}^\pm\rmd\xi_{\a \Delta}]\ e^{S_{KLMN}} 
\ee
where the above sum is constrained by $\sum_{\pm }\tilde{K}_\Delta^\pm(v')+\sum_{\pm }N_\Delta^\pm(v')=\sum_{\pm }\tilde{K}_\Delta^\pm(v)+\sum_{\pm }M_\Delta^\pm(v)\equiv \tilde{K}_\Delta$, $\sum_{\pm }L_\Delta^\pm(v')+\sum_{\pm }M_\Delta^\pm(v')=\sum_{\pm }L_\Delta^\pm(v)+\sum_{\pm }N_\Delta^\pm(v)\equiv L_\Delta$, and $\tilde{K}_\Delta^\pm(v)+L_\Delta^\pm(v)+M_\Delta^\pm(v)+N_\Delta^\pm(v)=2 J^\pm_\Delta $.
\be
S_{KLMN}&=&\sum_{\Delta, v,\pm}\Big[\,
\tilde{K}_\Delta^\pm(v)\ln \lag \xi_{\a \Delta}\lt|g_{v\a}^\pm{}^{-1}g_{v\b}^\pm \rt|\xi_{\b \Delta}\rag+
L_\Delta^\pm(v)\ln\lag J\xi_{\a \Delta}\lt|g_{v\a}^\pm{}^{-1}g_{v\b}^\pm \rt|J\xi_{\b \Delta}\rag+\nonumber\\
&&\quad +\ M_\Delta^\pm(v)\ln\lag \xi_{\a \Delta}\lt|g_{v\a}^\pm{}^{-1}g_{v\b}^\pm \rt|J\xi_{\b \Delta}\rag+
N_\Delta^\pm(v)\ln\lag J\xi_{\a \Delta}\lt|g_{v\a}^\pm{}^{-1}g_{v\b}^\pm \rt|\xi_{\b \Delta}\rag\ \Big].\\
w_\Delta&=&\mathop{\sideset{}{'}\sum_{\{k_f^\pm(v)\},\{l_f^\pm(v)\}}}_{\{m_f^\pm(v)\},\{n_f^\pm(v)\}}\mathop{\prod_{f\in\Delta}}_{f\neq f_0}\prod_v\lt[\frac{2j^\pm_f !}{k_f^\pm(v)!l_f^\pm(v)!m_f^\pm(v)!n_f^\pm(v)!}.
\frac{k_f!l_f!}{(2j_f+1)!}\rt]
\ee
The sum in $w_\Delta$ is constrained by $\sum_{\pm }k_f^\pm(v')+\sum_{\pm }n_f^\pm(v')=\sum_{\pm }k_f^\pm(v)+\sum_{\pm }m_f^\pm(v)\equiv k_f$, $\sum_{\pm }l_f^\pm(v')+\sum_{\pm }m_f^\pm(v')=\sum_{\pm }l_f^\pm(v)+\sum_{\pm }n_f^\pm(v)\equiv l_f$, $K_\Delta^\pm(v)=\sum_{f\neq f_0}k_f^\pm(v),\  L_\Delta^\pm(v)=\sum_{f\neq f_0}l_f^\pm(v),\  M_\Delta^\pm(v)=\sum_{f\neq f_0}m_f^\pm(v),\  N_\Delta^\pm(v)=\sum_{f\neq f_0}n_f^\pm(v)$.

The new action $S_{KLMN}$ is the old action $S$ in Eq.\Ref{Z} with $\xi_{\a f}$ ($f\in\Delta$) becoming either parallel $\xi_{\a f}= \xi_{\a\Delta}$ or anti-parallel $\xi_{\a f}= J\xi_{\a\Delta}$. Configurations with some $\xi_{\a f}$'s being parallel and others being anti-parallel have been discussed in Theorem \ref{SSSSS} for critical points of $S$. These critical points also appear in the new action. In contrast to $S$, here at least one of $K_\Delta^\pm(v), L_\Delta^\pm(v),M_\Delta^\pm(v),N_\Delta^\pm(v)$ has to be large, so it allows us apply the stationary phase approximation to the integral with the new action $S_{KLMN}$. The critical points in Theorem \ref{SSSSS} becomes useful here for computing integrals.

The integral $\int[\rmd g_{v\a}^\pm\rmd\xi_{\a \Delta}]\ e^{S_{KLMN}}$ has the following feature:
 
\begin{Lemma}\label{O1/N000}

$\int[\rmd g_{v\a}^\pm\rmd\xi_{\a \Delta}]\ e^{S_{KLMN}}$ prefers large $K_\Delta^\pm(v)$ or $ L_\Delta^\pm(v)$ and zero $M_\Delta^\pm(v),N_\Delta^\pm(v)$. $\int[\rmd g_{v\a}^\pm\rmd\xi_{\a \Delta}]\ e^{S_{KLMN}}$ with nonzero $M_\Delta^\pm(v),N_\Delta^\pm(v)$ is of $O(1/N)$ comparing to the integral with zero $M_\Delta^\pm(v),N_\Delta^\pm(v)$.

\end{Lemma}

\textbf{Proof:} Suppose $M_\Delta^\pm(v)$ is large (the argument of large $N_\Delta^\pm(v)$ is similar), 
\be
\lag \xi_{\a \Delta}\lt|g_{v\a}^\pm{}^{-1}g_{v\b}^\pm \rt|J\xi_{\b \Delta}\rag^{M_\Delta^\pm(v)}=e^{M_\Delta^\pm(v)\ln\lag \xi_{\a \Delta}\lt|g_{v\a}^\pm{}^{-1}g_{v\b}^\pm \rt|J\xi_{\b \Delta}\rag}
\ee
participates the integral over $\xi_{\a \Delta}$ (we interchange the integral of $\xi_{\a \Delta}$ and the finite sum in Eq,\Ref{klmn}). By the stationary phase analysis, this factor in the integrand lead to that critical points of the integral must satisfy
\be
g_{v\b}^\pm |J\xi_{\b \Delta}\rangle=e^{i\varphi_{\a v\b}^\pm}g_{v\a}^\pm \lt|\xi_{\a \Delta}\rag,\quad \text{i.e.}\quad \lag \xi_{\a \Delta}\lt|g_{v\a}^\pm{}^{-1}g_{v\b}^\pm \rt|\xi_{\b \Delta}\rag=0\label{xiJxi}
\ee
in order that the integrand is not suppressed exponentially. But the integral contains a factor contributed by $f_0$: $\langle \xi_{\a \Delta}|g_{v\a}^\pm{}^{-1}g_{v\b}^\pm |\xi_{\b \Delta}\rangle^{2j_{f_0}^\pm}$ which vanishes at the above critical points. Therefore the integral is of $O(1/N)$ by stationary phase analysis and in a neighborhood $D$ containing a single critical point $x_c$, 
\be
\int_D\rmd^nx\, a(x)\, e^{N S(x)}= \lt( \frac{2\pi}{N}\rt)^{n/2}\frac{1}{\sqrt{\det(- H)}}e^{N S(x_c)}\lt[a(x_c)+O\lt(\frac{1}{N}\rt)\rt],\label{formula111}
\ee
which is of $O(1/N)$ if $a(x_c)=0$. The same argument with critical equation Eq.\Ref{xiJxi} also applies to large $N_\Delta^\pm(v)$.


We cannot have e.g. both $K_\Delta^\pm(v) $ (or $L_\Delta^\pm(v)$) and $M_\Delta^\pm(v)$ (or $N_\Delta^\pm(v)$) large, otherwise the integral is suppressed exponentially. Indeed Eq.\Ref{xiJxi} is contradicting the 1st equation in Eq.\Ref{critical}, which is a critical equation from large $K_\Delta^\pm(v) $. The integrand is always suppressed exponentially if both $K_\Delta^\pm(v) $ (or $L_\Delta^\pm(v)$) and $M_\Delta^\pm(v)$ (or $N_\Delta^\pm(v)$) are large.

Therefore either $K_\Delta^\pm(v)$ or $ L_\Delta^\pm(v)$ has to be large, then the critical points must satisfy 
\be
g_{v\b}^\pm |\xi_{\b \Delta}\rangle=e^{i\varphi_{\a v\b}^\pm}g_{v\a}^\pm \lt|\xi_{\a \Delta}\rag,\quad\text{or}\quad g_{v\b}^\pm |J\xi_{\b \Delta}\rangle=e^{-i\varphi_{\a v\b}^\pm}g_{v\a}^\pm \lt|J\xi_{\a \Delta}\rag.
\ee
There is no contradiction between 2 equations since $J$ commutes with $g\in\Su$. Either one of them gives
\be
\lag \xi_{\a \Delta}\lt|g_{v\a}^\pm{}^{-1}g_{v\b}^\pm \rt|J\xi_{\b \Delta}\rag=\lag J\xi_{\a \Delta}\lt|g_{v\a}^\pm{}^{-1}g_{v\b}^\pm \rt|\xi_{\b \Delta}\rag=0
\ee
Then if $M_\Delta^\pm(v) $ or $N_\Delta^\pm(v)$ is nonzero, the integral is of $O(1/N)$ by the same reason as the above.\\
$\Box$


We set $M_\Delta^\pm(v)=N_\Delta^\pm(v)=0$ and define
\be
S_{KL}=\sum_{v,\Delta,\pm}\tilde{K}_\Delta^\pm(v)\ln \lag \xi_{\a \Delta}\lt|g_{v\a}^\pm{}^{-1}g_{v\b}^\pm \rt|\xi_{\b \Delta}\rag+\sum_{v,\Delta,\pm}{L}_\Delta^\pm(v)\ln\lag J\xi_{\a \Delta}\lt|g_{v\a}^\pm{}^{-1}g_{v\b}^\pm \rt|J\xi_{\b \Delta}\rag
\ee
$\tilde{K}_\Delta^\pm(v)$ and $L_\Delta^\pm(v)$ satisfy $\sum_{\pm }\tilde{K}_\Delta^\pm(v')=\sum_{\pm }\tilde{K}_\Delta^\pm(v)\equiv \tilde{K}_\Delta$, $\sum_{\pm }L_\Delta^\pm(v')=\sum_{\pm }L_\Delta^\pm(v)\equiv L_\Delta$, and $\tilde{K}_\Delta^\pm(v)+L_\Delta^\pm(v)=2 J^\pm_\Delta $.

Since $\mathrm{Re}(S_{KL})\leq 0$, the condition for preventing the integrand from being exponentially suppressed, $\mathrm{Re}(S_{KL})= 0$, is equivalent to 
\be
g_{v\b}^\pm |\xi_{\b \Delta}\rangle=e^{i\varphi_{\a v\b}^\pm}g_{v\a}^\pm \lt|\xi_{\a \Delta}\rag.\label{gxigxi1111}
\ee
The action $S_{KL}$ has several scaling parameters $\tilde{K}_\Delta^\pm(v),\tilde{L}_\Delta^\pm(v)$ which may not be all large. But Eq.\Ref{gxigxi1111} for all cases. 

When we compute $\delta_{\xi}S_{KL}$, we write $\delta\xi_{\a \Delta}=\eps_{\a \Delta}J\xi_{\a \Delta}+i\eta_{\a \Delta} \xi_{\a \Delta}$ and $\delta J\xi_{\a \Delta}=-\bar{\eps}_{\a \Delta}\xi_{\a \Delta}-i\eta_{\a\Delta} J \xi_{\a \Delta}$ where $\eps_{\a \Delta}\in\C$ and $\eta_{\a \Delta}\in\R$. The coefficient in front of $\xi_{\a \Delta}$ is purely imaginary because $\xi_{\a \Delta}$ is normalized. Since every $\xi_{\a \Delta}$ is shared by 2 terms with neighboring $v$'s
\be
&&\delta_{\xi_{\a \Delta}}S_{KL}
\nonumber\\
&=&\sum_{\pm}\lt[\tilde{K}_\Delta^\pm(v')\eps_{\a f} \frac{\lag{{\xi_{\b' f}}\big|(g^\pm_{v' \b'})^{-1}g^\pm_{v'\a}\big|J{\xi_{\a f}}}\rag}{\lag{{\xi_{\b' f}}\big|(g^\pm_{v'\b'})^{-1}g^\pm_{v'\a }\big|{\xi_{\a f}}}\rag}
+\tilde{K}_\Delta^\pm(v)\bar{\eps}_{\a f}\frac{\lag J{\xi_{\a f}}\big|(g^\pm_{v\a})^{-1}g^\pm_{v\b}\big|{\xi_{\b f}}\rag}{\lag{\xi_{\a f}}\big|(g^\pm_{v\a})^{-1}g^\pm_{v\b}\big|{\xi_{\b f}}\rag}+i\lt(\tilde{K}_\Delta^\pm(v')-\tilde{K}_\Delta^\pm(v)\rt)\eta_{\a \Delta}\rt]\nonumber\\
&-&\sum_{\pm}\lt[{L}_\Delta^\pm(v')\bar{\eps}_{\a f} \frac{\lag{{J\xi_{\b' f}}\big|(g^\pm_{v' \b'})^{-1}g^\pm_{v'\a}\big|{\xi_{\a f}}}\rag}{\lag{{J\xi_{\b' f}}\big|(g^\pm_{v'\b'})^{-1}g^\pm_{v'\a }\big|{J\xi_{\a f}}}\rag}
+{L}_\Delta^\pm(v){\eps}_{\a f}\frac{\lag {\xi_{\a f}}\big|(g^\pm_{v\a})^{-1}g^\pm_{v\b}\big|{J\xi_{\b f}}\rag}{\lag{J\xi_{\a f}}\big|(g^\pm_{v\a})^{-1}g^\pm_{v\b}\big|{J\xi_{\b f}}\rag}+i\lt({L}_\Delta^\pm(v')-{L}_\Delta^\pm(v)\rt)\eta_{\a \Delta}\rt]\nonumber\\
&=&0
\ee
by Eqs.\Ref{gxigxi1111}, and the orthogonality between $\xi,J\xi$.

For the derivative in $g_{v\a}^\pm$, we use $\delta {g_{v\a}^\pm}= \frac{i}{2} \theta_{v\a}^\pm\vec{\sig} {g_{v\a}^\pm}$ ($\theta_{v\a}\in\R$). At the critical point and by Eq.\Ref{critical},
\be
\delta_{g_{v\a}^\pm} S &=&\frac{i}{2} \theta_{v\a}^\pm\sum_{\Delta\subset\a}\kappa_{\a \Delta}(v) \lt(\tilde{K}_\Delta^\pm(v) \frac{\lag{{\xi_{\a f}}\big|(g^\pm_{v\a})^{-1}\vec{\sig} {g_{v\a}^\pm}\big|{\xi_{\a f}}}\rag}{\lag{{\xi_{\a f}}\big|(g^\pm_{v\a})^{-1}g^\pm_{v\a}\big|{\xi_{\a f}}}\rag}
+\tilde{L}_\Delta^\pm(v) \frac{\lag{{J\xi_{\a f}}\big|(g^\pm_{v\a})^{-1}\vec{\sig} {g_{v\a}^\pm}\big|{J\xi_{\a f}}}\rag}{\lag{{J\xi_{\a f}}\big|(g^\pm_{v\a})^{-1}g^\pm_{v\a}\big|{J\xi_{\a f}}}\rag}\rt)\nonumber\\
&=&\frac{i}{2} \theta_{v\a}^\pm\,(1\pm\g)\, g^\pm_{v\a}\cdot\sum_{\Delta\subset\a}\kappa_{\a\Delta}(v)\lt[\tilde{K}_\Delta^\pm(v)-\tilde{L}_\Delta^\pm(v)\rt]\vec{n}_{\a\Delta}\label{derivg}
\ee
where $\kappa_{\a\Delta}(v)=\pm 1$ satisfying $\kappa_{\a\Delta}(v)=-\kappa_{\a\Delta}(v')$ appears when $\partial_{g_{v\a}^\pm}$ acts on $g_{v\a}^\pm$ or $g_{v\a}^\pm{}^{-1}$. $\delta_{g_{v\a}^\pm} S =0$ is equivalent to
\be
\sum_{\Delta\subset\a}\kappa_{\a\Delta}(v)\lt[\tilde{K}_\Delta^\pm(v)-{L}_\Delta^\pm(v)\rt]\vec{n}_{\a\Delta}=0.\label{closure111}
\ee

However, there is a subtlety when $|\tilde{K}_\Delta^\pm(v)-{L}_\Delta^\pm(v)|$ is small. Notice that $\langle J \xi_{\a \Delta}|g_{v\a}^\pm{}^{-1}g_{v\b}^\pm |J \xi_{\b \Delta}\rangle$ is the complex conjugate of $\langle \xi_{\a \Delta}|g_{v\a}^\pm{}^{-1}g_{v\b}^\pm |\xi_{\b \Delta}\rangle$,
\be
S_{KL}=\sum_{v,\Delta,\pm}\lt[\tilde{K}_\Delta^\pm(v)-{L}_\Delta^\pm(v)\rt]\ln \lag \xi_{\a \Delta}\lt|g_{v\a}^\pm{}^{-1}g_{v\b}^\pm \rt|\xi_{\b \Delta}\rag+2\sum_{v,\Delta,\pm}{L}_\Delta^\pm(v)\ \mathrm{Re}\lt[\ln\lag J\xi_{\a \Delta}\lt|g_{v\a}^\pm{}^{-1}g_{v\b}^\pm \rt|J\xi_{\b \Delta}\rag\rt].\label{Re1111}
\ee
We assume $\tilde{K}_\Delta^\pm(v)\geq\tilde{L}_\Delta^\pm(v)$, while other cases can be work out analogously. If all $\tilde{K}_\Delta^\pm(v), \tilde{L}_\Delta^\pm(v)$ are all large at $v,\Delta$ but both $\tilde{K}_\Delta^\pm(v)-\tilde{L}_\Delta^\pm(v)$ is small, then the 1st term in Eq.\Ref{Re1111} is subleading, and the contribution from this $\Delta$ is negligible in Eq.\Ref{derivg}. Eq.\Ref{closure111} with one or more $\Delta$ absent corresponds to a semiclassically degenerate tetrahedron. 

Eq.\Ref{derivg} is valid when $\tilde{K}_\Delta^+(v)-\tilde{L}_\Delta^+(v)$ or/and $\tilde{K}_\Delta^-(v)-\tilde{L}_\Delta^-(v)$ is/are large for all involved $\Delta$'s. The number of parallel $\xi_{\a f}=\xi_{\a\Delta}$ is much greater than the number of anti-parallel $\xi_{\a f}=J\xi_{\a\Delta}$. In this case, $\tilde{L}_\Delta^+(v)\ll J_\Delta^+$ and $\tilde{K}_\Delta^+(v)\simeq J^+_\Delta$ (or/and $\tilde{L}_\Delta^-(v)\ll J_\Delta^-$ and $\tilde{K}_\Delta^-(v)\simeq J^-_\Delta$), we obtain the standard tetrahedron closure condition
\be
\sum_{\Delta\subset\a}J_\Delta \kappa_{\a\Delta}(v) \vec{n}_{\a\Delta}=0,\label{closure22}
\ee
and recover the critical equations as Eq.\Ref{critical}. The solutions of critical equations Eqs.\Ref{gxigxi1111} and \Ref{closure22} are the same as the situation with the parallel restriction imposed, and have been discussed in Section \ref{complex}. This result shows that critical points $(g_{v\a}^\pm,\xi_{\a\Delta})_c[J_\Delta]$, used extensively in Sections \ref{complex}, \ref{LQGstate}, and \ref{EE} indeed have nontrivial contributions in the stationary approximation of the amplitude $A(\ck)$ without the parallel restriction. 

Depending on the choice of $J_\Delta$, degenerate tetrahedra may still appear even when $\tilde{K}_\Delta^\pm(v)\gg{L}_\Delta^\pm(v)$, similar to the simplical EPRL/FK amplitude. But the discussion below Eq.\Ref{Re1111} shows that degenerate tetrahedra become generic in the present situation. The origin of these degenerate tetrahedra is the anti-parallel $\xi_{\a f}=J\xi_{\a\Delta}$ coming from integrating non-parallel $\xi_{\a f}$'s. The study of critical points with degenerate tetrahedra is beyond the scope of the present paper, so is postponed to future research. 

Although the integrals with nonzero $M_\Delta^\pm(v),N_\Delta^\pm(v)$ is of $O(1/N)$ comparing to the integrals with $M_\Delta^\pm(v)=N_\Delta^\pm(v)=0$, we can still perform the same stationary phase analysis to these integrals with small $M_\Delta^\pm(v),N_\Delta^\pm(v)$ by using Eq.\Ref{formula111}, where critical equations Eqs.\Ref{gxigxi1111} and \Ref{closure22} still applies. The dual situation with large $M_\Delta^\pm(v),N_\Delta^\pm(v)$ and small $\tilde{K}_\Delta^\pm(v),L_\Delta^\pm(v)$ can be analyzed in the similar way, by simply interchange the roles $M_\Delta^\pm(v),N_\Delta^\pm(v)\leftrightarrow\tilde{K}_\Delta^\pm(v),L_\Delta^\pm(v)$, and $\xi_{\a\Delta}\leftrightarrow J\xi_{\a\Delta}$ for some $\a$. The integral with all $M_\Delta^\pm(v),N_\Delta^\pm(v),\tilde{K}_\Delta^\pm(v),L_\Delta^\pm(v)$ large is suppressed exponentially as discussed in Lemma \ref{O1/N000}.

\section{Discussion and Outlook}

This paper explores the semiclassical behavior of LQG in small spins, and obtains promising results such as the entanglement entropy with thermodynamical analog and Regge geometries emerging from critical points in the stationary phase analysis. There are more interesting perspectives which should be investigated in the future.

In our work, we have seen the small-$j$ semiclassicality always relates to coarse-graining, e.g. a semiclassical Regge geometry with $J_\Delta$ as a macrostate is a collection of microstates $\{j_f\}$, and the entanglement entropy coarse-grains the microstates and gives an analog thermodynamical first law. Moreover, the EPRL-FK model with $J_\Delta$ as DOFs may be viewed as a coarse-grained effective theory whose fundamental fine-grained theory is the generalized spinfoam model with $j_f$ as DOFs. This result opens up a possibility that spinfoam models such as EPRL-FK might not be fundamental but rather coarse-grained effective theories emergent from some fine-grained theories which are more fundamental. In our work, we only consider to coarse-grain the face DOFs such as spins $j_f$, but do not consider to coarse-grain bulk DOFs such as intertwiners or spinfoam vertices in the fine-grained theory. It would be more interesting to coarse-grain/fine-grain these bulk DOFs (there have been some attempts in the literature, e.g. \cite{Bahr:2012qj, Dittrich:2016tys, Bahr:2017klw, Livine:2013gna,Livine:2016vhl,Bodendorfer:2016tky,Lang:2017beo,Eichhorn:2018phj}). It might be possible that there exists a fine-grained fundamental theory such that the EPRL-FK model emerges from coarse-graining both face and bulk DOFs. This anticipated fine-grained theory might closely relate to the continuum limit of spinfoam formulation.

As is mentioned in Section \ref{1stlaw}, the analog thermodynamical first law from the entanglement entropy is similar to the first law of LQG black hole in \cite{GP2011}. This similarity may orient us toward an explanation of black hole entropy from the entanglement entropy in spinfoam formulation. Understanding quantum black hole in spinfoam formulation or other full LQG framework is a long-standing open issue. Our work suggests a new routine toward formulating black hole in spinfoam. The idea is to consider spinfoam amplitude on a 4-manifold as a subregion in a black hole spacetime such as the Kruskal spacetime, and the spatial boundary $\Sig$ to be the spatial slice at the moment $T=0$ of time reflection symmetry. We may set the critical point $(g_{v\a}^\pm,\xi_{\a\Delta})_c[J_\Delta]$ to correspond to a discrete Kruskal geometry (in this subregion). $\Sig$ can be subdivided by the horizon (bifurcate sphere) in to $A$ and $\bA$. So we can compute the entanglement R\'enyi entropy $S_n(A)$ similar as this work. This computation has to be carried out in the Lorentzian spinfoam model, but the derivation and result should be carried over. Then the thermodynamical first law from $S_n(A)$ should be directly relate to the black hole thermodynamics.  

It would be interesting to relate the entanglement entropy from spinfoam to Jacobson's proposal \cite{Jacobson:2015hqa}: The semiclassical Einstein equation can be derived from $\delta S(A)=0$ where $S(A)$ is the entanglement entropy and satisfies the area-law. We hope to relate the entanglement entropy derived here to recent works \cite{Han:2018fmu,Han:2017xwo} which relate spinfoam amplitude to Einstein equation.  

There are other interesting questions on the semiclassical analysis of the fine-grained spinfoam model $A(\ck)$, e.g. how to understand the critical points with degenerate tetrahedra and their 4d geometrical interpretation. It would also be interesting if a semiclassical state $\psi$ could be defined with the fine-grained spinfoam model without imposing the parallel restriction, and still could be applied to computing entanglement entropy.


\section*{Acknowledgements}

I acknowledge Ling-Yan Hung for motivating me to study the entanglement entropy in spinfoam LQG, and Hongguang Liu for fruitful discussions on this aspect. I acknowledge Ivan Agullo for reminding me the possibility of semiclassicality with small spins, and acknowledge Andrea Dapor and Klaus Liegener for their hospitality during my visit at Louisiana State University. This work receives support from the US National Science Foundation through grant PHY-1602867 and PHY-1912278, and Start-up Grant at Florida Atlantic University, USA. 


\appendix

\section{Face Amplitude}\label{face}

We follow the choice of face amplitude in \cite{face}. The spinfoam amplitude in holonomy representation gives 
\be
\psi(\vec{U})=\sum_{\vec{j},\vec{i}}\prod_f\dim(j_f)\prod_v A_v(j_f, i_\a) T_{\vec{j},\vec{i}}(\vec{U})
\ee 
in terms of normalized intertwiners $\langle i_\a,i_\a'\rangle=\delta_{i,i'}$. $\vec{U}$ are boundary SU(2) holonomies. All face amplitudes are $\dim(j_f)=2j_f+1$ at internal and boundary $f$. The boundary state (neglecting the contracted indices)
\be
T_{\vec{j},\vec{i}}(\vec{U})=\prod_{\text{boundary}\ f}R^{j_f}(U_f)\prod_{\text{boundary}\ \a} i_\a
\ee 
is the boundary spin-network basis whose normalization is given by 
\be
\langle R_{mn}^j,R_{m'n'}^{j'}\rangle=\frac{1}{\dim(j)}\delta_{j,j'}\delta_{mm'}\delta_{nn'} .
\ee 

In terms of coherent intertwiners,  
\be
\psi(\vec{U})=\sum_{\vec{j}}\prod_f\dim(j_f)\int\underline{\rmd}\vec{\xi}\prod_v A_v(\vec{j}, \vec{\xi}) T_{\vec{j},\vec{\xi}}(\vec{U})
\ee 
where $T_{\vec{j},\vec{\xi}}(\vec{U})$ is given by replacing $i_\a$ in $T_{\vec{j},\vec{i}}(\vec{U})$ with coherent intertwiners. But every integral $\int\underline{\rmd}{\xi}_{\a f}=\dim(j_f)\int\rmd \xi_{\a f}$ by the resolution of identity for coherent states $\dim(j)\int\rmd \xi |j,\xi\rangle\langle j,\xi|=1 $ where $\rmd\xi$ is the normalized measure on the unit sphere. $A(\ck)$ in Eq.\Ref{Z} computes the coefficients in front of $T_{\vec{j},\vec{\xi}}(\vec{U})$, so gives
\be
&A_f({j_f})=A_\Delta(j_f)=(2j_f+1)^{n_v(\Delta)+1}& \text{for internal $f$},\nonumber\\
&A_f({j_f})=A_\Delta(j_f)=(2j_f+1)^{n_v(\Delta)+2}& \text{for boundary $f$}.
\ee


\bibliographystyle{jhep}

\bibliography{muxin}

\end{document}